\documentclass[12pt, a4paper]{article}
\pdfoutput=1
\usepackage{amsmath,amssymb, graphicx} 
\usepackage[english]{babel}
\usepackage{bm,url}
\usepackage[margin=0.89in, centering]{geometry}
\usepackage{enumitem}

\usepackage[compact,small]{titlesec}
\usepackage{lmodern}
\usepackage{natbib}
\usepackage{nicefrac}
\usepackage{booktabs}
\usepackage{multirow, footnote}
\usepackage{color}
\usepackage[hidelinks]{hyperref}
\usepackage{authblk}
\usepackage{scrextend}
\usepackage[table]{xcolor}
\usepackage{longtable, caption, subcaption, array, threeparttable} 
\usepackage{adjustbox}
\usepackage{setspace}
\usepackage{gensymb}

\usepackage{tikz}
\usepackage{rotating}

\usepackage{setspace} 

\usepackage{mathtools}
\usepackage{blkarray, bigstrut}

\usepackage{multibib} 
\newcites{latex}{References}

\usetikzlibrary{mindmap}

\addto\extrasenglish{%
}

\usepackage{tabularx}

\usepackage{pifont}
\newcommand{\cmark}{\ding{51}}%
\newcommand{\xmark}{\ding{55}}%

\newtheorem{hypothesis}{Hypothesis}

\begin{document}

\title{A Framework to Monitor the Effects of External Shocks on Housing Markets}

\date{\vspace{-1.1cm}}

\author[a,b]{Anja M. Hahn}
\author[a,b]{Sanela Omerovic}
\author[c,a]{Sofie R. Waltl}
\affil[a]{Vienna University of Economics and Business, Department of Economics\\ 
Welthandelsplatz 1, 1020 Vienna, Austria}
\affil[b]{DataScience Service GmbH, Neubaugasse 56, 1070 Vienna, Austria}
\affil[c]{University of Cambridge, Department for Land Economy\\
17 Mill Lane, Cambridge CB21RX, UK}

\clearpage\maketitle
\thispagestyle{empty}

\begin{spacing}{0.9}

\begin{abstract}
We develop a framework to holistically test for and monitor the impact of different types of events affecting a country's housing market, yet originating from housing-external sources. We classify events along three dimensions leading to testable hypotheses: prices versus quantities, supply versus demand, and immediate versus gradually evolving. 
These dimensions translate into guidance about which data type, statistical measure and testing strategy should be used.
To perform such test suitable statistical models are needed which we implement as a hierarchical hedonic price model and a complementary count model. These models are amended by regime and contextual variables as suggested by our classification strategy.
We apply this framework to the Austrian real estate market together with three disruptive events triggered by the COVID-19 pandemic, a policy tightening mortgage lending standards, as well as the cost-of-living crisis that came along with increased financing costs.
The tests yield the expected results and, by that, some housing market puzzles are resolved. Deviating from the prior classification exercise means that some developments would have been undetected.
Further, adopting our framework consistently when performing empirical research on residential real estate would lead to better comparable research results and, by that, would allow researchers to draw meta-conclusions from the bulk of studies available across time and space.
\end{abstract}

\begin{footnotesize}
\end{footnotesize}

\vfill
\begin{footnotesize}
\textbf{Notes and Acknowledgements:} We gratefully acknowledge funding through the \emph{OeNB Anniversary Fund, Grant No. 18767 (LocHouse)} as well as data provision by \emph{DataScience Service GmbH}, \emph{Google Community Mobility Reports} and the \emph{Austrian Federal Ministry of Social Affairs, Health, Care and Consumer Protection}. We thank Wolfgang Brunauer, Rainer Schulz and Karin Wagner for useful comments and information, as well as participants in the workshop \emph{Residential Housing Markets: Perceptions and Measurement} at LISER, the \emph{2nd Workshop on Residential Housing Markets: Connecting Housing Researchers in Austria} at WU Vienna, the \emph{2022 NOeG Winter Workshop}, the \emph{29$^{th}$ European Real Estate Society (ERES) Annual Conference} at University College London, the \emph{2023 NOeG Annual Meeting} at the University of Salzburg, the \emph{Austrian Statistics Days 2023} at Statistics Austria, the \emph{14th Geoffrey J.D. Hewings Regional Economics Workshop} at the Austrian Institute of Economic Research (WIFO), the \emph{$(LIS)^2ER$ Workshop on Housing Policy and Wealth Inequality} at the Luxembourg Income Study (LIS), the \emph{ESCoE Conference on Economic Measurement 2024} in Manchester, the \emph{4th Workshop on Rent Control and Other Housing Policies} in Turin, the \emph{3rd Workshop on Residential Housing Markets: A Market in Distress and Potential Solutions} at WU Vienna, and the \emph{38th IARIW General Conference} in London, as well as seminar participants at the \emph{University of Aberdeen} and the \emph{University of Regensburg}.
\end{footnotesize}

\end{spacing}

\newpage

\doublespacing
\section{Introduction}\label{sec:intro}

Real estate is known for capitalisation effects of many sorts. For instance, a large body of literature finds capitalisation of location-bound changes relating to new local transport infrastructure \citep[see][for further references]{mcmillen2004reaction,sharma2018does}, amenities and dis-amenities of the distance to large-scale infrastructure like, e.g., railway stations, airports or waste disposal sites \citep[see, for instance,][for a meta-study of the effect of airport noise on house prices]{nelson2004meta} or location-restricted amenities like school zones \citep{black1999better, ries2010school}.

Related yet different are changes in the policy setting making some types of properties more or less valuable in terms of sales or rent value achievable on the market \citep[see, for instance,][]{sommer2018implications}.
Besides that, real estate is also an important investment asset and is by that, due to the common use of mortgages to finance purchases, closely related to capital market conditions \citep{adams2010macroeconomic}.

Further capitalization channels refer to intangible effects related to, for instance, changes in tastes for certain (locational) amenities \citep{chen2016changing, gupta2022flattening} or systematic behavioural changes as seen since the outbreak of the COVID-19 pandemic, which meant a sudden increase in mobile working \citep{wang2024hedonic}. 

Yet, no clear guidance is followed when testing for or tracking capitalization over time. This hampers comparative research across single studies and establishing sound stylized facts. 
In this article, we argue that considering the dimensions of data choice and the applied econometric testing technique is key to drawing sensible and accurate conclusions. Thus, we argue that these dimensions should form part of any identification strategy. Further, we complement the currently main modelling approach to describe changes in prices by a counterpart that models changes in the trade volume, i.e., quantity effects.

We therefore propose a framework differentiating shocks along three dimensions: prices versus quantities, immediate versus gradually evolving, and led by the supply (sellers) or demand side. These dimensions allow us to derive predictions about what type of model, measurement technique, and data is expected to first identify changes in the market.

\begin{figure}[h]
\begin{center}
  \caption{$(A)$ versus $(A^B)$}  \label{fig:index_ads_brokered_ads} 
  \includegraphics[width=0.56\textwidth]{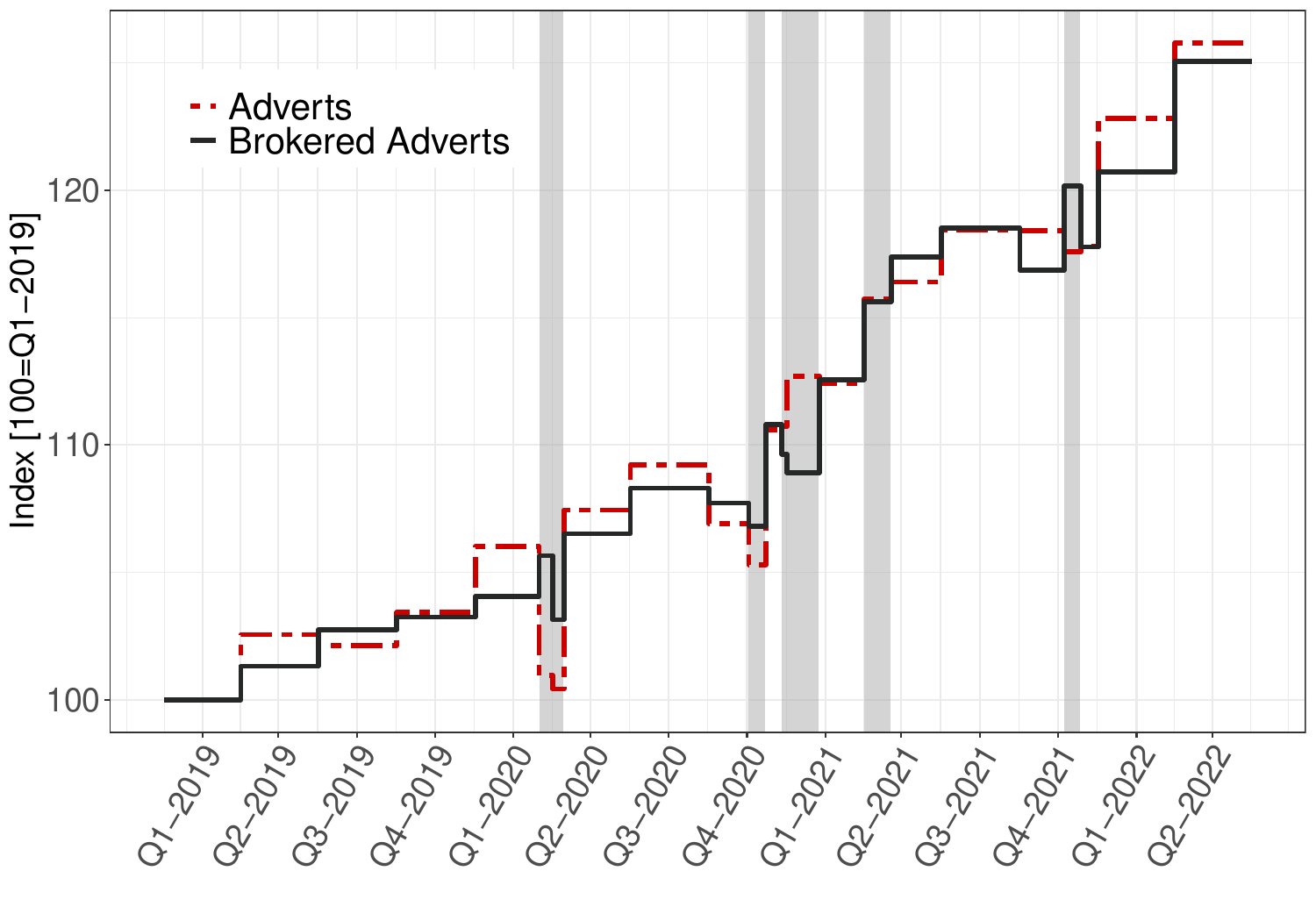}
    \end{center}
\begin{footnotesize}
    \emph{Notes:} The Figure depicts time dummy price indices for adverts $(A)$ versus brokered adverts $(A^B)$. We use quarterly time dummies including separate levels for lockdown periods.
\end{footnotesize}
\end{figure}

But, if not carefully considering these dimensions, what could actually go wrong? 
For instance, upon the outbreak of the COVID-19 pandemic, it was speculated which consequences this shock might have on different markets. With regard to the housing market, it was hypothesised that the first lockdown may constitute the prelude to a deep recession adversely also affecting housing markets. As visible in \autoref{fig:index_ads_brokered_ads} such negative sentiments are indeed clearly visible during the initial strict lockdown when assessing advertised prices: a severe price drop perfectly coinciding with the first publication date of advertisements suggests strong and immediate capitalisation of this information. However, if one would have looked at final transaction prices such a drop is visible too yet does not coincide with the initial stay-at-home ordinance and also appears much less severe. Thus, depending on which data was used to assess consequences of lockdowns on real estate prices, one may have come to different, if not even contradicting, conclusions. 

When assessing changes in tastes for certain property characteristics, we also need to distinguish between how sellers price in certain attributes and how buyers may do that. Such a change may not be immediately affecting prices but will over time as more and more buyers will bid for properties offering these more and more valued amenities. Testing for such changes using advertisement data is doomed to fail, as sellers only tend to react to market signals but do rarely initiate them. If, for instance, we wanted to test whether the pandemic experience had led to increased demand for properties in non-urban areas, we need to test for a gradual effect affecting both, final transaction prices and quantities.

If, on the other hand, a new policy tightening lending standards crowds out a share of potential buyers, this likely affects predominately the number of transactions and to a certain degree final prices, but -- at least in the short-run -- not at all quantities of advertised properties and advertised prices. 

Indeed, we find that the number of advertised properties, which purely measures supply-side activities, reacts only little on the announcement of stricter lending criteria, while the number of final transactions, which also reflects demand-side sentiments, reacts strongly. 

Standards may also help explaining some surprising or even counter-intuitive phenomena reported: For instance, \citelatex{stay-at-home-order} measure the effect of lockdowns (`stay-at-home orders') during the initial phase of the pandemic on final sales prices, new listings and time-on-market. While during lockdowns time-on-market is significantly longer and the number of new listings significantly reduced, there are virtually no differences on sales prices and the number of concluded deals -- both measured via final transactions data. This suggests that data choices matter and clearly tell different stories.

Guidelines, thus, would be important to improve approaches applied to test for capitalisation in housing markets and enable large-scale comparative research and gain meta-results that hold across time and space.

We apply our framework to the Austrian residential housing market covering the period between 1st of January 2019 and 31st of March 2023. During this period, the market was affected by mainly three shocks external to the housing market: the COVID-19 pandemic, the implementation of strict bank lending standards for private mortgages as well as the cost of living crisis also meaning higher financing costs. 
We then classify each shock along the three dimensions constituting our classification framework, and therefrom derive hypotheses.

For testing these hypotheses, we make use of a collection of real estate micro-data tracking different stages of the marketing and transaction process (adverts, brokered adverts and notary deeds), and representing the entire Austrian residential housing market. 

We develop a hedonic price model fit to model an entire country's housing market within a single model yet allowing for different price trajectories and variability by sub-markets. Thus, we model prices as hierarchical linear model with three nested levels: the individual property level, districts, and federal states. Additionally, we propose a count model complementary to hedonic price models used to test for quantity effects within a statistical model. We demonstrate that counts follow a Negative Binomial distribution suggesting a Generalized Linear Model (GLM). Further, we develop a suitable trigonometric transformation to capture seasonality in counts.
Both, the price and count models are subsequently amended with regime variables indicating for each specific event the start and end as well as any potentially important date in between, as well as contextual variables measuring an event's intensity over time and space. Contextual variables are chosen to correlate with either normative or positive behavioural aspects of events. 

While immediate effects are identified using main effects associated with regime or contextual variables, gradually evolving ones rely on segment-specific indices that measure deviations from a global trend for revealing heterogeneity across locations, dwelling characteristics, or price segments similar to attempts by \citelatex{francke2000efficient} and \citelatex{francke2004hierarchical} for variation across property characteristics and location, and \citelatex{mcmillen2014local} and \citelatex{waltl2019variation} for further modelling variation across price segments.

This set-up is fit to test for the manifold consequences of the three events presented for illustrating the methods and provides detailed insights into the Austrian housing market.

The remainder of this article is structured as follows: \autoref{sec:theory} develops our testing framework and derives hypotheses. Section \ref{sec:data} presents the data used and \autoref{sec:econometric_strategy} introduces the econometric modelling and the identification strategy applied. Section \ref{sec:empirical} carries out the empirical application, and \autoref{sec:conclusions} concludes. The appendix provides further details.

\section{Classification to Events and Hypotheses}\label{sec:theory}

\subsection{Classification of Events}\label{sec:classification}

We demonstrate positive and negative capitalisation effects related to reduced economic activity as a consequence of the pandemic, changes in affordability due to changes in bank lending criteria and increased cost-of-borrowing and non-housing consumption due to the inflationary period and associated interest hikes.

These three events mean a change in regime as they are longer lasting and fundamentally affecting market operations. Yet, they differ in \emph{how} exactly they impact real estate markets: First, it could be a change in actions set by \emph{owners/sellers} or \emph{buyers} which we label a supply or a demand channel. A second characterizing difference is the speed of capitalisation, which we classify in \emph{immediate} and \emph{gradually evolving}. Regarding the first, we follow \citelatex{anenberg2024volatility} and interpret the number of buyers and the number of sellers active on the market as fundamental measures of demand and supply.

In sum, we assess implications along the following three dimensions:
\begin{itemize}[noitemsep,topsep=0pt]
    \item[(PQ)] prices versus quantities,
    \item[(IG)] immediate versus gradually evolving effects and
    \item[(SD)] supply versus the demand side.
\end{itemize}

This leads to expectations about effects measurable via prices, quantities, or both.
Accordingly, \autoref{sec:pandemic}, \autoref{sec:BLS} and \autoref{sec:COL} discuss events in this light and derive event-specific hypotheses from this classification.

We translate this theoretical framework into concrete testing recommendations.
Generally, properties are observed on the market at two points in time, namely upon (initial) advertisement and upon transaction (see \autoref{sec:data} for details). Advertisements reflect supply-side sentiments, while data on registered transactions report the overlapping sentiments (if existent) of buyers (demand) and sellers (supply). Depending on which side of the market is immediately affected by a regime-switch, we thus also allude to the expected timing (immediate or gradual) and preferable data type for measuring the effect.

\begin{table}[h]
 \begin{center}
\caption{Summary of Testing Strategy}
    \label{tab:testing_strategy}
    \begin{small}
\begin{tabularx}{\textwidth}{l c ccc c X}
\toprule
\toprule
&&  \multicolumn{3}{c}{Data Type} && \multicolumn{1}{c}{Details}  \\
&& $(A)$ & $(D)$ & $(A^B)$       && \\
\cmidrule{3-5} \cmidrule{7-7}
Assessed Variable\\
\quad Quantities && (\checkmark) & \checkmark & . && $(D)$ constitute a complete repository of transactions while $(A^B)$ is a subset of transactions relying on real estate agents and is thus not considered. $(A)$ indicates the volume the supply-side is willing to transact. Nonetheless, changes in the number of records in $(A)$ and $(A^B)$ can be used as a relative measure of transaction volume.   \\
\quad Prices && \checkmark & . & \checkmark && $(A)$ reports supply side price expectations, $(A^B)$ reports realised market prices, i.e., joint supply- and demand-side sentiments. $(D)$ is ill-suited to track price effects as these data lack relevant hedonic characteristics. \\
Effect Type \\
\quad Immediate && \checkmark  & . & . && \small{Testing Strategy: Splitting of Time Dummies by Regime Variable(s)}\\
\quad Gradual   && . & \checkmark & \checkmark && \small{Testing Strategy: Interaction of Contextual Variable Measuring the Intensity with Time Dummies}\\
Leading Party \\
\quad Supply Side &&  \checkmark & (\checkmark) & (\checkmark) && Effects are immediately visible in $(A)$ and with a time-lag also in the other sources. The effect size should be more pronounced in $(A)$ than the other data sets.\\
\quad Demand Side &&  . & (\checkmark) & (\checkmark) && Effects are expected to be visible with a time lag regardless of which data was used; the lag should be larger for $(D)$ than for $(A^B)$.\\
\bottomrule
\bottomrule
\end{tabularx}
\end{small}
 \end{center}
     \footnotesize
\emph{Notes:} The table summaries the testing strategies applied. `\checkmark' indicates a preferable strategy, `(\checkmark)' a valid but sub-optimal strategy and `.' an unsuited strategy. 
\end{table}

Change driven by either the supply- or demand-side is associated with the different types of data available to us as again summarized in \autoref{tab:testing_strategy}. Details about each data set are reported in \autoref{sec:data}.

As indicated, implications on prices and quantities can both be measured using either advertisements $(A)$, brokered advertisements $(A^B)$ or notary deeds (´final transactions') $(D)$ in general, however there is a clear hierarchy in terms of suitability of each data set.
$(A)$ and $(A^B)$ are better suited for measuring price effects as they naturally contain an ample set of hedonic characteristics (used for marketing the property) which is crucial for filtering price effects net of the naturally varying mix of characteristics across time and space. $(D)$, in contrast, usually contains a very limited set of characteristics as information is extracted from registered purchase contracts.%
\footnote{Austrian civil law, however, does not require a purchase contract to contain a detailed description of the property and the principle ``freedom of contract'' generally allows the involved parties to freely agree on a contract format and clauses added. Also, notaries need not add substantial information content when registering the purchase in the land registry.}
This means that $(D)$ is not well suited to study price effects.

In contrary, $(D)$ is perfectly suited to measure quantity effects as the data constitute a complete count of all concluded transactions. $(A)$ and $(A^B)$, while not complete, can be used to measure relative changes in quantities. Thereby, $(A)$ measures quantity effects originating on the supply side whereas $(A^B)$ and $(D)$ again measure changes in overlapping supply- and demand-side sentiments.   

To measure immediate effects, $(A)$ is best suited as they report changes in the very first step of a transaction process \citep[see also][]{anenberg2017more}. $(A^B)$ and $(D)$ refer to the very last step and thus are affected by any changes in the market with a delay. Furthermore, $(D)$ is additionally delayed as the registration with notaries consumes time whereas real estate agents indicate the purchase upon offer acceptance.

Real estate can be used for measuring gradual effects by splitting time dummies according to the regime timing within the hedonic price model or by tracking counts over time. To adjust for seasonality in counts, we show counts relative to the value in the same quarter of the preceding year. Both approaches yield time-dependent (and thus potentially gradually evolving) regime effects. However, it is important to note that a longer delay is expected for $(D)$ as a consequence of lengthy transaction processes.
The data set $(A)$ emerges from the actions of willing sellers and thus reflects almost exclusively supply-side sentiments. $(D)$ and $(A^B)$, in contrast, reflect the common -- ``market'' -- sentiments.

In practise, our real estate data come with two shortcomings:
They do not objectively report changes exclusively stemming from the demand side.
Further, $(A)$ come with a rich set of hedonic characteristics, while $(D)$ report a sheer minimum demanded by notaries. However, $(A)$ only represents a subset of all transactions, namely those marketed online by agents.
To accommodate for some of these shortcomings, we introduce an additional intermediate data set that emerges when relying on annotations by real estate agents: they usually set the status of an advertisement to ``brokered'' when an agreement between the seller and a buyer has been achieved. While these data -- referred to by $(A^B)$ here -- contain all the characteristics extracted from advertisements that allows us to estimate stable price models, they do not feature the attractive property of completeness necessary for solid count models.

\subsection{The pandemic}\label{sec:pandemic}
The beginning of the COVID-19 pandemic meant a general negative shock, as homes ready to sell abruptly could no longer be matched to willing buyers and large uncertainty further decelerated market activity. The very beginning of the pandemic -- which we identify by the start of the first lockdown -- meant such a sudden freezing of housing markets globally. All steps of the transaction process were affected. 

This is also supported by stated expectations of brokers and real estate agents surveyed during the early period of the pandemic by the broker network RE/MAX Austria:%
\footnote{The survey was sent out on April 4, 2020 and participants had one week to respond. Across all questions between 351 and 355 responses were collected, which means an overall coverage rate of about 65\%. See \citelatex{immo_aktuell} for details.}
The vast majority reports a reduction of successful deals during the initial lockdown. As top reasons for this downturn, professionals name (i) the increased general insecurity, (ii) (obligatory) short-time work, unemployment or income losses of self-employed, and (iii) the impossibility of having in-person viewing appointments. 
Asked about their one-year-ahead sales and rent price expectations, the majority of professionals expect stagnant or even falling levels for all property types.
In the very long-run, there may be increased turnover due to the market catching up eventually.

In-line with these stated expectations, we predict a stark shrinkage of the number of transacted dwellings during the initial period. This concerns all types of dwellings and all steps of the transaction process. Thus, all three data sets $(A)$, $(D)$ and $(A^B)$ appear suited.

\begin{hypothesis}[Pandemic Quantity Effects] \label{hyp:quant.pandemic} 
Quantity effects triggered by dampened economic activity during the pandemic are expected to vary in the following way:
\begin{enumerate}[noitemsep,topsep=0pt]
    \item During the initial general lockdown following the break-out of the COVID-19 pandemic, the sudden slow-down of all human interactions is expected to lead to a significant shrinkage of successfully transacted dwellings. 
    \item Severe phases of the pandemic generally lead to the expectation of reduced economic activity and thus also reduced numbers of transactions in the housing market.
   While the adaptation of business modalities to a ``new normal'' means a \emph{weaker response to later virus-intense periods} following the initial lockdowns, the delays in all steps of the production and transaction processes are expected to lead to longer lasting lower numbers of realised transactions only fading out slowly. In the long-run, however, new demand and a catch-up process of foregone transactions may lead to increased volume.
\end{enumerate}
Quantity effects are comprehensively yet delayed measured by $(D)$. \emph{Changes} in transaction volume triggered by a general slowdown or a decrease in supply is immediately visible in $(A)$ and with a time lag also in $(D)$, while changes in demand is only visible in $(D)$.
\end{hypothesis}

Reduced economic activity during lockdowns may mean that only sellers in need of selling quickly actively participate in transactions. This implies price drops during such periods. The severity of such drops may diminish over time due to adaptation of business strategies. In contrast, the pandemic experience may have meant a change in needs and desires regarding one's housing situation and the lockdown period also meant high saving rates for the upper middle class suggesting increased demand following the initial COVID-shock.

During the initial lockdown, people spent much more time in their own premises than ever before. Meeting in public spaces as well as spending time at the work place was restricted meaning that many more activities than usual were carried out at home \citep[see, for example,][]{bick2023work}. While this may have meant a relaxing time for white-collar workers living in large houses, such a situation was more challenging for people living in small urban apartments potentially not even offering open space facilities and thus access to private recreational areas.
Even after the lockdown period, this change towards more flexible work patterns may turn out to become a permanent one \citep{phillips2020working} for some -- particularly white collar workers that can work (partly) remotely. In fact, in this sense the pandemic signalled the start of a new era:
For people preferring rural life moving out of the city became feasible. Such changes affecting living and working conditions may lead to a gradual shift in demand and thus also relative prices for amenities supporting new working styles and generally non-urban and less dense locations \citep[as documented for the US by][]{gupta2022flattening, liu2021impact}.

\begin{hypothesis}[Pandemic Price Effects] \label{hyp:prices.pandemic}
We distinguish between immediate and gradual price effects representing a slow-down of economic activity and a shift of preferences, respectively. We expect
\begin{enumerate}[noitemsep,topsep=0pt]
    \item immediate price drops during periods of restrictions or reduced activity $(A)$ and price increases thereafter.
    \item gradually increasing demand and thus increasing relative prices for apartments offering open space amenities $(A^B)$.
    \item gradually increasing relative prices for properties in non-urban areas due to increased demand $(A^B)$.
    \item gradually decreasing relative prices for studios and micro-apartments due to shrinking demand $(A^B)$.
\end{enumerate}
\end{hypothesis}

\subsection{Bank lending standards}\label{sec:BLS}
Ever since 2016, experts part of the \emph{Financial Market Stability Board (FSMG)} and the OeNB had inquired at the Ministry of Finance for thus far missing macro-prudential tools that would allow supervisors to monitor and -- if necessary -- intervene in lending practices to private households for the sake of preventing systemic risks potentially emerging in the course of a housing boom.%
\footnote{In 2016, the FSMG disseminated a report stating this inquiry for the first time: 
``[...] Based on analyses by the OeNB, a note was sent to the Federal Minister of Finance in spring to create new instruments. The FMSG sees the preventive expansion of the macroprudential toolbox, such as limits on the loan-to-value ratio, the debt-ratio or the debt-service ratio when granting new mortgages, as necessary to be able to act in the event of a real estate price boom that is fraught with systemic risks [...].'' 
\citep{FSMG2016}.} %
In 2018, concrete quantitative measures were communicated, which were -- however -- only recommendations but no enforceable regulations.  
The policy, labelled as \emph{Kreditinstitute-Immobilienfinanzierungsma\ss nahmen-Verordnung (KIM-VO)}, was finally announced during a press conference of the FSMG on 13 December 2021%
\footnote{See \url{https://www.fmsg.at/en/publications/press-releases/2021/30th-meeting.html}, last accessed in September 2024.} %
and enacted as of August 2022.%
\footnote{See the original law text from the Federal Gazette (BGBl. II Nr. 230/2022) under
\url{https://www.ris.bka.gv.at/Dokumente/BgblAuth/BGBLA_2022_II_230/BGBLA_2022_II_230.pdfsig}, last accessed in August 2023.} %
Concretely, the KIM-VO foresees 
\begin{itemize}[noitemsep,topsep=0pt]
    \item[(i)] a maximum mortgage duration of 35 years,
    \item[(ii)] a maximum loan-to-value ratio (LTV)%
\footnote{LTV is defined as the amount borrowed relative to the value of the property purchased.} %
of 90\% (meaning after fees and taxes roughly 80\% of the sales price), and
    \item[(iii)] a maximum debt-service ratio (DSR)%
\footnote{DSR is defined as the monthly amount of debt service payments (interest plus amortisations) relative to disposable household income.} of 40\%.
\end{itemize}

A household may thus be pushed off the housing ladder due to one or more constraints: the duration restriction sets an upper age limit, the LTV restriction requires a certain minimum wealth position upon mortgage application (wealth effect), and the DSR restriction demands a minimum current income (income effect). The three criteria jointly determine a household-specific maximum affordable house price and thus also an enforced (enlarged) restriction of the segment of the housing market a household can search in. 
Thus, we expect a drop in the number of concluded transactions following the enactment of this policy. This should have different implications across the price distribution as the restrictions are likely only binding for poorer households.

Forward-looking home-seekers may have hurried up securing a deal \emph{between} the announcement (December 2021)
and enactment (August 2022) of the policy. Once enacted, the number of concluded transactions is expected to gradually fall. The number of adverts may not necessarily decrease instantaneously as only the demand side is directly affected and advertising behaviour reflects predominantly supply-side beliefs and sentiments that may only adjust with a time-lag after observing the drop in demand.

Tighter lending standards mean that a positive share of potential home-buyers are excluded from the market and pushed off the housing ladder. These are those not fulfilling the lending criteria but previously would have had the chance to make use of the credit market more extensively to fulfil their envisioned home-ownership plans.
Thus, a smaller number of actors on the demand side is expected.

The lower share of actors also means less competition between prospective buyers in the medium to long term. Thus, prospective buyers are not expected to bid up prices but may even be able to negotiate a lower price with sellers in need of selling fast. Thus, after an expected initial immediate stagnation of overall price levels, this period is likely followed by falling prices, which in turn is only dampened by sellers' loss aversion. The latter is indeed expected to have significant downwards cushioning effects as private sellers in the housing market as well as developers have been found to be particularly strongly affected by loss aversion and downward price corrections may thus only be observed with a delay \citep[see][for a survey]{bao2017loss}.

\begin{hypothesis}[Bank Lending Standards Quantity Effects] \label{hyp:quant.lending} 
Tightened requirements to obtain a mortgage mean that the group of buyers eligible for a mortgage financing the purchase shrinks. This concerns both, potential buyers lacking sufficient wealth to meet the LTV requirements, and/or sufficient income meeting the DSR requirements within the maximum duration allowed.
Thus, $(D)$ are expected to gradually fall.
    \begin{itemize}[noitemsep,topsep=0pt]
        \item As stricter lending standards set an upper limit to prices affordable to prospective buyers that need a mortgage to finance the purchase. This means overall decreased demand and hence decreased transactions. 
        \item Heterogeneity across the price distribution may emerge as properties targeted to lower classes may be transacted less frequently while properties at the upper end of the price distribution may be effected to a lesser extend.
    \end{itemize}
\end{hypothesis}

\begin{hypothesis}[Bank Lending Standards Price Effects] \label{hyp:price.lending} Price effects triggered by changed bank lending standards are the consequence of crowding-out effects: A smaller number of actors is bidding for dwellings. 
This shift in the market power of the demand side mechanically leads to a gradual decrease in prices measured via $(A^B)$ and to a lesser extend $(A)$. Again, only households in need of a mortgage are concerned. Thus, real estate at the upper end of the price distribution targeted to wealthy households may be less affected by this restriction.   
\end{hypothesis}

\subsection{Cost of Living Crisis}\label{sec:COL}

The inflationary period starting in mid-2021 meant both, a tighter budget for consumers as well as a general more pessimistic economic outlook. 
Regarding consumers' tighter budget, Austrian labour law and institutional practise generally foresees compensation for a loss in purchasing power of consumers via wage adjustments. This is implemented in the Austrian wage-adjustment mechanisms: once per year, adjustments to wages are collectively negotiated between representatives of employers and employees separately by industry. Usually, the average CPI-inflation over the \emph{past} 12 months acts as a benchmark. Hence, employees and workers effectively lose purchasing power with every additional month the current inflation rate exceeds wage increases in the same month \citep[see][for details]{inflation_wages}.
A more pessimistic outlook is confirmed by survey data \citep[\emph{Austrian Corona Panel Project (ACPP)} reported by][]{Corona_blog}: Between October 2021 and March 2022 the share of people stating a negative outlook increased by 25pp from 37\% to 62\%.%
\footnote{Economic outlooks were measured via the following question: ``Compared to the current situation, how will the economic situation in Austria develop in the near future?'' For measuring a negative outlook, we combined respondents stating that they expect slightly or much worse economic situation within the upcoming 12 months.
} 

In such an environment, large and difficult-to-reverse investments may be postponed. Purchasing a dwelling is indeed such a large and often once-in-a-lifetime purchase. As \citelatex{knotek2011households} put it: ``While households know their current income and wealth, the uncertainty shock may lead them to believe that the distribution of future income and wealth prospects favours more extreme outcomes [...]. As a result, the uncertainty shock amplifies households’ caution, making them less likely to act. That is, the real option value of waiting temporarily increases. As a consequence, households optimally respond by reducing their purchases of houses and durable goods below normal levels.''

In addition to a change in sentiments and budgetary restrictions, mortgage costs also rose as inflation-targeting central banks (including the ECB) started to raise policy rates to dampen inflationary pressure. This triggers a credit channel effect \citep{bernanke1995inside, iacoviello2008credit}: Through the monetary transmission process, policy rates were gradually passed on to private lenders, thus increasing mortgage rates. Further, the deterioration of asset values meaning devaluation of collateral makes mortgages even more expensive and less available to households. As the majority of housing purchases make use of mortgages,%
\footnote{\label{foot:albacete}\citelatex{albacete2009} use survey data to estimate the importance of external financing for the purchase of a main residence in Austria: while roughly a quarter of owner-occupiers acquired their home without external financing, the vast majority (74\%) relied at least partly on borrowing for financing the purchase.} %
the demand for housing decreases as fewer households can access sufficient credit \citep{mishkin2007}, which puts downwards pressure on house prices. 
Both described mechanisms mean decreased demand, i.e., effects led by the demand side. 

Private owners in particular have been shown to be affected by (nominal) loss aversion and adopting high reservation prices, which they are often unwilling to underscore. In other words, willingness-to-pay and willingness-to-accept diverge leading to fewer transactions and downward price stickiness, i.e., the downward price pressure may not be passed through during an economic downturn. These mechanisms have been described and confirmed in the theoretical and empirical literature \citep{adams2010macroeconomic}.

A technical detail in the way inflation is measured in Europe, namely the exclusion of owner-occupied housing costs in the harmonised index of consumer prices (HICP) as discussed in \citelatex{hill2023owner}, would further lead to the assumption that in periods with large volatility in house prices, a RPPI would lead the HICP and particularly turning points may be first visible in RPPI and only thereafter in inflation rates.

As discussed above, we have no data at our disposal directly reflecting changes in demand-side sentiments. Translating this to the data types that are available to us, we would expect no price effects in $(A)$, and no or only small effects in $(A^B)$ reflecting joint buyer-seller sentiments. As discussed in \citelatex{adams2010macroeconomic} the price effect may only emerge in real yet not nominal prices.
Similarly, we expect strong quantity effects measured by data reflecting joint buyer-seller decision making, i.e., $(D)$. In contrast, the pure supply-side quantity measure $(A)$ is expected to be much less or even not at all responsive to rising inflation. 
All these considerations lead to \autoref{hyp:quantities.inflation} and \autoref{hyp:prices.inflation}.

\begin{hypothesis}[Cost of Living Crisis Quantity Effects]\label{hyp:quantities.inflation}
Rising inflation and interest rates are expected to be only indirectly observable due to a lack of direct information on changes in buyers' behaviour. Concretely, we expect
\begin{enumerate}[noitemsep,topsep=0pt]
    \item drops in concluded transactions $(D)$.
    \item no or only small decreases in the number of advertisements $(A)$.
 \item that declines in $(D)$ are negatively correlated with changes in the consumer price index (inflation).
    \item that declines in $(D)$ are negatively correlated with interest hikes by the ECB as well as realised mortgage interest rates.
\end{enumerate}
\end{hypothesis}

\begin{hypothesis}[Cost of Living Crisis Price Effects] \label{hyp:prices.inflation}
Effects are expected to be visible with a time-lag due to a lag of direct information on changes in buyers' behaviour. We thus rely, as a second-best option, on $(A^B)$ yet expect changes to be visible with a lag. Concretely, we expect
\begin{enumerate}[noitemsep,topsep=0pt]
    \item stagnation or even drops in prices once interest rates started to rise $(A^B)$. The effect is weaker or even absent for $(A)$.
        \item that gradually evolving slowdown or stagnation measured using $(A^B)$ is negatively correlated with mortgage rates from the start of interest hikes onwards. The effect is weaker or even absent for $(A)$.%
    \item that the gradually evolving slowdown or stagnation measured using $(A^B)$ leads changes in the consumer price index from the start of interest hikes onwards. The effect is weaker or even absent for $(A)$.

\end{enumerate}
\end{hypothesis}

\section{Data}
\label{sec:data}
\subsection{Residential Real Estate Data}\label{sec:data.RE}
\begin{figure}[h]
    \centering
        \caption{Data Overview}
    \label{fig:data_overview}
    \includegraphics[width=0.75\textwidth]{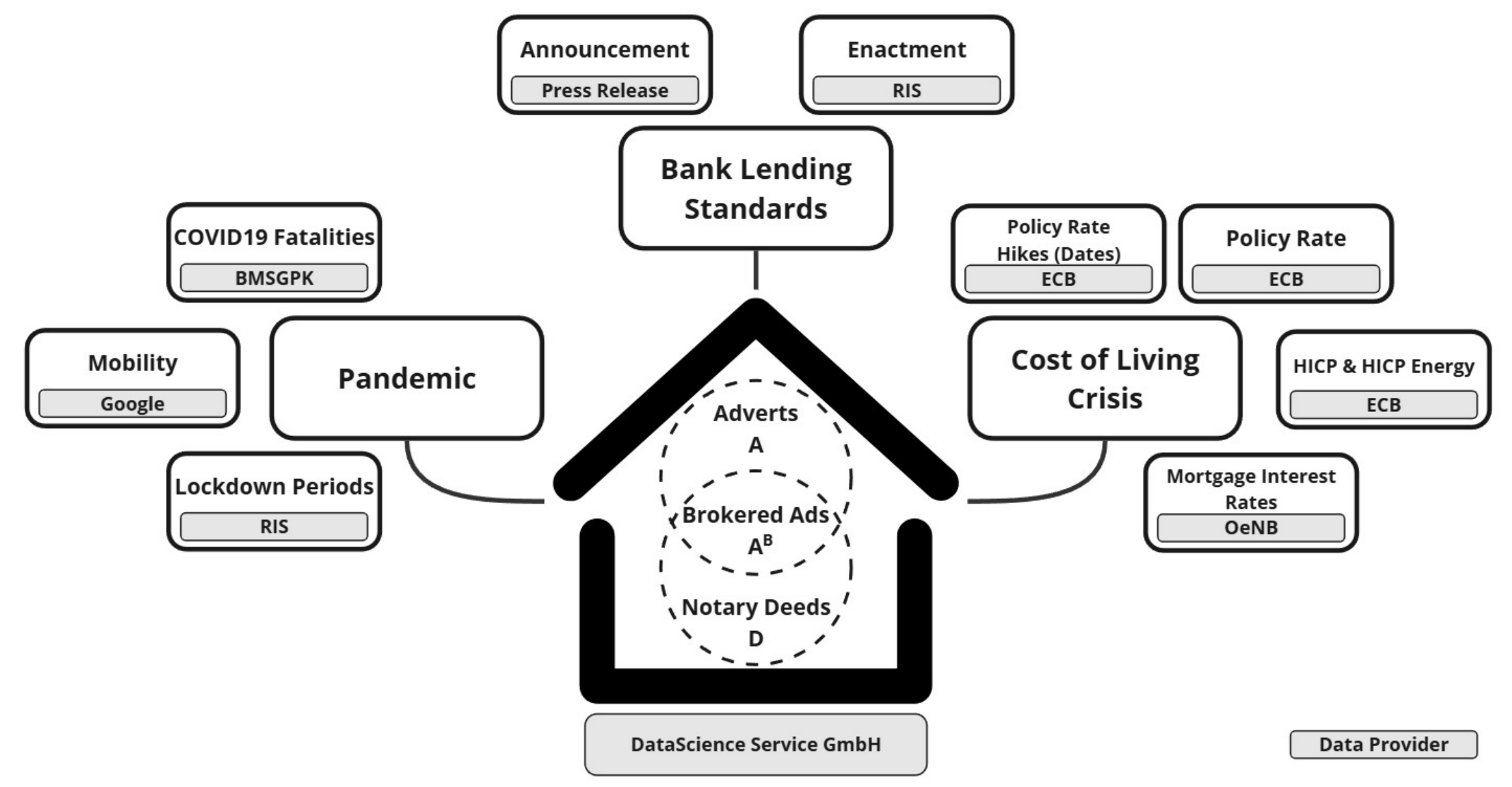}
\end{figure}

\autoref{fig:data_overview} provides an overview of all data used which include three types of real estate data and several external sources for identifying events.
For our price model, we rely on two real estate data sets that both come with ample hedonic characteristics: adverts $(A)$ and brokered adverts $(A^B)$. While the first is a classic collection of web adverts, the second one is based on the first one yet comes with an identifier set by real estate agents upon successful sale. Thus, the data is representative for brokered sales (see \autoref{sec:data.brokers}) yet not the universe of transactions.
The count model aims to measure total market activity (``trade volume''), and therefore completeness is crucial. This is achieved by sourcing data from the land registry maintained by district courts and documenting all properties in Austria together with associated ownership rights. The data $(D)$ is summarised in \autoref{sec:data.contracts}.

\subsubsection{$(A)$ and $(A^B)$: Brokers' Database}\label{sec:data.brokers}

We source data collected by the real estate data company \textit{DataScience Service GmbH} maintaining entries of real estate agents into an online system used for advertising properties. Once a deal is sealed, agents would set an advert as `brokered' and the final price is added. Thus, all hedonic characteristics having been added to an advert are recovered and linkable to the transaction. 

We divide the raw data set into adverts $(A)$ and brokered adverts $(A^B)$ based on the type of price (advertised vs. realised) and date (date of initial advertisement vs. date of agreement).
The data comes with a comprehensive set of standard hedonic variables. Summary statistics are reported in \autoref{tab:summarystats_cat} and \autoref{tab:summarystats_num} in the Appendix. 
We trim the data at EUR 5 Mio. and EUR 100,000.  
This yields 66,529 observations for $(A)$ and 40,182 observations for $(A^B)$.  
Across all data used, the number of observations per state and year are strongly positively correlated, and numbers also correlate with population counts as shown in \autoref{tab:dat_corr}. Both findings suggest no geographical selection bias.

\begin{table}[ht]
\begin{center}
\caption{Data Coverage: Pearson Correlation Coefficients}
\label{tab:dat_corr}
\begin{tabular}{r | rrrr|}
\multicolumn{1}{c}{ } & $(D)$ & $(A^B)$ & $(A)$ & \multicolumn{1}{c}{Population} \\ 
  \cline{2-5}
$(D)$ & -- & $0.91^{***}$ & $0.78^{***}$ & $0.70^{***}$ \\ 
  $(A^B)$ & $0.91^{***}$ & -- & $0.94^{***}$ & $0.75^{***}$ \\ 
  $(A)$ & $0.78^{***}$ & $0.94^{***}$ & -- & $0.68^{***}$ \\ 
  Population & $0.70^{***}$ & $0.75^{***}$ & $0.68^{***}$ & -- \\ 
   \cline{2-5}
\end{tabular}
\end{center}
\begin{footnotesize}
    \emph{Notes:} The table reports Pearson correlation coefficients between data counts of our real estate data sets -- notary deeds ($D$), brokered adverts ($A^B$), adverts ($A$) -- and population data (number of persons registered). The numbers of observations per year and federal state level are compared. Significance of the two tailed correlation test is indicated as follows: $^{***}p<0.001$; $^{**}p<0.01$; $^{*}p<0.05$.
\end{footnotesize}
\end{table}

The relative shares of observations per state are similar for $(A)$ and $(A^B)$, however there are noticeable shifts: While $50.5\%$ of $(A)$ are observed in Austria's single-largest city Vienna, this share amounts to just $31.9\%$ in $(A^B)$. In contrast, in all other states the shares observed in $(A^B)$ are larger than in $(A)$. In-line with that, $(A)$ contains disproportionally more apartments, more units located in areas classified as ``urban'' and on average smaller units (see \autoref{tab:summarystats_cat} and \autoref{tab:summarystats_num}). Thus, all models contain several locational identifiers and the price model further allows for heterogeneity in variances across locations.

\subsubsection{$(D)$: Notary Deeds}\label{sec:data.contracts}
The notary deeds collected in the land registry \emph{per construction} cover the entire market. We access these data via \textit{DataScience Service GmbH}. 
As with most land registries, the deeds contain only very few hedonic characteristics as purchase contracts are not standardised and attributes are not systematically listed. Still, the data entail information on officially recorded realised sales prices, the date of contract signing, the type of the dwelling (single-family house or apartment), and the address. \autoref{tab:summary_deeds} reports summary statistics. In-line with $(A)$ and $(A^B)$, we trim deeds at EUR 100,000 and EUR 5 Mio.
Over the period of analysis (1 January 2019 -- 31 March 2023), there were 264,257 transactions (apartments incl. duplexes and terraced houses, and single-family houses) recorded, thus substantially exceeding numbers in $(A)$ and $(A^B)$. 

Population size appears to be a main driver of the number of deeds. Correlating the number recorded to population size yields a Pearson correlation coefficient of $0.70^{***}$ as reported in \autoref{tab:dat_corr} and graphically shown \autoref{fig:map_deeds_pop} in the Appendix.

While the registry is complete, price information is not timely. This is a consequence of the fact that a purchase is registered after price negotiations and the time lag could sometimes be quite large. 
More importantly, deeds also contain transactions under favourable conditions occurring between relatives and friends. As Austria collects a property transfer tax with the recorded price as tax base, there are strong incentives to keep the recorded price low. \citelatex{kopczuk2015mansion} show evidence that a transfer taxes may introduce market distortions via price bunching just below the tax threshold. In Austria there is no such threshold, but the total tax amount proportionally increases with the recorded purchase price. Thus, while we expect no bunching effects, we cannot exclude lower reported prices.
To avoid distortions by any such effects, we exclusively use deeds for measuring quantity effects but not for prices.

\subsection{External Data}\label{sec:data.external}

\begin{figure}[h]
\begin{center}
\captionsetup{font=large}
\caption{Regimes in Austria} \label{fig:regimes} 
\includegraphics[width=0.8\textwidth]{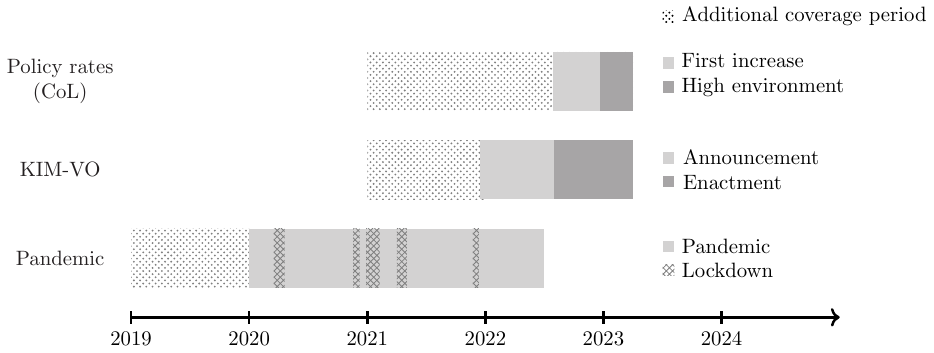}
\end{center}
\begin{scriptsize}
\emph{Notes:} The figure shows the three regimes, including the timing of lockdowns on federal or state level. In addition, we only account for strict lockdowns, i.e., we do not consider light lockdowns or lockdowns applying only to unvaccinated residents. Exact dates are reported in \autoref{tab:timeline} and \autoref{tab:lockdown_periods}.
\end{scriptsize}
\end{figure}

Regimes are determined by normative actions and lead to a start and end date as well as potentially a further distinctive intermediate date (\autoref{fig:regimes}).
To identify the intensity of each event over time and space, we further use contextual variables that correlate with some aspect of the event as summarized in \autoref{tab:external_data}.

\subsubsection{Pandemic}\label{sec:data.pandemic}

Normatively, we rely on COVID-19 regulations (i.e., lockdowns). 
The positive viewpoint, by contrast, focuses on what is observable as a response by assessing people's behaviour; precisely, mobility patterns.
Furthermore, we add measures of the intensity of the pandemic in the form of the number of fatalities caused by COVID-19.

The \emph{normative} perspective allows us to look at defined periods characterised by strong interventions in public life. The start and end of restrictive regulations were published in the Federal Gazette as COVID-19 Protective Measures Ordinances (``COVID-19-Schutzmaßnahmen\-verordnungen'') and COVID-19 Relaxation Ordinances (``COVID-19-Lockerungsverordnung''), respectively, and summarised in \citelatex{stoeger2021}.
We link real estate data to the timing and geographical coverage of lockdown mandates (see \autoref{tab:lockdown_periods} and \autoref{fig:regimes}). This yields indicators per lockdown, which can also be pooled to an overall lockdown variable.

Mobility is tracked by \textit{Google Mobility Data}%
\footnote{See \url{https://www.google.com/covid19/mobility/} (last accessed in November 2022) for more information. We access these data via the \textit{COVID-19 Data Hub} \citep{guidotti2020}, see \url{https://covid19datahub.io} (last accessed in November 2022).} %
on a daily basis by districts. We use data on mobility related to workplace travel, which is measured as a percentage change from a respective baseline weekday (the median between 3 January 2020 and 6 February 2020). Since these separate baselines do not account for possibly inherent differences between weekdays, we compute a moving average with a 15-day window.

The first COVID-19 related death in Austria was reported on 12 March 2020. Thus, we set fatalities to zero before this date and thereafter count the number of new fatalities per 1,000 inhabitants per district provided by the \textit{Austrian Federal Ministry of Social Affairs, Health, Care and Consumer Protection} and accessed via the \textit{COVID-19 Data Hub}.

\begin{figure}[h]
\begin{center}
            \caption{Google Workplace Mobility and COVID-19 Fatalities}
    \label{fig:mobility}
    \includegraphics[width=0.8\textwidth]{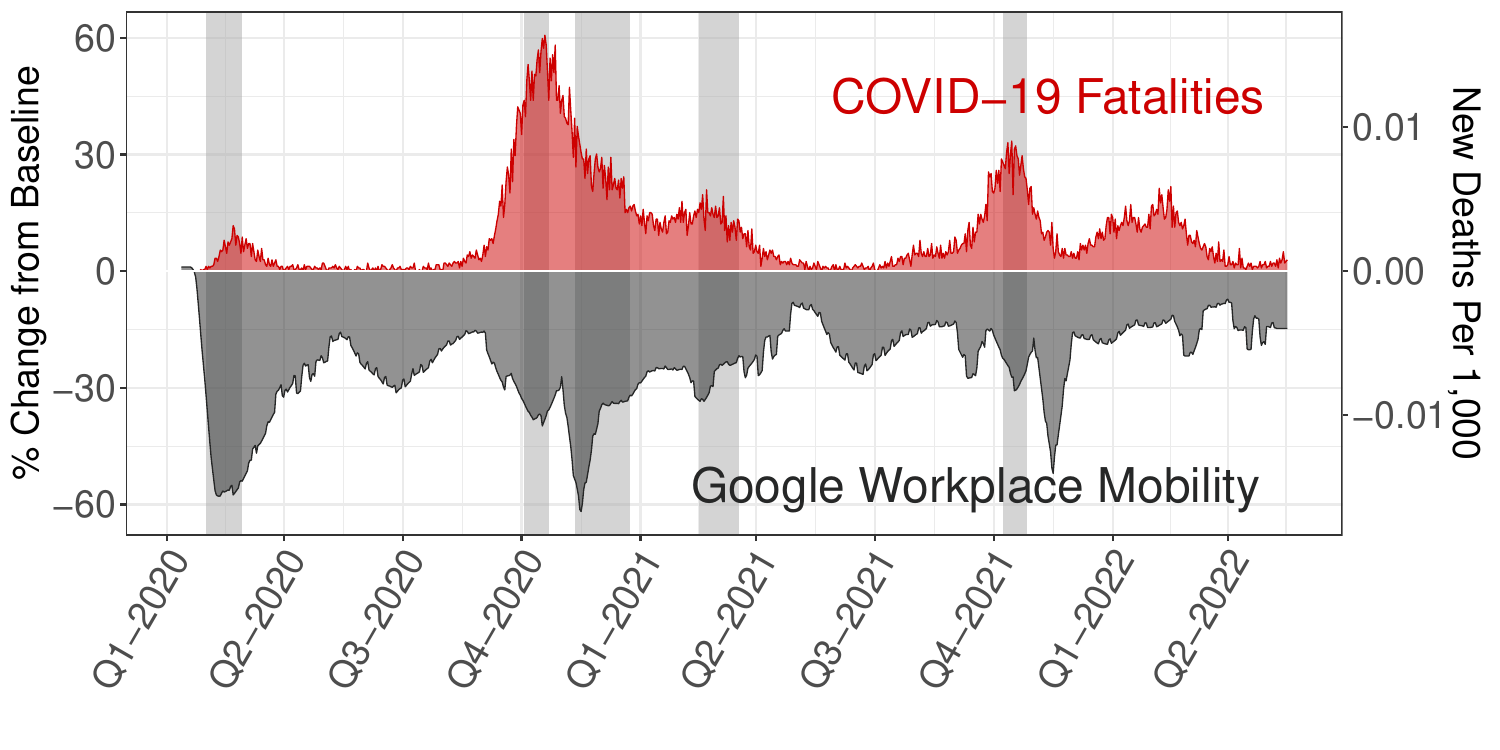}
        \end{center}
\begin{footnotesize}
    \emph{Notes:} The figure shows a 15-days window moving average of Google workplace mobility and the daily counts of new COVID-19-related deaths. Lockdown periods are highlighted.\\
    \emph{Source:} Google and Austrian Federal Ministry of Social Affairs, Health, Care and Consumer Protection
\end{footnotesize}
\end{figure}

\subsubsection{Bank Lending Standards}\label{sec:data.bank}

To identify effects of the newly introduced bank lending standards, we rely on the policy's announcement and enactment dates as normative measures to create three periods: the pre-announcement, announcement and enactment period.

\subsubsection{Cost of Living Crisis}\label{sec:data.inflation}
In mid 2022, prices started to rise globally and, as wages are not immediately and fully adjusted in most countries including Austria, these rises meant increased cost-of-living. Monetary policy reactions also increased the cost of lending and by that affecting home-buyers.

As positive measure, we use changes in the Austrian HICP, i.e., inflation rates. 
As a further positive measure, we make use of the average realised mortgage rates reported by commercial banks to the OeNB.%
\footnote{See \url{https://www.oenb.at/isawebstat/stabfrage/createReport?lang=EN&report=2.10}, last accessed in February 2024.}
Mortgage rates are volume-weighted, published on a monthly basis and can therefore be matched to our real estate data based again on time.

In response to surging prices, the ECB Governing Council announced a first rise in policy rates on 27 July 2022.%
\footnote{See \url{https://www.ecb.europa.eu/stats/policy_and_exchange_rates/key_ecb_interest_rates/html/index.en.html}, last accessed in March 2024.} 
By that, the rate on the deposit facility turned positive for the first time after more than eight years. From a normative perspective, this date thus marks the beginning of exacerbated financing costs and we interpret this date as regime start. 
Following this initial rise, rates were further stepwise increased and triggered rising mortgage rates charged by commercial banks and credit institutions. In accordance, we use the policy rate%
\footnote{A time lag is expected due to the monetary transmission process, but also, as discussed in \autoref{sec:classification}, due to the fact that this change first affects the demand side and the willingness of the supply side to adjust prices downwards may mean a delay in data reflecting the sentiments of the supply side $(A)$ and joint sentiments $(A^B)$. An empirical assessment suggests a three-months lag.} %
to normatively differentiate between different regimes (low, medium and high interest environment defined using the ECB's policy rate) and use average realised mortgage rates as a positive contextual quantity.

\section{Econometric Modelling Strategy}
\label{sec:econometric_strategy}

\subsection{Quasi-Hedonic Count Model}\label{sec:count_model}
We model weekly counts per object $Type$ (houses versus apartments), federal state $Loc1$, and the urban-rural classification $Loc2$ (see Appendix \ref{sec:stataut}) leading to $2 \times 9 \times 4 = 72$ cells.
Federal states represent a single spatial unit, which we further divide into finer locational clusters capturing within-state variation related to the degree of urbanisation, differences in the prevalent housing type, accessibility, or tourism hot spots.

As we model counts, we consider a \emph{Negative Binomial distribution (NB)}, which also captures overdispersion in the data.%
\footnote{Compared to the simplest count distribution -- the Poisson distribution -- the Negative Binomial distribution provides more flexibility emerging from two separable parameters.} %
Let $y_{t^w_i} \sim NB(\mu_i, a)$ be the number of weekly counts with mean $\mu_i := \mu_{t^w_i}$, time steps $t^w_i = t^w_1, t^w_2,\ldots$ and probability mass function 
\begin{equation}
    \mathbb{P}[y_{t^w_i} = k] = \frac{\Gamma(a+k)}{\Gamma(a) k!} \left(\frac{\mu_i}{\mu_i + a}\right)^k \left(\frac{a}{\mu_i + a}\right)^a,
\end{equation}
whereby $\Gamma(\cdot)$ denotes the Gamma function.

 As we naturally observe more housing transactions in more densely populated federal states (see \autoref{sec:data}), we align mean counts by including the population count per federal state $Pop^{Loc1}$ as exposure variable (see \autoref{sec:stataut}).

We model the logarithmic mean of $y_{t^w_i}$ divided by population totals via
\begin{equation}\label{quantmod}
    \log\left(\frac{\mu_i}{Pop_i^{Loc1}} \right) = \beta_0 + f_{cyclical}(\tilde{\beta}_1, \tilde{\beta}_2)+ 
    \beta_3 Time_i + \beta_4 Type_i + \beta_5 Loc1_i + \beta_6 Loc2_i.
\end{equation}
As regressors we use the categorical variables $Type$, $Loc1$, $Loc2$, time dummies $Time$ and a seasonal trend.
While seasonality in house prices is rare and subject to specific settings \citep[see, e.g.,][]{roed2024house, harding2003estimating}, the number of transactions has been shown to exhibit cyclical patterns \citep[see also][]{ngai2014hot}. This is also the case in Austria. Thus, we suggest including a periodic term defined via trigonometric functions for modelling such type of seasonality. 
Therefore, we consider a cosine function with amplitude $\alpha$ and shift parameter $\theta$ which we transform to a linear combination of a sine and a cosine function suitable to be added to a linear regression model, i.e.,  
\begin{eqnarray}
    f(t) &=& \alpha \cos\left( \frac{2\pi t}{T} - \theta \right) 
         = \underbrace{\alpha \cos \theta}_{=:\beta_1} \cos \left(\frac{2\pi t}{T}\right) + \underbrace{\alpha \sin \theta}_{=:\beta_2} \sin \left(\frac{2\pi t}{T}\right) \nonumber\\
         &=& \beta_1 \cos \left(\frac{2\pi t}{T}\right) + \beta_2 \sin \left(\frac{2\pi t}{T}\right) =: f_{cyclical}(\beta_1, \beta_2).  \label{eqn:cyc}
\end{eqnarray}
Extreme values are derived by $t^{max} = \nicefrac{T}{2\pi t} \cdot \arctan \left(\nicefrac{\beta_1}{\beta_2} \right)$ and $t^{min} = t^{max} + \theta$. 
To capture usual cyclical patterns and to prevent distortions from crisis regimes (see \autoref{sec:econometric_strategy}), we estimate the cyclical trend only in our base model and add the estimated trend in the form of an offset to all further models. We denote the estimated function by $f_{cyclical}(\tilde{\beta}_1, \tilde{\beta}_2)$ (see estimated coefficients in \autoref{tab:baseline_count}).

\subsection{Hierarchical Hedonic Price Model}\label{sec:hierarchical_model}

We estimate a hedonic equation as hierarchical linear model%
\footnote{Other commonly used terms for hierarchical (linear) models are (linear) mixed-effects models, mixed models or multilevel models. We stick to the term ``hierarchical model'' since this explicitly denotes the hierarchical (nested) data structure used.} 
\citep{raudenbush_bryk_hierarchical}
to capture the locational heterogeneity of the entire Austrian housing market.
Given the nested structure -- the national level is divided into federal states made up of districts -- the hierarchical model appears as natural choice for a country-wide model as each hierarchal cluster is supposed to share common features likely influencing prices -- be it due to geographical, political or legal specificities.
We model this structure by \textit{random effects} as source of variation, which are held constant at every entity yet vary between them. Thus, we account for the spatial heterogeneity of the country.
This set-up leads to a single testing equation yet controls for the rich variation in physical characteristics and geography informed by three administrative hierarchies.%
\footnote{While being somewhat neglected in economics, these models are popular in other quantitative fields such as psychology, medical and educational sciences \citep[][]{mcneish2019fixed}.}

The hierarchical model brings about the desirable feature of ``partial pooling,'' i.e., a balance between treating each geographical unit entirely independently and pooling all observations together. As such, not the cluster parameters themselves but their variance is estimated.

Precisely, we regress the logged price $p_{i,d,s}$ of dwelling $i$ located in district $d$ in federal state $s$ on a set of hedonic characteristics \citep{Rosen1974}, as well as regime and contextual variables used for identification. Naturally, hedonic characteristics are mainly measured at the property-level while regime and contextual variables are measured on the state- or federal level. We subsume both type of characteristics in matrices $X$. This yields model equations per hierarchy:
$\log p_{ids} = \beta_{0ds} +  \mathbf{X}_{1ids} \beta_{1ds} + \varepsilon_{0ids}$ for the property level, 
$\beta_{0ds} = \beta_{0s} +  \mathbf{X}_{2ds} \beta_{2s}  + \varepsilon_{0ds}$ for the districts level, and $\beta_{0s} = \beta_{0} +  \mathbf{X}_{3s} \beta_{3} + \varepsilon_{0s}$ for federal-states with associated error terms $\varepsilon_{0ids} \sim \mathcal{N}(0, \sigma^2_{\varepsilon_{0ids}})$, $\varepsilon_{0ds} \sim \mathcal{N}(0, \sigma^2_{\varepsilon_{0ds}})$ and $\varepsilon_{0s} \sim \mathcal{N}(0, \sigma^2_{\varepsilon_{0s}})$.
This set-up collapses to the single model equation
\begin{equation}\label{mod:model_equation}
\log p_{ids} = \beta_{0} +   \mathbf{X}_{1ids} \beta_{1ds} 
+  \mathbf{X}_{2ds} \beta_{2ds}  + 
\mathbf{X}_{3s} \beta_{3s} 
+  \varepsilon_i,
\end{equation}
with $\varepsilon_i = \varepsilon_{0ids} + \varepsilon_{0ds}  + \varepsilon_{0s}$. Per assumption, the error terms are independent and thus $\varepsilon_i \sim  \mathcal{N}(0, \sigma^2_{\varepsilon_{0ids}} + \sigma^2_{\varepsilon_{0ds}} +  \sigma^2_{\varepsilon_{0s}})$.

By not simply modelling locational variation as main effects but hierarchically, the model is fit to capture different sub-markets that may differ in both price levels and variability. 
This is suitable for our price model, as both general locational effects, as well as the degree and timing of consequences of events or regulations vary strongly across administrative boundaries. This is rather obvious in the case of the pandemic as the number of fatalities varies with location and also containment measures were partly implemented on an administrative regional level. While national policies and overall inflation do not vary cross-sectionally, their interaction with pre-existing boundaries of sub-markets may lead to differential impacts. The importance of these hierarchies are evident when assessing differences in price between federal states and districts as shown in \autoref{fig:price_bl}.

\subsection{Econometric Strategies for Identifying Effects within Hedonic Price and Quantity Models}\label{sec:identification}

For identifying effects within our price and count models, we make use of contextual variables presented in \autoref{sec:contextual}. Further, each crisis is associated with a specific time period identified using normative measures as discussed in \autoref{sec:timing}. These time-periods are differentiated using regime-identifiers.

\subsubsection{Contextual Event Variables}\label{sec:contextual}

Reconsidering the three dimensions for classifying events introduced in \autoref{sec:theory}, we discuss how to translate this scheme into econometric tests.  
First, dimension \textit{(PQ) - prices versus quantities} demands having at hand equal tools for testing price and quantity effects. This is achieved by separately modelling price and quantity dynamics for the entire country over time. 
For price effects, we make use of a classical hedonic pricing model  (developed and adapted to fit the Austrian context as presented in \autoref{sec:hierarchical_model}) that is amended by steering variables describing different kinds of events. Hedonic variables thus take over the role of controls, while the added contextual variables, that are selected to correlate with each event, take over the role of steering variables. The hierarchical structure allows us to simultaneously model movements for the entire country encompassing specificities of local housing markets. 

\begin{table}[h]
    \begin{center}
    \caption{Event Identifiers}
    \label{tab:external_data}
     \fontsize{9pt}{9pt}\selectfont
    \begin{tabularx}{\textwidth}{X c XXcX}
    \toprule
    \toprule
        Event && Indicator & Description & Type & Source \\
         \midrule
        Pandemic && Lockdowns & Timing of lockdowns & N & RIS \& \citelatex{stoeger2021}  \\
        Pandemic && Mobility & Mobility related to workplace travel  & P & Google Mobility Data  \\
       Pandemic &&  COVID-19 Fatalities  & Daily number of new COVID-19-related deaths divided by population counts per district & P & Austrian Federal Ministry of Social Affairs, Health, Care and Consumer Protection  \\
        Bank Lending Standards && Announcement and Enactment & Timing of the policy's announcement and enactment & N & RIS and Press Release \\
    Cost of Living Crisis &&  Policy Rate and Dates of Announced Changes & ECB policy rate (deposit facility) and dates of changes & N & ECB \\ 
       Cost of Living Crisis && HICP & Annual changes in the harmonised consumer price index for Austria & P & ECB\\
        Cost of Living Crisis && Mortgage Interest Rates  & Changes in average granted interest rates for new mortgages & P & OeNB \\ 
    \bottomrule
    \bottomrule
    \end{tabularx}
    \end{center}
    \footnotesize
\emph{Notes:} Type distinguishes between normative (N) and positive (P) classifiers. Law texts are (if not stated otherwise) retrieved from the Federal Gazette published in the Austrian Legal Information System (RIS).
\end{table}

For testing quantity effects, we model weekly quantities per federal state as presented in \autoref{sec:count_model}. Analogously to hedonic price models, quantities are regressed on economic variables explaining differences in local levels of market turnover.
Again, these variables take over the role of controls, while we focus on interpreting regime and contextual variables.

By varying the way how we include these contextual variables into our regression models, we can differentiate effects along the second dimension: \textit{(IG) immediate versus gradually evolving effects}.
This is achieved by either adding contextual variables as main effects, i.e., equally affecting all properties observed at or after the event start, or interacted with time-dummies, i.e., modelling the effect dynamically. By that we can detect periods of constant effect sizes, as well as periods of intensifying or diminishing impacts. 
For instance, lockdowns are immediately effective and have an end date, while changes in tastes are expected to emerge gradually. Likewise, indirect effects related to inflation or changes in lending practises are expected to gradually impact prices and quantities yet are present for a long(er) period.

In accordance with the effect types, we present immediate and gradually evolving effects differently: While the former are reported as main effects, the latter are presented as indices. Indices for prices are constructed by interacting contextual variables with time-dummies or evaluating indices separately by regime periods similar to hedonic time-dummy indices \citep[see][for a taxonomy]{hill2013hedonic, handbook}. This can be done to study overall gradual effects or heterogenous ones by allowing for segment-specific time dummies. 
This approach means a visual representation of the evolution of prices. Price indices are usually normalised to the event start and thus the difference in percent between event start and any other future time period can thus directly be inferred from this representation. 
We also introduce a new concept of quantity indices, which compare counts over time relative to the same period in the preceding year to account for seasonality. Again, a general index can be constructed yet also segment-specific ones to detect heterogeneity in time-varying market activity.

Finally, our framework differentiates events by market side initially affected: \emph{(SD) -- supply-side versus demand-side}. This yields to predictions about the type of data from which we expect first signals of change. Therefore, we exploit our three data sets that track properties at different stages of the transaction process as explained in \autoref{tab:testing_strategy}. In practise, the differentiation between supply- and demand-side reactions is achieved by relying on different data feeding our regression models. The appropriate data is thus specified as part of each hypothesis in \autoref{sec:theory}.

\subsubsection{Event Timing}\label{sec:timing}

\autoref{fig:regimes} 
summaries the timing of events that also informs the time periods covered by our models for testing. We use the exact day of observing a dwelling (advertising, agreement or registration date) to link it with regimes. Detailed dates are provided in \autoref{tab:timeline}  in the Appendix.

In case of the pandemic and the bank lending standards, there are normative end dates specified:
The COVID-related restrictions and reporting obligations have been lifted by 30 June 2023 which we consider as legal expiration date for the COVID-pandemics. Here, however, we set 30 June 2022 as end date as we observe a post-COVID normalisation and a decrease in COVID-19 related fatalities: mobility converged rapidly to the pre-COVID level and for many employees home office became permanent. 
With regard to the bank lending standards, we split the regime into the announcement period (13.12.2021--01.08.2022), and enactment period ending with the planned abolishing of the policy (01.08.2022--30.06.2025).

Regarding the cost-of-living crisis, we consider announced changes in the ECB's policy rate to define the regime: We separate the regime periods by the very first announced policy rate hike on 27 July 2022 where the deposit facility was increased from -0.5\% to 0\% and the 21 December 2022 when the rate was lifted to 2\%. We define the end of this period as the announcement of the first interest decrease on 12 June 2024.


\section{Empirical Results}\label{sec:empirical}

\subsection{Pandemic}\label{sec:results:pandemic}
\subsubsection{Hypothesis 1: Pandemic Quantity Effects}\label{sec:hyp.1}

\begin{table}[h!]
\begin{center}
\caption{Pandemic: Quantity Effects $(D)$ and $(A)$}
\label{tab:pandemic_count}
\renewcommand{\arraystretch}{0.7}
\adjustbox{max width=\textwidth}{%
\begin{tabular}{l ccccc c ccccc}
\toprule
\toprule
 & $(D)$ & $(D)$ & $(D)$ & $(D)$ & $(D)$ && $(A)$ & $(A)$ & $(A)$ & $(A)$ & $(A)$ \\
& $(1)$ & $(2)$ & $(3)$ & $(4)$ & $(5)$ &&  $(6)$ & $(7)$ & $(8)$ & $(9)$ & $(10)$\\
 \cmidrule{2-6} \cmidrule{8-12}
COVID $\times$ Lockdown (Ref.: Pre-COVID)                          & $-0.077^{**}$  &                &                &                &             &   & $-0.116^{***}$ &                &                &                &                \\
                                                  & $(0.024)$      &                &                &                &             &   & $(0.034)$      &                &                &                &                \\
COVID $\times$ No Lockdown                          & $0.005$        &                &                &                &           &     & $-0.103^{***}$ &                &                &                &                \\
                                                  & $(0.015)$      &                &                &                &             &   & $(0.022)$      &                &                &                &                \\
Reduced Workplace Mobility                        &                & $-1.133^{***}$ &                &                &                &                &                & $-0.295^{**}$  &                &                &                \\
                                                  &                & $(0.076)$      &                &                &                &                &                & $(0.103)$      &                &                &                \\
                                                  Reduced Workplace Mobility $\times$ Lockdown      &                &                & $-1.146^{***}$ &                &                &                &                &                & $-0.294^{**}$  &                &                \\
                                                  &                &                & $(0.076)$      &                &                &                &                &                & $(0.103)$      &                &                \\
Reduced Workplace Mobility $\times$ No Lockdown   &                &                & $-0.987^{***}$ &                &                &                &                &                & $-0.314^{*}$   &                &                \\
                                                  &                &                & $(0.089)$      &                &                &                &                &                & $(0.124)$      &                &                \\
Mortality                                         &                &                &                & $0.280^{***}$  &                &                &                &                &                & $0.355^{***}$  &                \\
                                                  &                &                &                & $(0.008)$      &                &                &                &                &                & $(0.013)$      &                \\
Mortality $\times$ Lockdown                       &                &                &                &                & $0.187^{***}$  &                &                &                &                &                & $0.234^{***}$  \\
                                                  &                &                &                &                & $(0.011)$      &                &                &                &                &                & $(0.016)$      \\
                                                  Mortality $\times$ No Lockdown                    &                &                &                &                & $0.360^{***}$  &                &                &                &                &                & $0.511^{***}$  \\
                                                  &                &                &                &                & $(0.010)$      &                &                &                &                &                & $(0.020)$      \\
\midrule
Housing Type                                  & \checkmark     & \checkmark      & \checkmark      & \checkmark  & \checkmark  && \checkmark     & \checkmark      & \checkmark      & \checkmark  & \checkmark   \\ 
Time Dummies                                  & \xmark         & \checkmark      & \checkmark      & \checkmark    & \checkmark && \xmark    &  \checkmark       & \checkmark      & \checkmark  & \checkmark \\
Cyclical Trend                              & \checkmark         & \checkmark      & \checkmark      & \checkmark   & \checkmark & &\checkmark     & \checkmark       & \checkmark      & \checkmark  & \checkmark\\
Location Fixed Effects                        & \checkmark     & \checkmark      & \checkmark      & \checkmark   & \checkmark && \checkmark    &  \checkmark      & \checkmark      & \checkmark  & \checkmark \\
\midrule
BIC                                               & $84,803$  & $55,764$    & $55,763$    & $55,251$    & $55,177$   && $36,628$  & $24,053$    & $24,062$    & $23,634$    & $23,575$    \\
Num. of obs.                                         & $12,132$        & $8,057$         & $8,057$         & $8,057$         & $8,057$         && $7,627$         & $5,057$         & $5,057$         & $5,057$         & $5,057$         \\
\bottomrule
\bottomrule
\multicolumn{11}{l}{\footnotesize{$^{***}p<0.001$; $^{**}p<0.01$; $^{*}p<0.05$}}
\end{tabular}
}
\end{center}
\begin{scriptsize}
    \emph{Notes:} Immediate effects on the number of transactions due to COVID-19, reduced workplace mobility and COVID-19-related mortality. Time periods assessed are shorter for models including workplace mobility or COVID-19 mortality rates due to data availability (see \autoref{tab:timeline}).\\
    \emph{Source:} Land Registry and DataScience Service GmbH
\end{scriptsize}
\end{table}

We measure quantity effects by adding COVID-identifiers to the baseline model (\ref{quantmod}).  
Since we use population on a federal state level as exposure variable (as outlined in \autoref{sec:count_model}), we interpret the results with regard to the logged number of transactions per capita. Therefore, the exponents of the estimated coefficients translate to incidence rate ratios between two categories. 
Model (1) and (6) use a three-dimensional factor to distinguish between the pre-COVID phase, the pandemic phase during lockdowns, and the pandemic phase outside of lockdown measures being in place. According to model (6), the pandemic phase was generally associated with a reduced volume of adverts while the lockdown phases attract an even lower coefficient. This is not the case for $(D)$: Model (1) suggests that only lockdown phases are associated with a negative coefficient. This comparison suggests that the longer-lasting COVID-19 effects were  rather a supply-side issue than a longer-lasting decrease in demand.

\begin{figure}[h]
\begin{center}
\captionsetup{font=large}
\caption{Pandemic: Split Quarterly Quantity Effects} \label{fig:quarters_count_model} 
\includegraphics[width=0.8\textwidth]{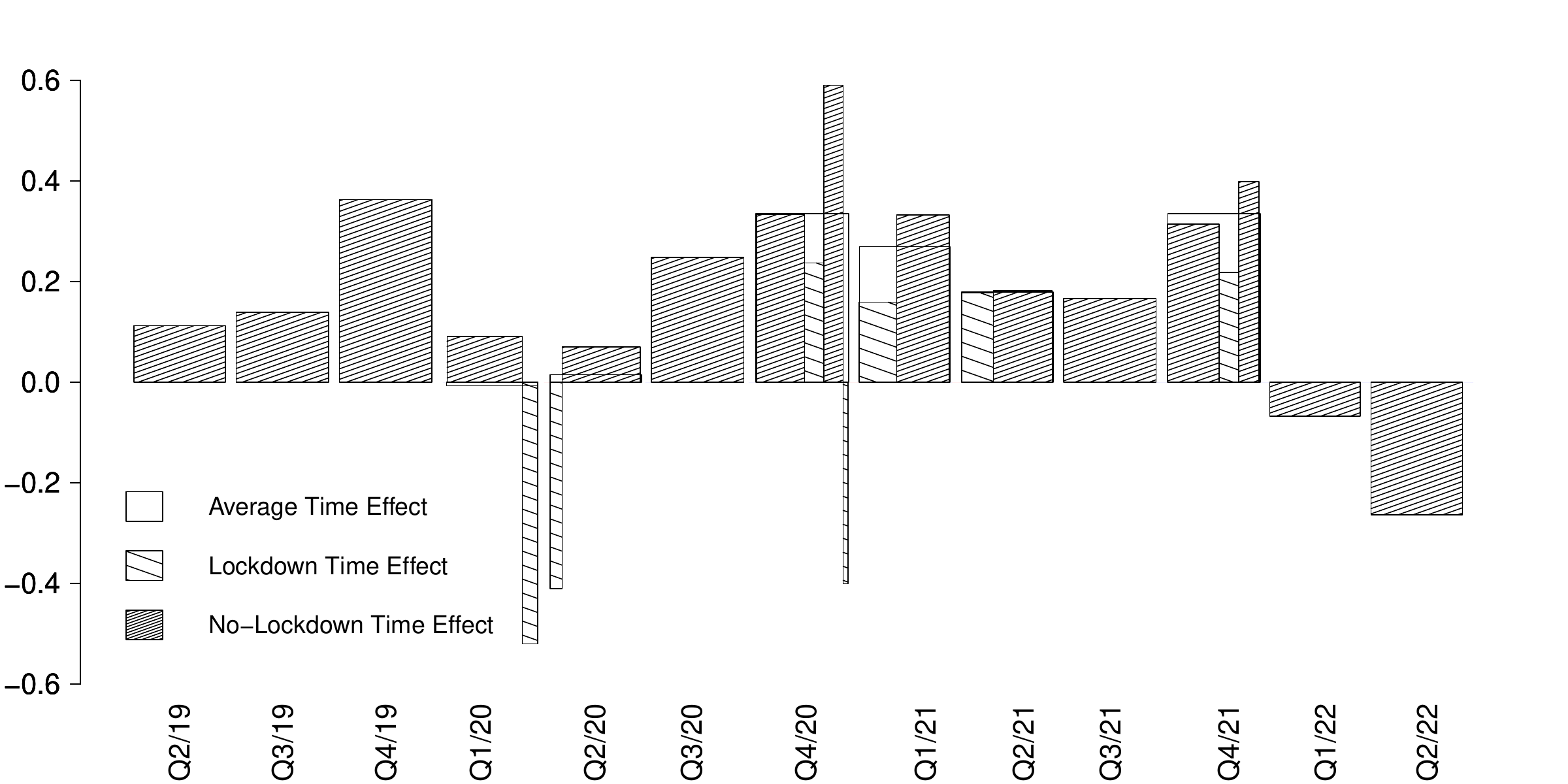}
\end{center}
\begin{scriptsize}
\emph{Notes:} The figure depicts split estimated time dummies. The splitting is obtained by interacting quarterly time-dummies with lockdown periods.\\
    \emph{Source:} Land Registry  
\end{scriptsize}
\end{figure}

For a closer look on the evolution of these effects, we incorporate instead of broad COVID-periods rather quarterly time-dummies, which we interact with exact lockdown periods. This means that, for instance, two transactions both occurring during the same quarter yet one of them would be recorded on a day with mandated restrictions in place and the other one not they would be assigned to two separate time dummies. Irrespective of that, at the end of each calendar quarter a new time period starts. These ``split quarterly quantity effects'' are compared to a model with standard quarterly time-dummies.
\autoref{fig:quarters_count_model} depicts results. The dashed bars refer to lockdown periods and the filled ones to periods without such restrictions. In comparison to average quarterly time effects (white bars), lockdown periods generally associated with fewer recorded deeds. However, following the large downturn during the initial lockdown, the general trend shows similarly large numbers of transactions as before the pandemic from already soon after the first lock-down onward. These findings indeed suggest, that there was a general slow down associated with the outbreak of COVID-19 and effects are mainly associated with the initial lockdown yet were not long-lasting as stated in Hypothesis 1.

Besides measuring the intensity of the pandemic via lockdown periods themselves, we use two further contextual variable therefore to test for the robustness of results. First, we assess changes in workplace-related mobility. 
Model (2) reports a negative correlation between workplace-related mobility and trade volume, which is -- as suggested by model (3) -- more severe during lockdown periods. 
Model (4) refers to the effect of COVID-19 related fatalities which we test again separately for the pandemic subregimes in model (5).
The findings are in-line with our results for lockdowns: the first lockdown had the strongest (negative) impact on the real estate market while their effects fade out thereafter. In contrast, the number of fatalities were low at the beginning of the pandemic and peaked only in winter 2020/2021 as shown in \autoref{fig:mobility}. Thus, we can conclude that not the severity of the pandemic but rather the intensity of legal restrictions had a negative effect on the trade volume. 
The remaining results for $(A)$ show a very similar pattern (Models (7)-(10) in \autoref{tab:pandemic_count}).

\subsubsection{Hypothesis 2: Pandemic Price Effects}\label{sec:hyp.2}

\begin{figure}
\begin{center}    
  \caption{Pandemic: Immediate Price Effects $(A)$ and $(A^B)$} 
  \label{fig:price_covid_immediate}
    \begin{subfigure}[t]{0.45\textwidth}
    \centering
    \caption{$(A)$}
      \label{fig:price_covid_immediate_A}
\includegraphics[width=\textwidth]{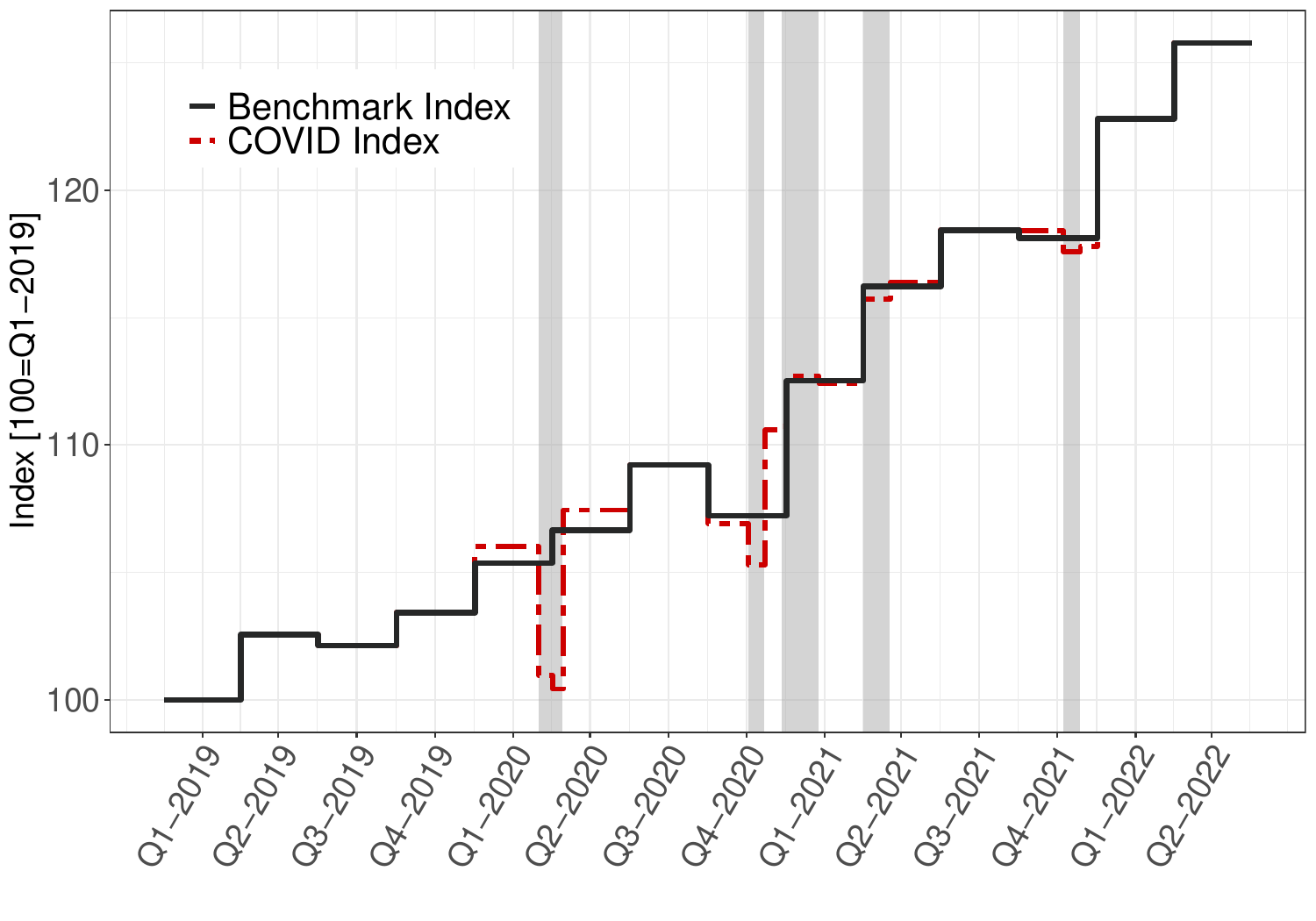}
\end{subfigure}
\hfill
\begin{subfigure}[t]{0.45\textwidth}
\centering
    \caption{$(A^B)$}
          \label{fig:price_covid_immediate_AB}
    \includegraphics[width=\textwidth]{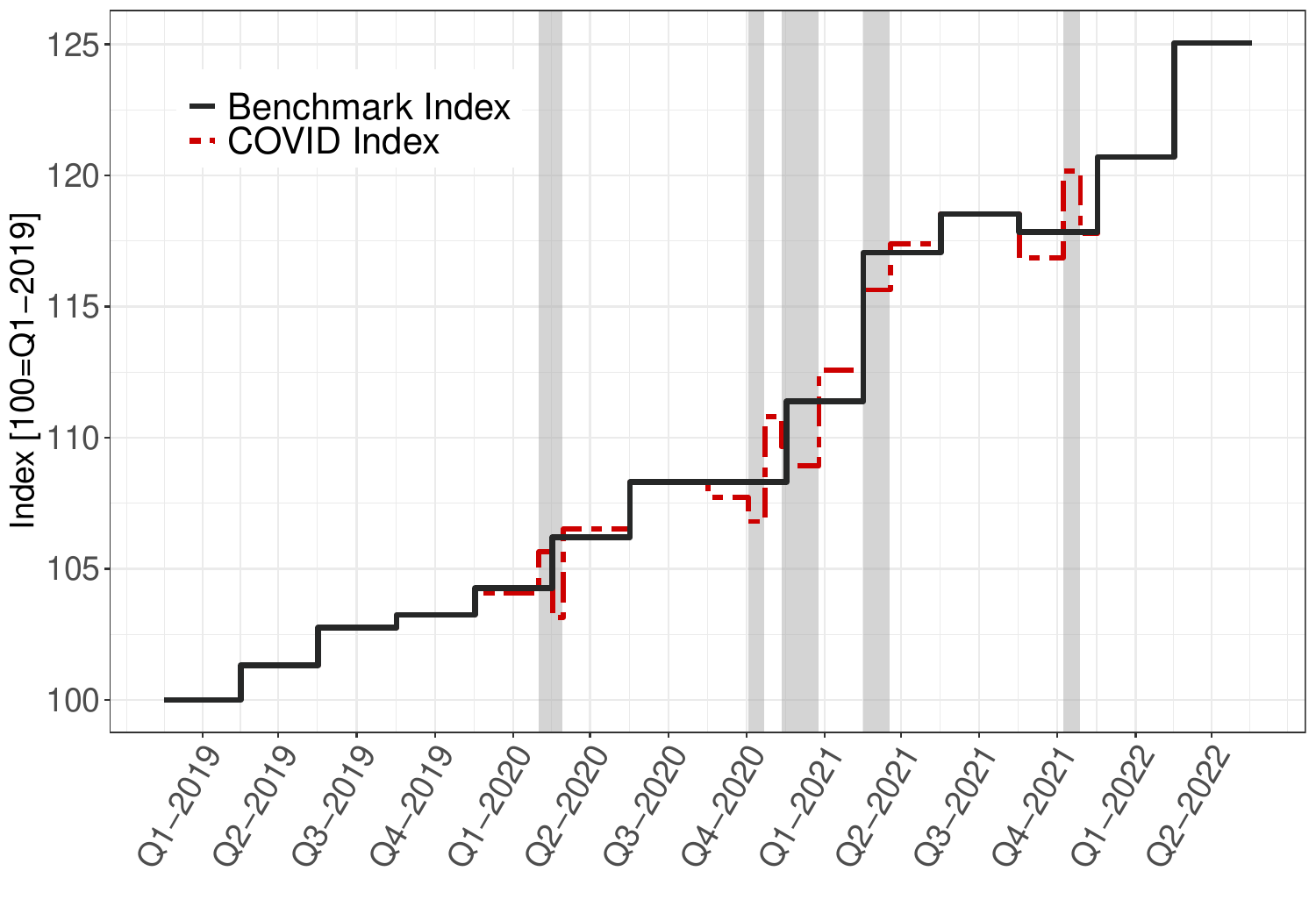}
\end{subfigure}
\end{center}
\begin{footnotesize}
    \emph{Notes:} The Figures depict immediate price effects for adverts $(A)$ (\autoref{fig:price_covid_immediate_A}) and $(A^B)$ brokered adverts (\autoref{fig:price_covid_immediate_AB}). Lockdown periods are indicated in grey. The benchmark index is a quarterly hedonic time dummy index whereas the COVID-index splits quarters inline with lockdown periods.\\
    When splitting time-dummies, we depict indices as step function to make visible the exact timing of price changes. In case of no splitting, we use the usual way of plotting indices based on linear interpolation. 
    \emph{Source:} Authors' own calculations based on DataScience Service GmbH data.
\end{footnotesize}
\end{figure}

\autoref{hyp:prices.pandemic} differentiates between immediate (item 1) and gradual (items 2-4) price effects.
Item 1 alludes to the impact of lockdowns.
Similar as in the count model, we allow for precisely timed lockdown effects by splitting quarterly time dummies and, by that, aligning them with the exact legal duration of mandated lockdowns.
We find a sizeable price drop during the first lockdown in $(A)$ indicating an immediate yet short-lived reaction of the supply side (see \autoref{fig:price_covid_immediate_A}). $(A^B)$ shown in \autoref{fig:price_covid_immediate_AB} exhibit smaller effect sizes, and slightly delayed price decreases during the first lockdown. 
Subsequently, the second lockdown brought about a drop in advertised prices, yet the effect is smaller in magnitude. Later lockdowns are not associated with significant price drops any more. 
Numerical results support this: When introducing a three-level factor differentiating the COVID period into a lockdown and a no-lockdown phase with the pre-COVID phase as baseline in \autoref{tab:reduced_mob_prices}, we find that the overall COVID period coincided with a housing boom but was slowed-down during intermitting lockdown phases. Again, $(A^B)$ identifies larger positive price effects than $(A)$ suggesting effects were driven by the demand side.

\begin{figure}[h!]
\begin{center}
\captionsetup{font=large}
  \caption{Pandemic: Gradual Price Effects $(A)$ and $(A^B)$} 
  \label{fig:pandemic_gradual_price}
  \begin{subfigure}[t]{0.4\textwidth}
    \centering
    \caption{Open Space -- $(A)$}
\includegraphics[width=\textwidth]{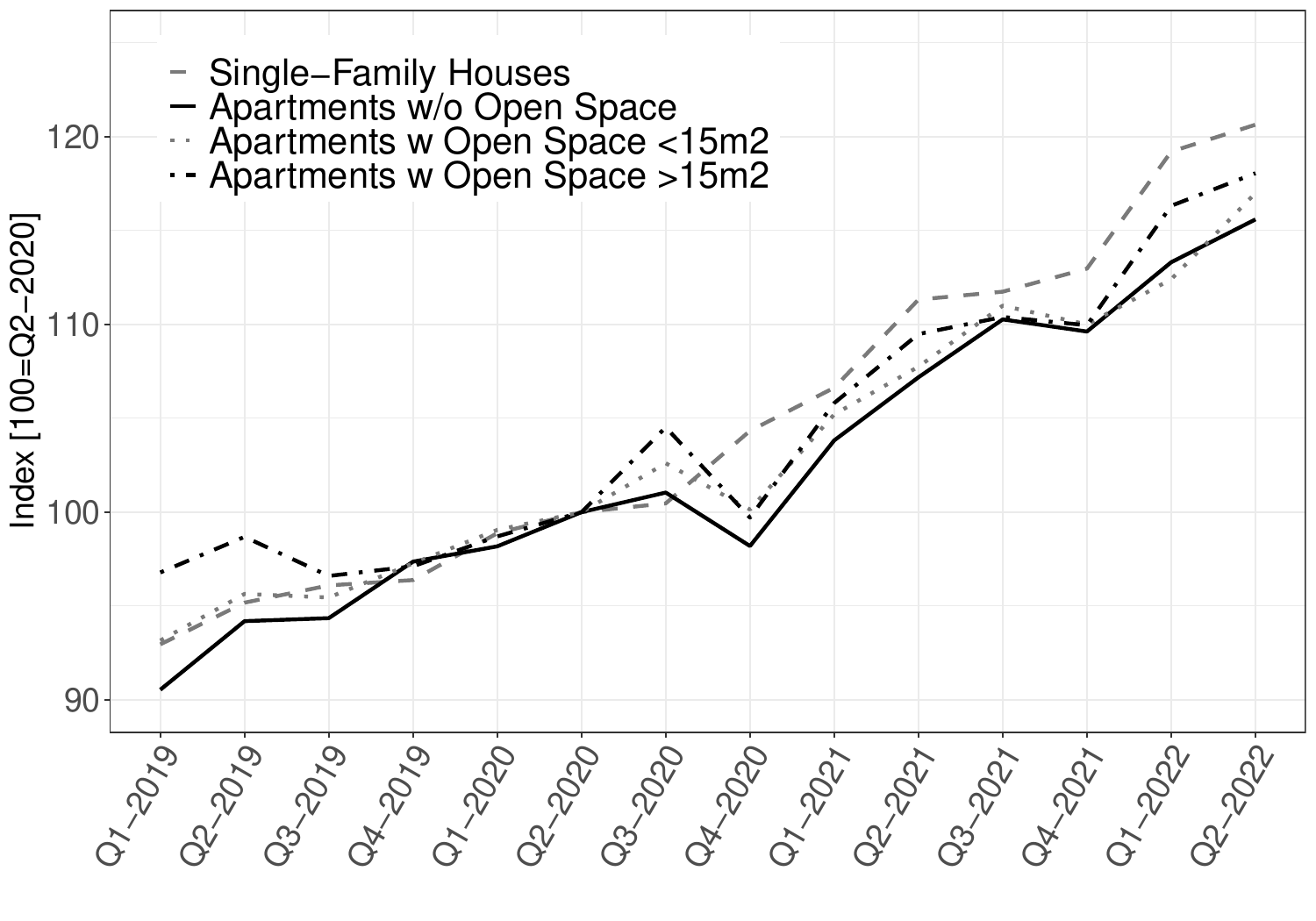}
\label{fig:pandemic_gradual_price_open_space_A}
\end{subfigure}
 \hfill
  \begin{subfigure}[t]{0.4\textwidth}
  \caption{Open Space -- $(A^B)$}
\includegraphics[width=\textwidth]{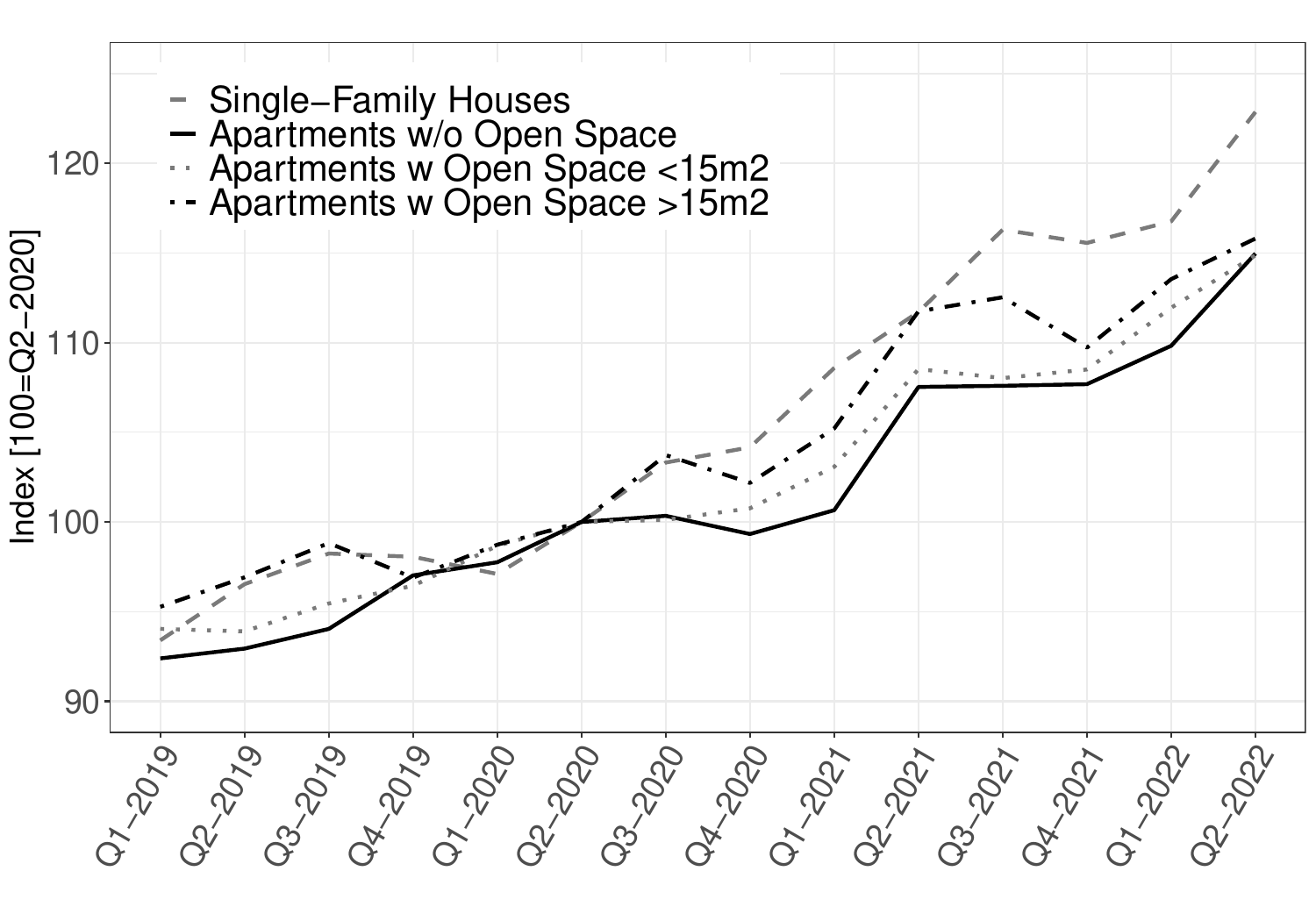}
\label{fig:pandemic_gradual_price_open_space_AB}
\end{subfigure}
\hfill
  \begin{subfigure}[t]{0.4\textwidth}
    \centering
    \caption{Degree of Urbanisation -- $(A)$}
    \label{fig:pandemic_gradual_price_urbanisation_A}
  \includegraphics[width=\textwidth]{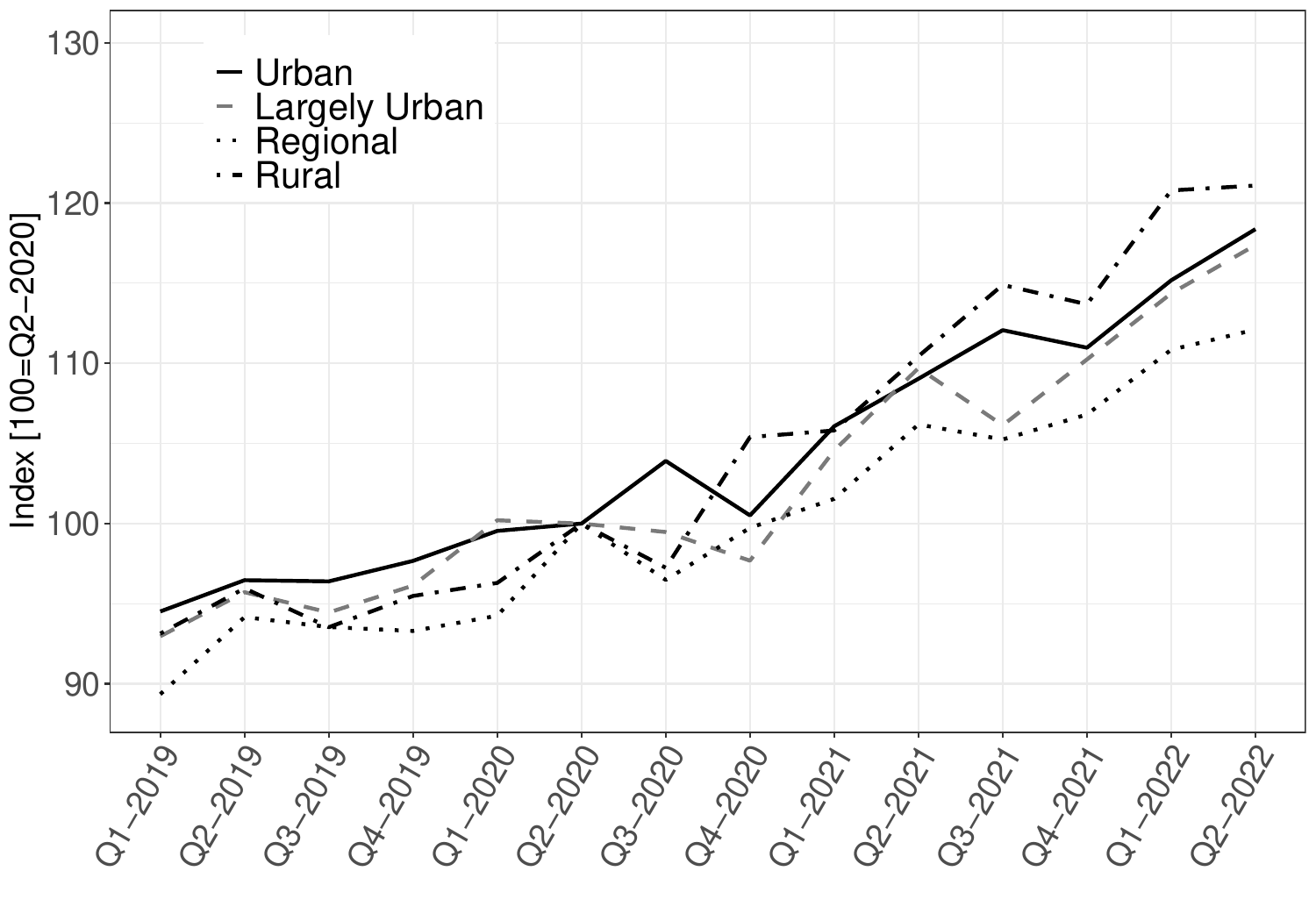}
\end{subfigure}
 \hfill
  \begin{subfigure}[t]{0.4\textwidth}
  \caption{Degree of Urbanisation -- $(A^B)$}
      \label{fig:pandemic_gradual_price_urbanisation_AB}
\includegraphics[width=\textwidth]{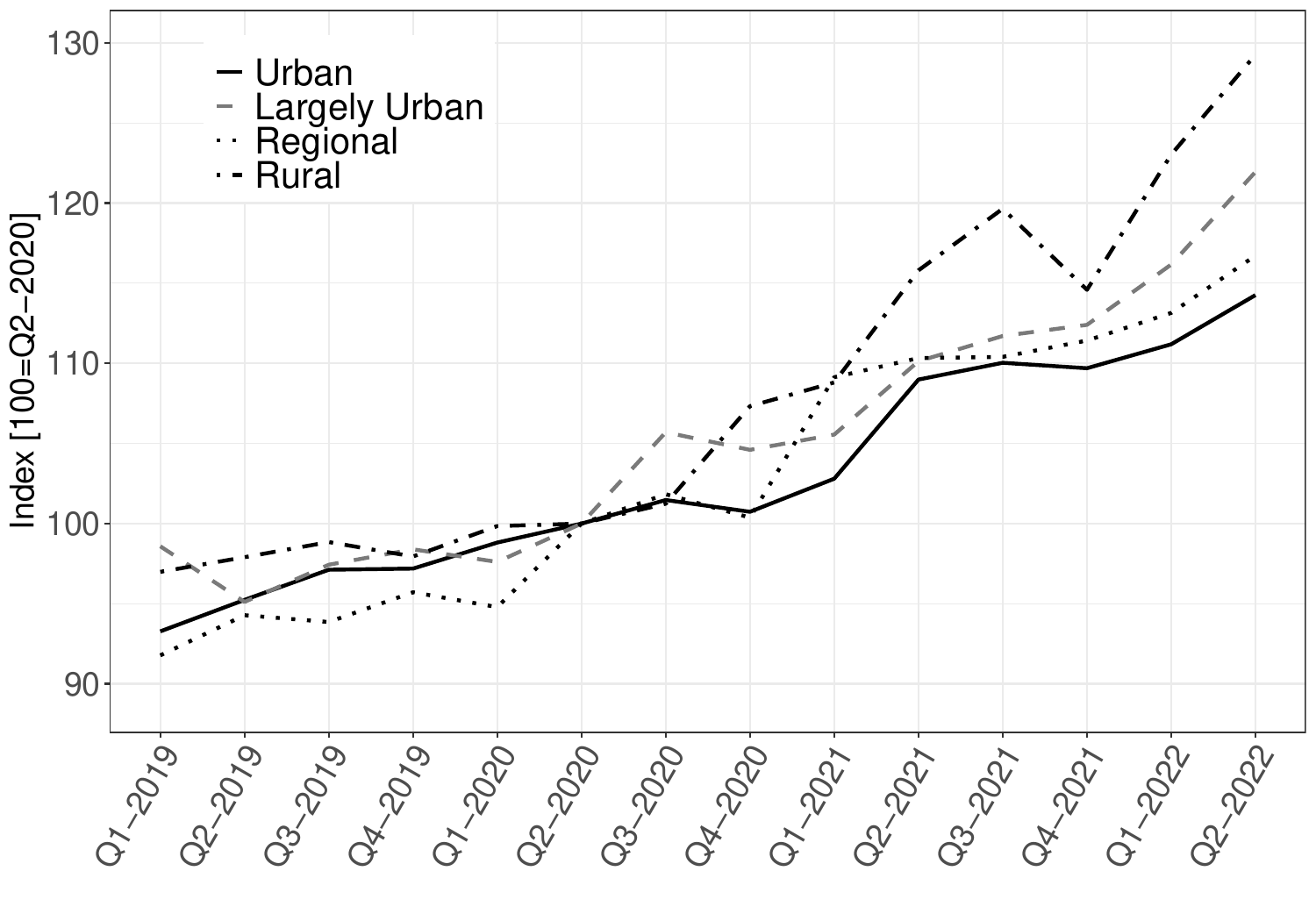}
\end{subfigure}

  \begin{subfigure}[t]{0.4\textwidth}
    \centering
    \caption{Dwelling Type -- $(A)$}
        \label{fig:pandemic_gradual_price_micro_A}

  \includegraphics[width=\textwidth]{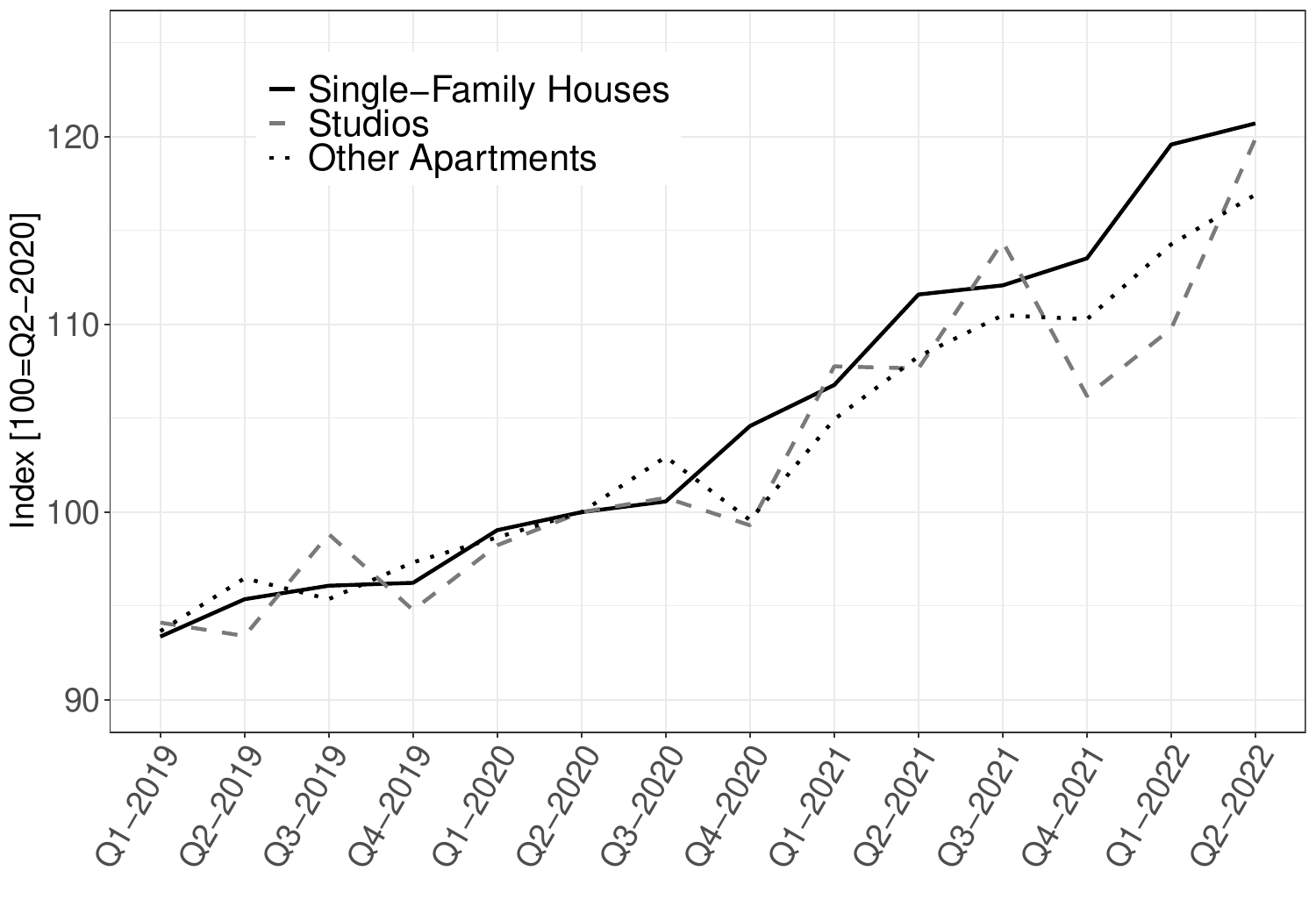}

\end{subfigure}
 \hfill
  \begin{subfigure}[t]{0.4\textwidth}
  \caption{Dwelling Type -- $(A^B)$}
          \label{fig:pandemic_gradual_price_micro_AB}
\includegraphics[width=\textwidth]{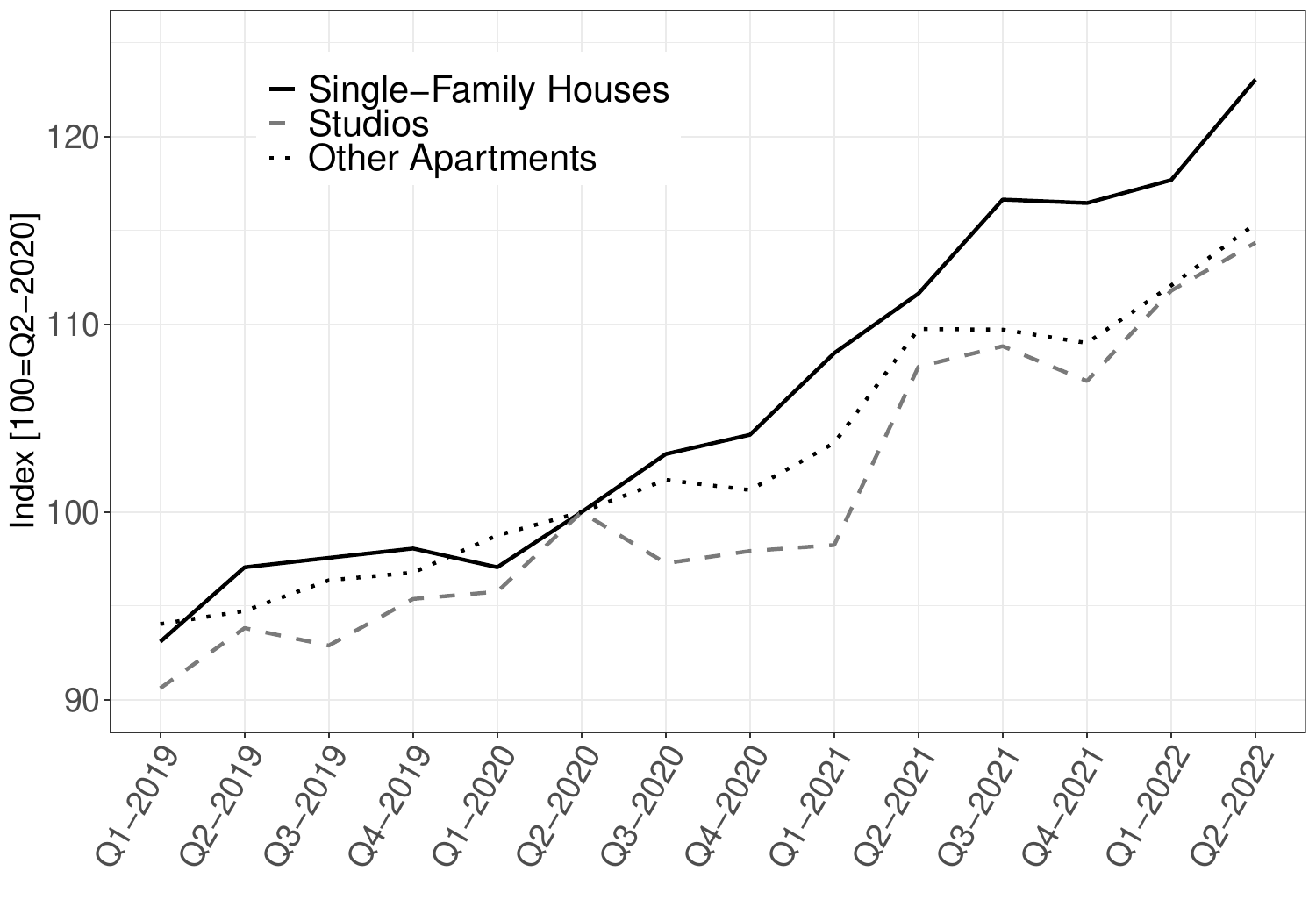}
\end{subfigure}
\end{center}
\begin{scriptsize}
\emph{Notes:} Indices obtained from time-variant coefficients associated with selected hedonic characteristics retrieved from price models fed by $(A)$ and $(A^B)$, respectively. These test gradual price effects predicted in \autoref{hyp:prices.pandemic}.
\end{scriptsize}
\end{figure}

These findings reinforce item 1 of \autoref{hyp:prices.pandemic}: There is an immediate price drop measured by $(A)$ that is mainly driven by the first lockdown. The general insecurity and stringent stay-at-home mandate during this initial phase of the pandemic is, thus, immediately manifested in advertised prices. Given that these patterns are not reiterated when testing for the same effects on $(A^B)$, we classify these effects as immediate yet transitory supply-side reactions.%
\footnote{As the effect is immediate in nature and short-lived, we cannot rule out that the demand-side had initially equally negative price expectations. The period of low price increases following the first lockdown could be considered as a hint towards this interpretation.}
Overall, the pandemic was rather associated with increasing prices (see \autoref{tab:reduced_mob_prices}, Model (1)) with the first lockdowns just being short-lived interim periods. Similarly, measuring reduced economic activity via reduced workplace mobility is also negatively related to prices confirmed by Models (2) and (7), yet this overall effect is again particularly driven by lockdown periods as shown in Models (3) and (8). The severity of the pandemic measured via the mortality rate has no sizeable impact on any type of prices according to Models (4), (5), (9) and (10).
The overall positive effect seemed to have been driven by increased demand as indicated by the slightly larger coefficients associated with the overall COVID period in Model (6) estimated on $(A^B)$ than in Model (1) estimated on $(A)$, as well as the smaller negative or even non-negative effects associated with reduced mobility or the number of fatalities when combined with $(A^B)$ rather than $(A)$. That price increases are rather driven by the demand side is also in-line with our results for counts: As shown in \autoref{tab:pandemic_count}, during the pandemic the number of advertisements strongly decreased while final transactions decrease less leading to more price pressure. This is in-line with findings for the US by \citelatex{gamber2023stuck} showing that more time spent at home led to increased demand for housing.

Next, we test for gradually evolving price effects associated with changes in preferred types of properties associated with changing working patterns and living conditions over the course of the pandemic.
As the driving forces are suspected to be the demand side but not the supply side, \autoref{hyp:prices.pandemic} suggests $(A^B)$. Reversely, we would expect that these effects are less clearly visible in data reflecting supply-side sentiments, i.e., $(A)$.

\autoref{hyp:prices.pandemic} suggests three types of changes:
First, we test for increased relative prices for apartments offering open space amenities. The results in \autoref{fig:pandemic_gradual_price_open_space_A} and \autoref{fig:pandemic_gradual_price_open_space_AB} show that prices of apartments without such facilities are outperformed by those offering them. The larger relative increase of properties with such amenities is driven by earlier periods of the pandemic and towards the end of the pandemic, prices of apartments without open space eventually catch up -- documented in $(A^B)$. 
This suggest a transitory effect. 
At the same time, prices for single-family homes, which usually offer some kind of open space, experience overall the largest increases, which appears to be a permanent one visible in $(A^B)$ over the entire time period.
Overall, this confirms part 2 of \autoref{hyp:prices.pandemic}.

The third part of \autoref{hyp:prices.pandemic} predicts gradually increasing relative prices for properties in non-urban areas. According to $(A^B)$, increases in urban regions are lower than in all other locations with highest price increases in rural areas (\autoref{fig:pandemic_gradual_price_urbanisation_AB}). 
For $(A)$, however, we do not find proof of such clearly inverted developments. Over the pandemic, increases in $(A)$ for urban properties were still second-highest after rural ones (\autoref{fig:pandemic_gradual_price_urbanisation_A}). From this contraindicating results between $(A)$ and $(A^B)$, we deduct that this change was predominantly driven by the demand side, which confirms the third part of this hypothesis.

The last part of \autoref{hyp:prices.pandemic} predicts that relative prices for studios and micro-apartments lost in value relative to larger units. This was argued to be a gradual effect triggered by shrinking demand. Again, $(A^B)$ exhibit a sharp downward trend for studios (\autoref{fig:pandemic_gradual_price_micro_AB}) while this is not visible for $(A)$ (\autoref{fig:pandemic_gradual_price_micro_A}).
This difference is particularly strong during the initial phase of the pandemic: prices of studios experienced a decrease, followed by a short period of stagnation. Single-family houses are the clear winners of the pandemic. Thus, we confirm also the last part of \autoref{hyp:prices.pandemic}.


\subsection{Bank Lending Standards}
\subsubsection{Hypothesis 3: Bank Lending Standards Quantity Effects}

\autoref{table:count_kim_coefficients} reports the results for the announcement and enactment of new bank lending standards.
Model $(1)$ tests for the effect of the policy's enactment. Final prices drop by a factor of $\exp(-0.309) = 0.73 <1$. 
To assess whether the total effect had already started to evolve from the policy's announcement in December 2021 onwards,%
\footnote{This period is also characterised by the the beginning of the inflationary period. We isolate this effect by focusing on specific contextual variables in \autoref{sec:results.prices.inflation}.} %
model $(2)$ distinguishes between a pre-announcement, announcement and enactment period. The sheer announcement came along with a slow-down in the market in comparison to the pre-announcement period. Yet, volumes declined even more upon enactment. Overall, however, we find that quantities did already decline before the policy was even announced likely because of by then already excessive prices.

\begin{table}[h]
\begin{center}
\caption{Bank Lending Standards: Quantity Effects $(D)$ and $(A)$}
\label{table:count_kim_coefficients}
\scriptsize
\resizebox{0.95\linewidth}{!}{
\begin{tabular}{l c c c c c  c cc}
\toprule\toprule
 & $(D)$ & $(D)$ & $(D)$ && $(A)$ & $(A)$ & $(A)$ \\ 
 & $(1)$ & $(2)$ & $(3)$  && $(4)$ & $(5)$ & $(6)$  \\
\cmidrule{2-4} \cmidrule{6-8}
KIM-VO (enactment)           & $-0.309^{***}$ &                &                && $-0.095^{***}$ &                &                \\
                             & $(0.019)$      &                &                && $(0.026)$      &                &                \\
KIM-VO (announcement)        &                & $-0.383^{***}$ &                &&                & $-0.118^{***}$ &                \\
                             &                & $(0.021)$      &                &&                & $(0.029)$      &                \\
KIM-VO (enactment)           &                & $-0.448^{***}$ &                &&                & $-0.141^{***}$ &                \\
                             &                & $(0.020)$      &                &&                & $(0.028)$      &                \\
Q2 2021 (Pre-KIM-VO)         &                &                & $-0.091^{**}$  &&                &                & $-0.164^{***}$ \\
                             &                &                & $(0.035)$      &&                &                & $(0.049)$      \\
Q3 2021 (Pre-KIM-VO)         &                &                & $-0.106^{**}$  &&                &                & $-0.059$       \\
                             &                &                & $(0.035)$      &&                &                & $(0.049)$      \\
Q4 2021 (Pre-KIM-VO)         &                &                & $0.007$        &&                &                & $-0.014$       \\
                             &                &                & $(0.037)$      &&                &                & $(0.053)$      \\
Q4 2021 (KIM-VO announcement)&                &                & $0.135^{*}$    &&                &                & $-0.241^{*}$   \\
                             &                &                & $(0.059)$      &&                &                & $(0.096)$      \\
Q1 2022 (KIM-VO announcement)&                &                & $-0.339^{***}$ &&                &                & $-0.076$       \\
                             &                &                & $(0.036)$      &&                &                & $(0.050)$      \\
Q2 2022 (KIM-VO announcement)&                &                & $-0.536^{***}$ &&                &                & $-0.188^{***}$ \\
                             &                &                & $(0.036)$      &&                &                & $(0.050)$      \\
Q3 2022 (KIM-VO announcement)&                &                & $-1.009^{***}$ &&                &                & $-0.511^{***}$ \\
                             &                &                & $(0.051)$      &&                &                & $(0.073)$      \\
Q3 2022 (KIM-VO enactment)   &                &                & $-0.480^{***}$ &&                &                & $-0.285^{***}$ \\
                             &                &                & $(0.040)$      &&                &                & $(0.058)$      \\
Q4 2022 (KIM-VO enactment)   &                &                & $-0.371^{***}$ &&                &                & $-0.252^{***}$ \\
                             &                &                & $(0.035)$      &&                &                & $(0.052)$      \\
Q1 2023 (KIM-VO enactment)   &                &                & $-0.673^{***}$ &&                &                & $-0.125^{*}$   \\
                             &                &                & $(0.037)$      &&                &                & $(0.049)$      \\
\midrule
Housing type                                  & \checkmark     & \checkmark      & \checkmark       && \checkmark   & \checkmark     & \checkmark        \\ 
Time Dummies                                  & \xmark         & \xmark      & \checkmark       &   & \xmark  & \xmark    & \checkmark         \\
Cyclical Trend                               & \checkmark         & \checkmark      & \checkmark      && \checkmark & \checkmark     & \checkmark        \\
Location Fixed Effects                        & \checkmark     & \checkmark      & \checkmark       && \checkmark  & \checkmark     & \checkmark         \\
\midrule
BIC                          & $52,994$    & $52,670$    & $52,348$    && $23,126$    & $23,118$    & $23,127$    \\
Num. obs.                    & $7,878$         & $7,878$         & $7,878$        & & $5,036$         & $5,036$         & $5,036$         \\
\bottomrule
\bottomrule
\multicolumn{5}{l}{\scriptsize{$^{***}p<0.001$; $^{**}p<0.01$; $^{*}p<0.05$}}
\end{tabular}
}
\end{center}
\tiny
\emph{Notes:} Time horizon: 01.01.2022 - 31.03.2023.
\end{table}

Model $(3)$ splits the periods into quarters.  
In Q1 2022, the number of transactions underscores that of preceding quarters. As already evident from Model $(2)$, transactions were generally declining, yet numbers declined at higher rates upon the announcement and decreases remained large and highly significant over the announcement and enactment periods.

We argue that the policy mainly affects the demand, which should thus react first. To test this, we re-estimate the same specification on data reflecting pure supply-side actions, i.e., $(A)$, and contrast results against $(D)$. We expect that effects are smaller or even null, and reactions manifest only in response to decreased demand, i.e., with a time lag. 
For adverts, we find an overall enactment effect below unity: $\exp(-0.095) = 0.91<1$ according to Model $(4)$ and $\exp(-0.141) = 0.87<1$ according to Model $(5)$. Thus, they are not null, yet effect sizes are much smaller than corresponding results from $(D)$, which indicates that the demand side pulled down quantities.

\begin{figure}[h]
\begin{center}
\captionsetup{font=large}
  \caption{Bank Lending Standards: Gradual Quantity Effects $(D)$ and $(A)$} \label{fig:KIM_gradual_counts}
  \begin{subfigure}[t]{0.45\textwidth}
  \caption{$(D)$}
  \label{fig:KIM_gradual_deeds}
\includegraphics[width=\textwidth]{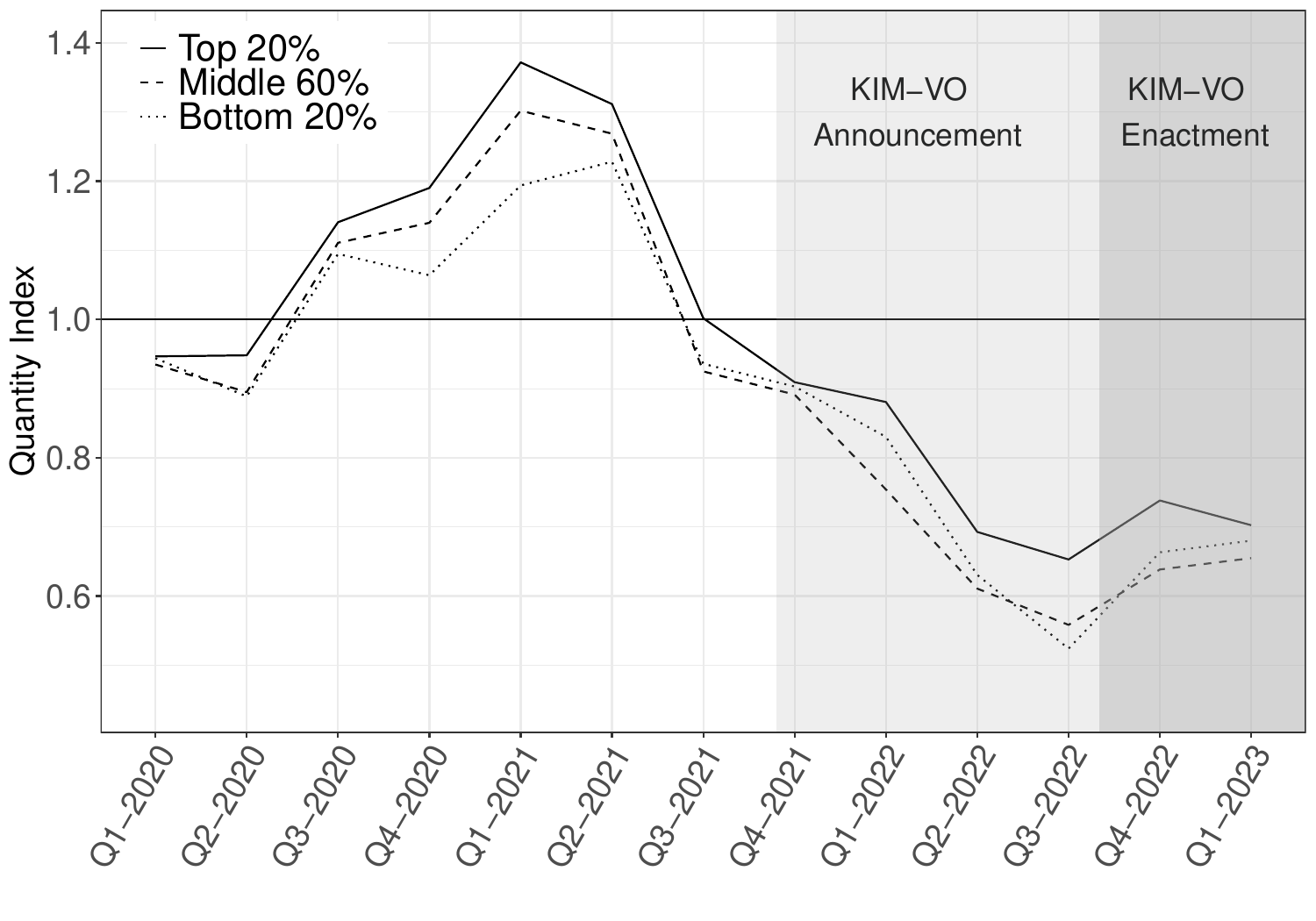}
  \end{subfigure}
  \hfill
\begin{subfigure}[t]{0.45\textwidth}
\caption{$(A)$}
\label{fig:KIM_gradual_adverts}
\includegraphics[width=\textwidth]{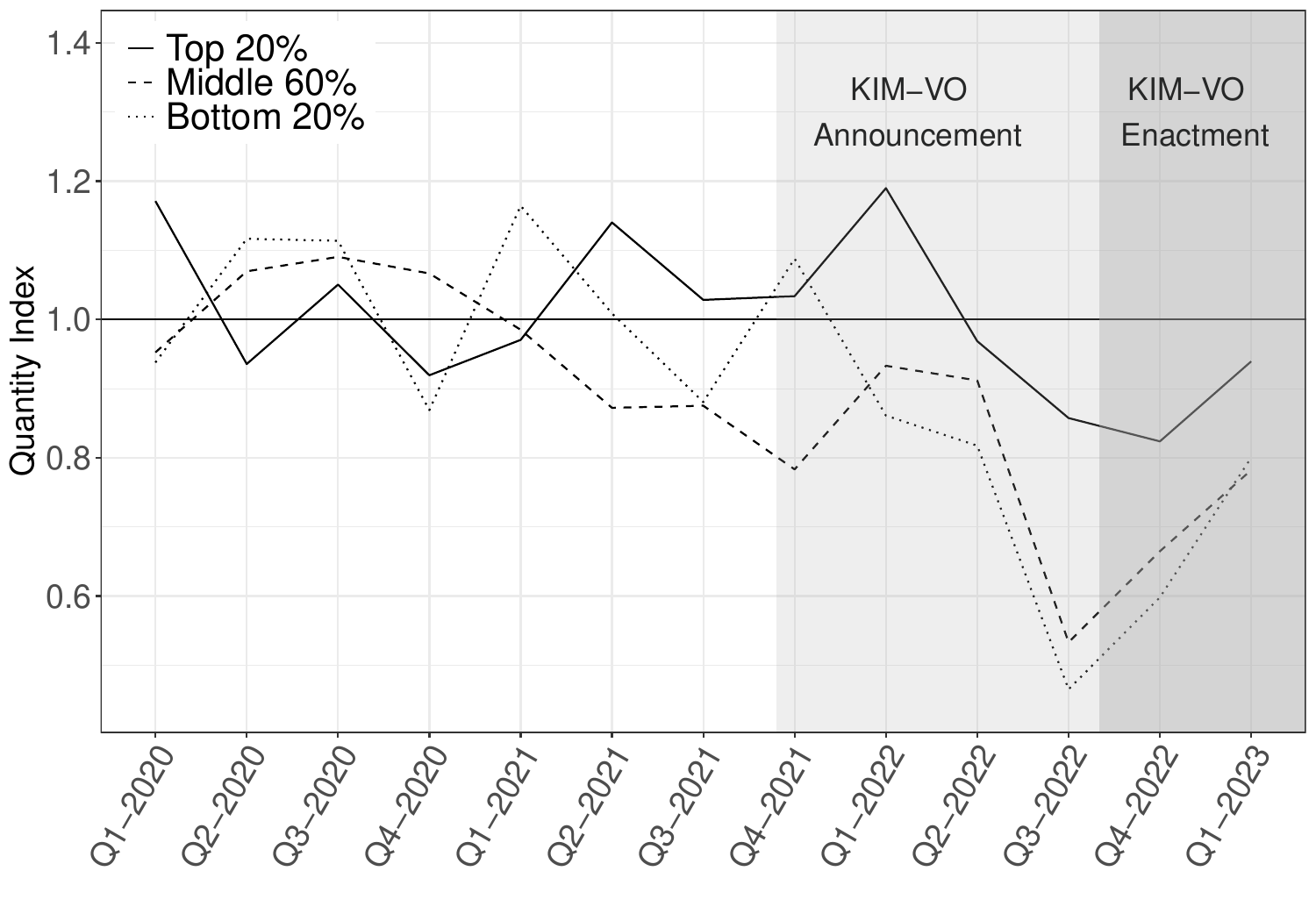}
\end{subfigure}
   \end{center}
\begin{scriptsize}
\emph{Notes:} The Figure shows gradually evolving quantity effects by market segments as quarterly count indices expressing changes relative to the same quarter in the preceding year. The bold line shows parity, i.e., no change to the previous year.
\end{scriptsize}
\end{figure}

We further hypothesized about heterogeneity across price segments as expensive properties may be less affected than properties targeted to the economically less well-off, i.e., lower-priced properties. 
Therefore, we construct quantity indices for different types of properties as explained in \autoref{sec:contextual}. These are achieved by counting deeds by various selection criteria and dividing counts by the number recorded in the same quarter of the previous year. By that, we account for seasonality, which has been shown above to be crucial.

\autoref{fig:KIM_gradual_counts} show results:
We sort districts by average price per square meeter between 01.01.2019 and 30.03.2023 in $(A^B)$ and contrast counts across three segments: the top 20\% most and least expensive ones and the middle segment.%
\footnote{\autoref{fig:gradual_counts_II} in the Appendix also shows results for the ten most expensive districts (which are in descending order Vienna 1st district (Innere Stadt), Vienna 8th district (Josefstadt), Vienna  19th district (Döbling), and Vienna 7th district (Neubau), Kitzbühel, Vienna 2nd district (Leopoldstadt), Vienna 9th district (Alsergrund), Vienna 4th district (Wieden), Vienna 6th district (Mariahilf), Innsbruck-Land), and the 10 least expensive ones (which are in descending order Oberwart, Melk, Scheibbs, Waidhofen an der Thaya, Lilienfeld, Oberpullendorf, Horn, Gmünd, Zwettl, Güssing). Results are similar.}
The index constructed from $(D)$ shown in \autoref{fig:KIM_gradual_deeds} was below unity from Q3-2021 onward and thus even before the KIM-VO had been announced and the speed of increasing counts had already decelerated from the first halve of 2021 onwards. This holds true quite homogenously across all price segments. 
It is only noticeable that the top price segment experienced a smaller decrease in the transaction volume than the rest of the price distribution in-line with the second part of Hypothesis 3.

To test whether these developments were rather triggered by the supply-side or the demand-side, we repeat the same analysis for $(A)$.
\autoref{fig:KIM_gradual_adverts} shows indeed a delayed and more heterogenous effect over segments: numbers were stickier for longer and particularly in the top segment there were still increases in the number of posted advertisements up until Q2-2022. This indicates that the demand-side was central in the slow-down. Further, the fact that final transactions had already started to fall quite dramatically long before the KIM-VO had been announced, this indicates that it was not primality this policy that triggered these reductions in transactions but rather decreased demand from Q1-2022 on -- a period that was characterised by record-high house prices as shown in \autoref{sec:results.prices.inflation}.

Although the general trend of decreased demand had started before the announcement already, the policy may have potentially cushioned the severity of the slow-down as forward-looking buyers may have invested slightly more in housing as they otherwise would have done: this is suggested by the smaller rate of decrease during the announcement period than during the pre-announcement period as evident from steeper lines, i.e., a lower gradient, in the pre-announcement period than thereafter as visible in \autoref{fig:KIM_gradual_deeds}.

\subsubsection{Hypothesis 4: Bank Lending Standards Price Effects}\label{sec:results.price.lending}

\begin{figure}[h]
\begin{center}
\captionsetup{font=large}
  \caption{Bank Lending Standards: Gradual Price Effects} 
  \begin{subfigure}[t]{0.47\textwidth}
  \caption{$(A)$: Split Time-Dummies}
\includegraphics[width=\textwidth]  {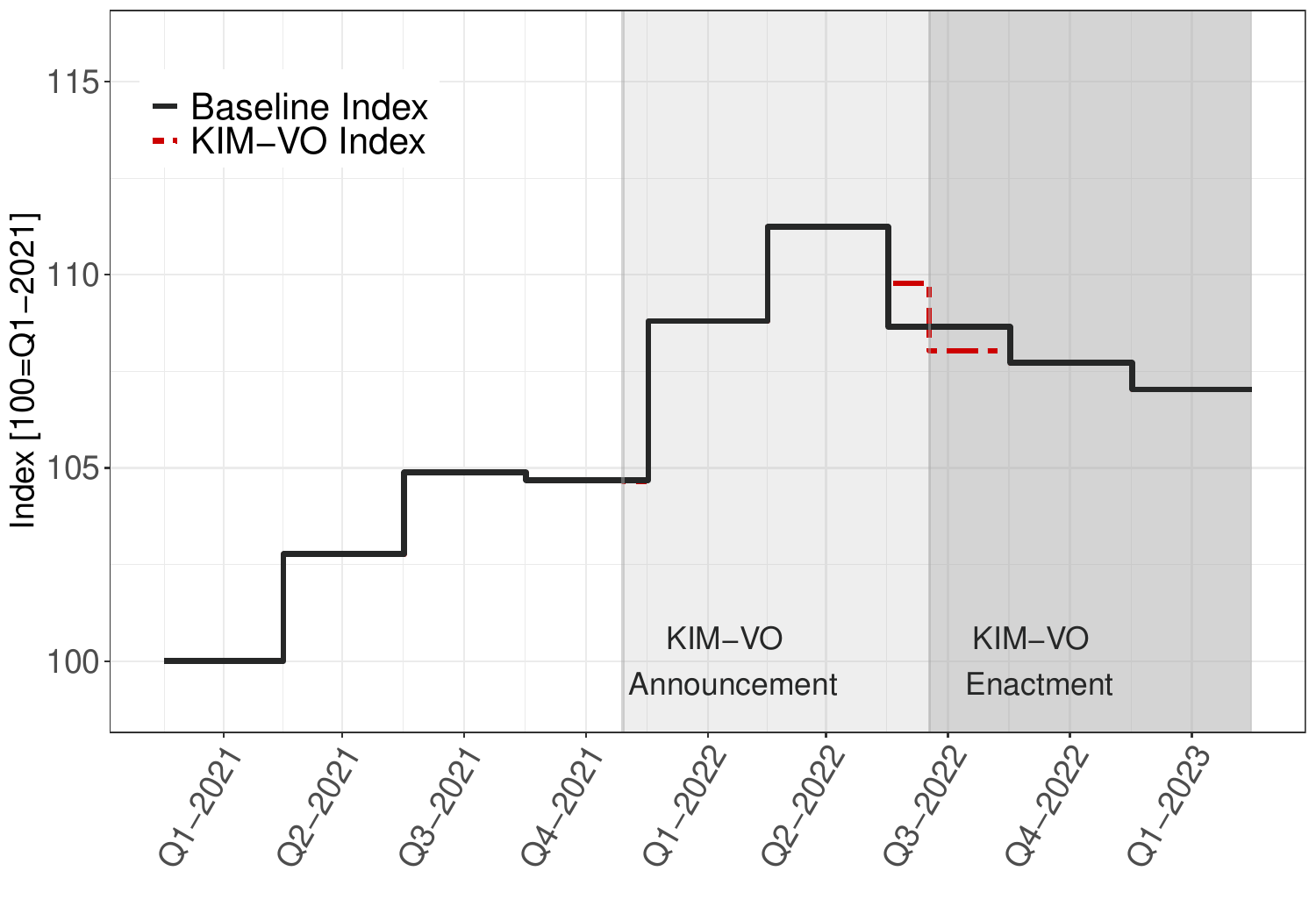}
\label{fig:KIM_gradual_A}
  \end{subfigure}
\begin{subfigure}[t]{0.47\textwidth}
  \caption{$(A^B)$: Split Time-Dummies}
  \label{fig:KIM_gradual_AB}
 \includegraphics[width=\textwidth]{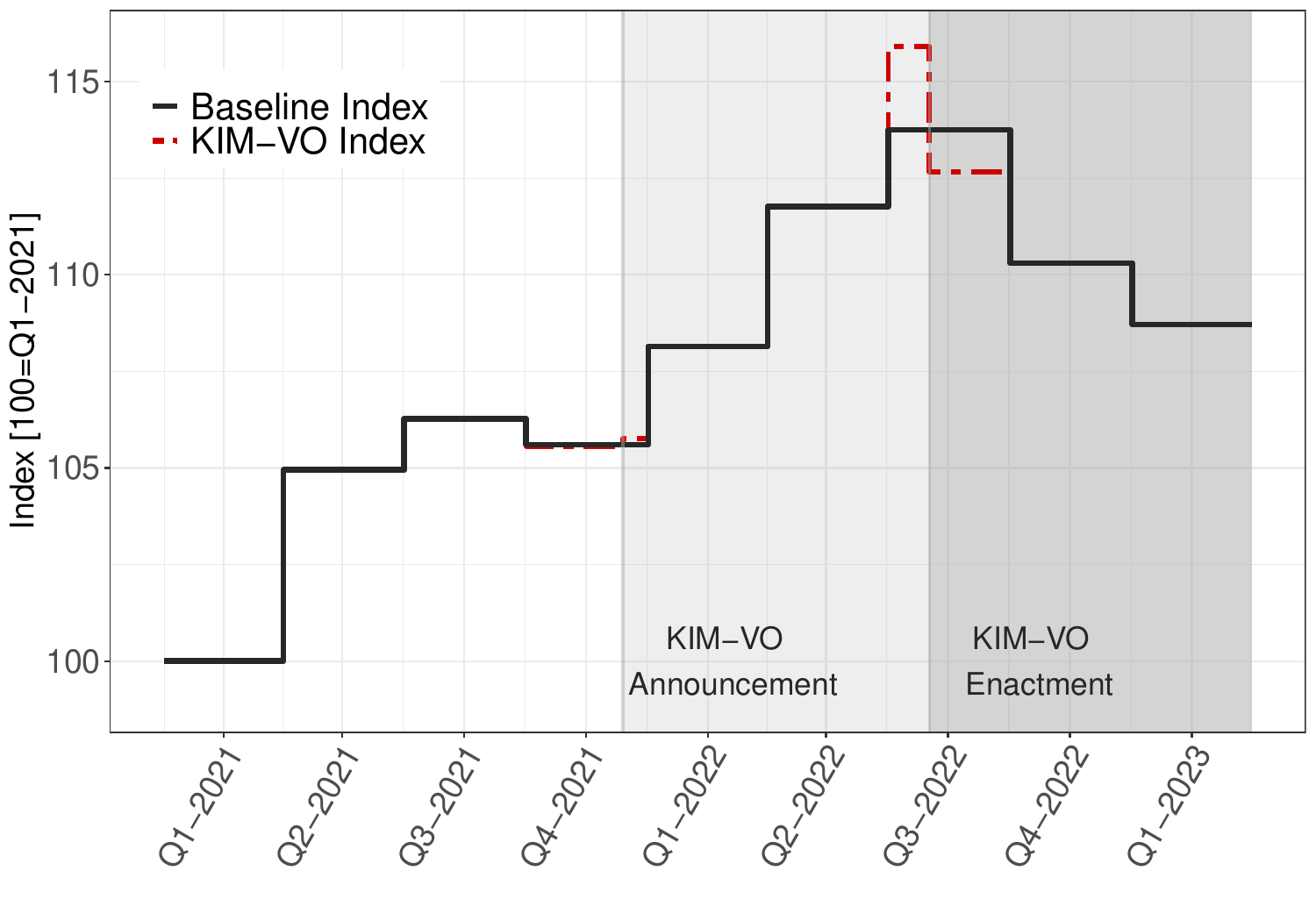}   
\end{subfigure}
  \begin{subfigure}[t]{0.47\textwidth}
  \caption{$(A)$: Effects by Price Segment}
\includegraphics[width=\textwidth]  {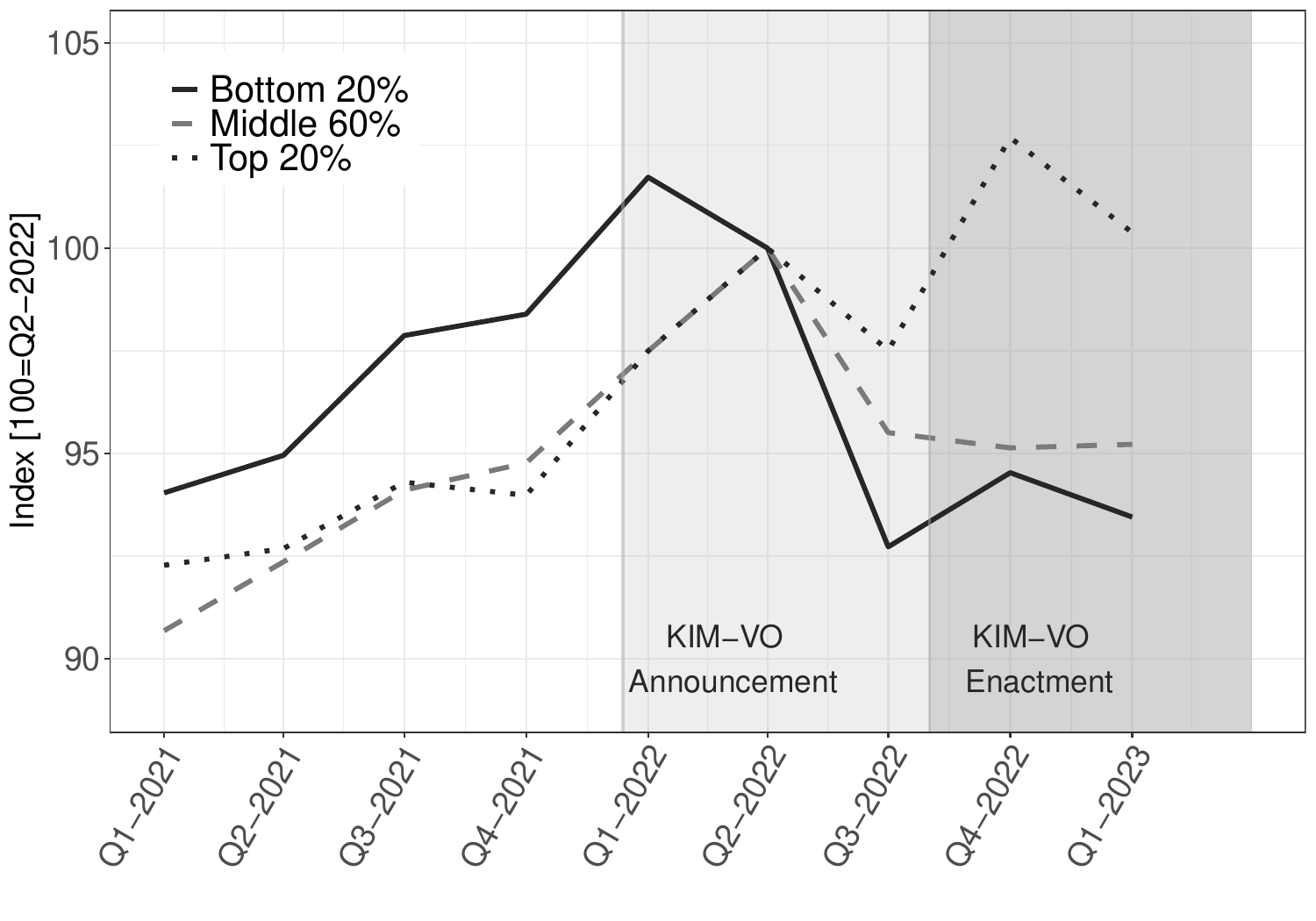}
  \label{fig:price_segments_A}
  \end{subfigure}
\begin{subfigure}[t]{0.47\textwidth}
  \caption{$(A^B)$: Effects by Price Segment}
\includegraphics[width=\textwidth]{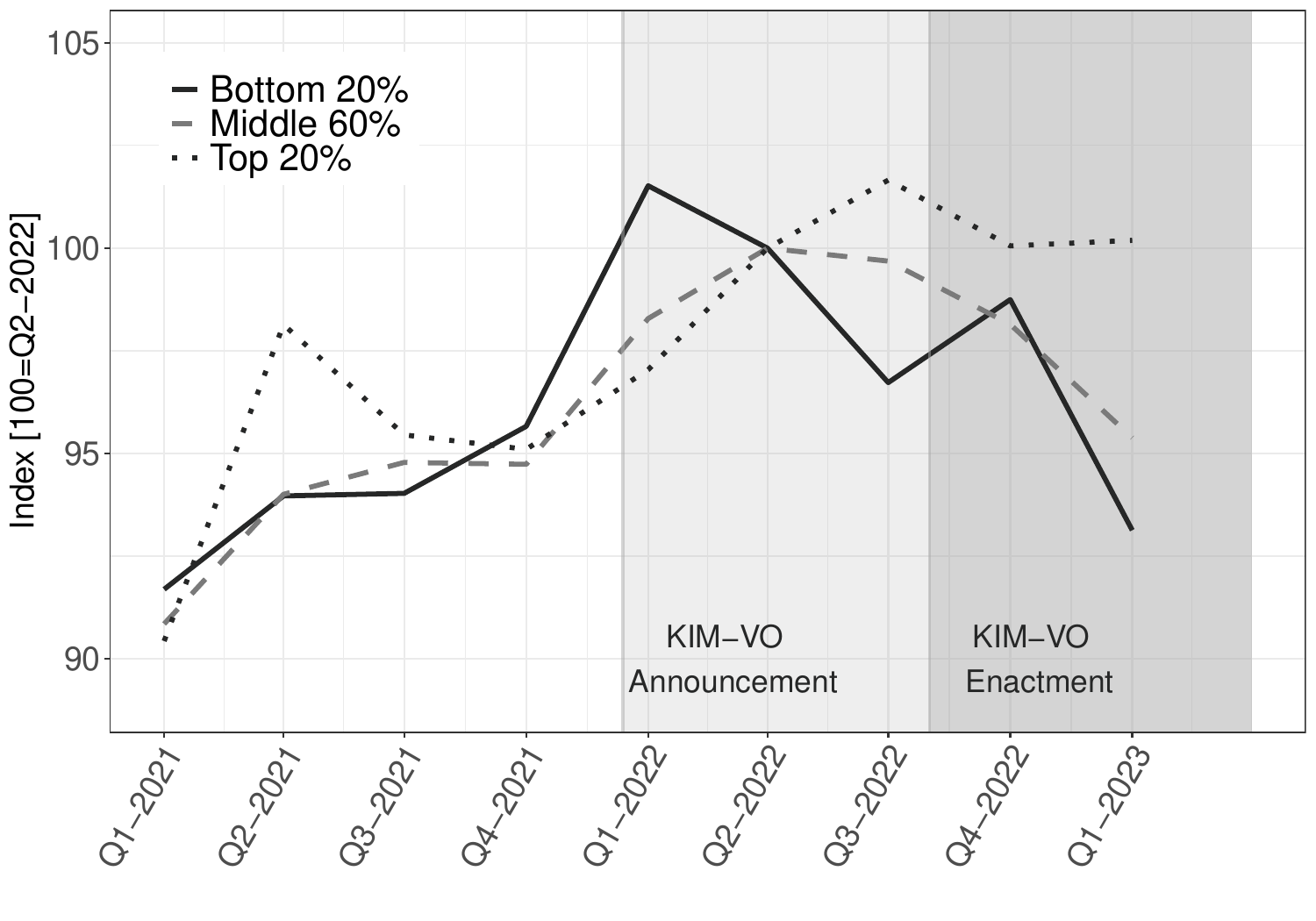}  
  \label{fig:price_segments_AB}
\end{subfigure}
\end{center}
\begin{scriptsize}
\emph{Notes:} The figure shows gradually evolving price effects. \autoref{fig:KIM_gradual_A} and \autoref{fig:KIM_gradual_AB} show time-dummy price indices where periods are split along the exact announcement and enactment date of the KIM-VO. \autoref{fig:price_segments_A} and \autoref{fig:price_segments_AB} depict price indices per price segment distinguishing the ``bottom 20\%'', ``middle 60\%'' and ``top 20\%'' based on their position within the observed quarter-specific price per square meter distribution. Approaches evaluating indices per object type (apartment or house) or region (urban, largely urban, regional, rural), respectively, yield very similar results.  
\end{scriptsize}
\end{figure}

While the KIM-VO seemed to not have been the primary source of the slow-down in housing market activity, results also indicate that the rate of decrease was slightly lower during the announcement period than in the immediate period before. This may have had impacts on prices.
Further, the implementation of new bank lending standards also meant a mechanical crowding-out of a share of households not meeting the stricter requirements any more.
 
This rise and fall in the number of actors bidding for dwellings suggests elevated prices before the enactment and decreasing prices or at least reduced growth rates thereafter. This is indeed what we find: \autoref{fig:KIM_gradual_A} and \autoref{fig:KIM_gradual_AB} show that from the announcement onward, both $(A)$ and $(A^B)$ increased at an elevated pace, peaking at enactment. Subsequently, $(A^B)$ drop. $(A)$ show a similar yet less pronounced pattern indicating that the change was driven by the demand-side. When splitting time dummies to exactly align with the enactment, it becomes evident that the drops indeed coincide almost perfectly with the enactment.

We hypothesized that only buyers strongly dependent on mortgages to finance a purchase would be affected by the policy. These are, however, the majority of households: As noted in \autoref{foot:albacete}, only roughly a quarter of all households in 2009 purchased their main residence without a mortgage.
We use this to zoom into the underlying dynamics: Assuming perfect sorting, i.e., the price of dwellings purchased by households correlates with their affluence, implies that purchase prices strongly correlate with buyers' income and wealth. The demand for lower-priced dwellings (inline with the share of purchases with mortgages identified here as the bottom 80\%) should thus decrease upon enactment. Between announcement and enactment, there may thus be extra demand driven by forward-looking agents suggesting elevated prices for lower-priced dwellings. Upon enactment, we would expect falling prices in all but the top price segments. 

These predictions are confirmed: \autoref{fig:price_segments_A} and \autoref{fig:price_segments_AB} differentiate price trajectories by price segments (top 20\% of observed prices, the middle 60\%, and the bottom 20\%). While the bottom and middle segments -- associated with buyers likely relying on mortgages -- saw reduced prices upon enactment, prices in the top segment remained constant though having also shown increases before enactment. While we argue that the demand-side leads, the pattern is visible in both $(A)$ and $(A^B)$ suggesting that the supply-side may have been aware of and willing to adjust to this trend. Numerical results are reported in \autoref{tab:lending_price_effects}.

\subsection{Cost of Living Crisis}
\subsubsection{Hypothesis 5: Cost of Living Crisis Quantity Effects} \label{sec:results.quantities.inflation}

\begin{table}[h]
\begin{center}
\caption{Cost of Living Crisis: Quantity Effects $(D)$ and $(A)$}
\label{tab:coefficients_count_col}
\resizebox{0.75\linewidth}{!}{
\begin{tabular}{l c c c c c}
\toprule \toprule
& $(D)$ & $(D)$  &&  $(A)$ & $(A)$ \\\
 & $(1)$ & $(2)$    && $(3)$ & $(4)$ \\
\cmidrule{2-3} \cmidrule{5-6}
Q1 2021 $\times$ HICP            & $0.197^{***}$  &              &  & $0.133^{*}$    &                \\
                               & $(0.043)$      &                && $(0.060)$      &                \\
Q2 2021 $\times$ HICP            & $0.082^{**}$   &              &  & $0.009$        &                \\
                               & $(0.027)$      &                && $(0.037)$      &                \\
Q3 2021 $\times$ HICP            & $0.065^{**}$   &              &  & $-0.023$       &                \\
                               & $(0.023)$      &                && $(0.031)$      &                \\
Q4 2021 $\times$ HICP            & $0.088^{***}$  &              &  & $-0.016$       &                \\
                               & $(0.018)$      &                && $(0.025)$      &                \\
Q1 2022 $\times$ HICP            & $-0.003$       &              &  & $0.022$        &                \\
                               & $(0.012)$      &                && $(0.017)$      &                \\
Q2 2022 $\times$ HICP             & $-0.031^{***}$ &             &   & $0.000$        &                \\
                               & $(0.009)$      &                && $(0.012)$      &                \\
Q3 2022 $\times$ HICP            & $-0.031^{***}$ &              &  & $-0.037^{***}$ &                \\
                               & $(0.007)$      &                && $(0.010)$      &                \\
Q4 2022 $\times$ HICP            & $-0.005$       &              &  & $-0.024^{**}$  &                \\
                               & $(0.006)$      &                && $(0.009)$      &                \\
Q1 2023 $\times$ HICP            & $-0.035^{***}$ &              &  & $0.006$        &                \\
                               & $(0.007)$      &                && $(0.009)$      &                \\
Mortgage Rate (3m lag) $\times$ Low policy rate environment  &                & $0.136$        &&                &                \\
                               &                & $(0.104)$      &                &            &    \\
Mortgage Rate (3m lag) $\times$ First policy rate increase &                & $-0.029$       &  &              &                \\
                               &                & $(0.072)$      &                &             &   \\
Mortgage Rate (3m lag) $\times$ High policy rate environment &                & $-0.113^{*}$   & &               &                \\
                               &                & $(0.047)$      &                &                \\
Mortgage Rate $\times$ Low policy rate environment      &                &                &   &             & $-0.084$       \\
                               &                &                &                && $(0.072)$      \\
Mortgage Rate $\times$ First policy rate increase      &                &         &       &                & $-0.155^{***}$ \\
                               &                &                &                && $(0.039)$      \\
Mortgage Rate $\times$ High policy rate environment   &                &          &      &                & $-0.022$       \\
                               &                &                &                && $(0.029)$      \\
\midrule
Housing type                                  & \checkmark     & \checkmark    & & \checkmark      & \checkmark     \\ 
Time Dummies                                  & \xmark         & \xmark      && \xmark      & \xmark   \\
Cyclical Trend                              & \checkmark         & \checkmark  &   & \checkmark      & \checkmark    \\
Location Fixed Effects                        & \checkmark     & \checkmark     & & \checkmark      & \checkmark      \\
\midrule
BIC                            & $52,435$    & $52,940$    && $23,125$    & $23,189$    \\
Num. obs.                      & $7,878$         & $7,878$       &  & $5,036$         & $5,036$         \\

\bottomrule \bottomrule
\multicolumn{3}{l}{\scriptsize{$^{***}p<0.001$; $^{**}p<0.01$; $^{*}p<0.05$}}
\end{tabular}
}
\end{center}
\vspace{-0.4cm}
{\scriptsize
\emph{Notes:} 
Models (1) and (3) report coefficients associated with time-dummies interacted with changes in the Austrian HICP. Further, models (2) and (4) include average newly granted mortgage rates per sub-regime. Due to the delay between fixing an interest rate and recording a transaction, a lag of three months is respected for $(D)$. Results are robust to changes in the lag length (2-4 months). 
}
\end{table}

As already seen in \autoref{fig:KIM_gradual_counts}, counts started to decrease even before the KIM-VO was announced, which was also long before the first policy rate hike (see \autoref{fig:gradual_counts_II} in the Appendix). It is evident that deeds fell first and are only followed by adverts with a time lag. The largest decreases are found in on average cheaper areas. Again, the effect seems to be led by the demand-side as $(D)$ lead $(A)$. This is confirmed when incorporating the full Austrian HICP into our count model in
\autoref{tab:coefficients_count_col}.
The HICP shown in columns (1) and (3) correlates with counts as expected: increasing general price levels put a downward pressure on the number of transactions. From Q2-2022 onward, the HICP is consistently negatively and significantly correlated with counts. 
As buyers' budgets are tightening, the effect is predominantly visible in $(D)$ indirectly also measuring buyers' actions, yet less so in $(A)$ which purely describes sellers' actions.%
\footnote{The supply-side is also affected by increased prices, yet this may predominantly affect the planning of new developments and should thus result in decreased advertising in the future.}

We use mortgage rates as another positive measure. Results are less conclusive but point again into the direction of an indirect relationship between trade volume and cost -- here, the cost of borrowing: Model (2) measures an increasingly more negative and finally significant correlation between increases in realised mortgage rates and the number of deeds $(D)$ across policy rate regimes. 
An initially insignificant correlation, finally turns negative and slightly statistically significant when policy rates were at their peak.
In contrast, $(A)$ shown in Model (4), though indicating a negative effect, do not follow a consistent pattern over time.

\subsubsection{Hypothesis 6: Cost of Living Crisis Price Effects}\label{sec:results.prices.inflation}

\begin{figure}[h]
\begin{center}    
  \caption{Cost of Living Crisis: Price Effects $(A)$ and $(A^B)$}\label{fig:COL_prices}
    \begin{subfigure}[t]{0.45\textwidth}
    \caption{Normalisation to First Regime Switch}
    \centering
      \label{fig:COL_prices_1}
\includegraphics[width=\textwidth]{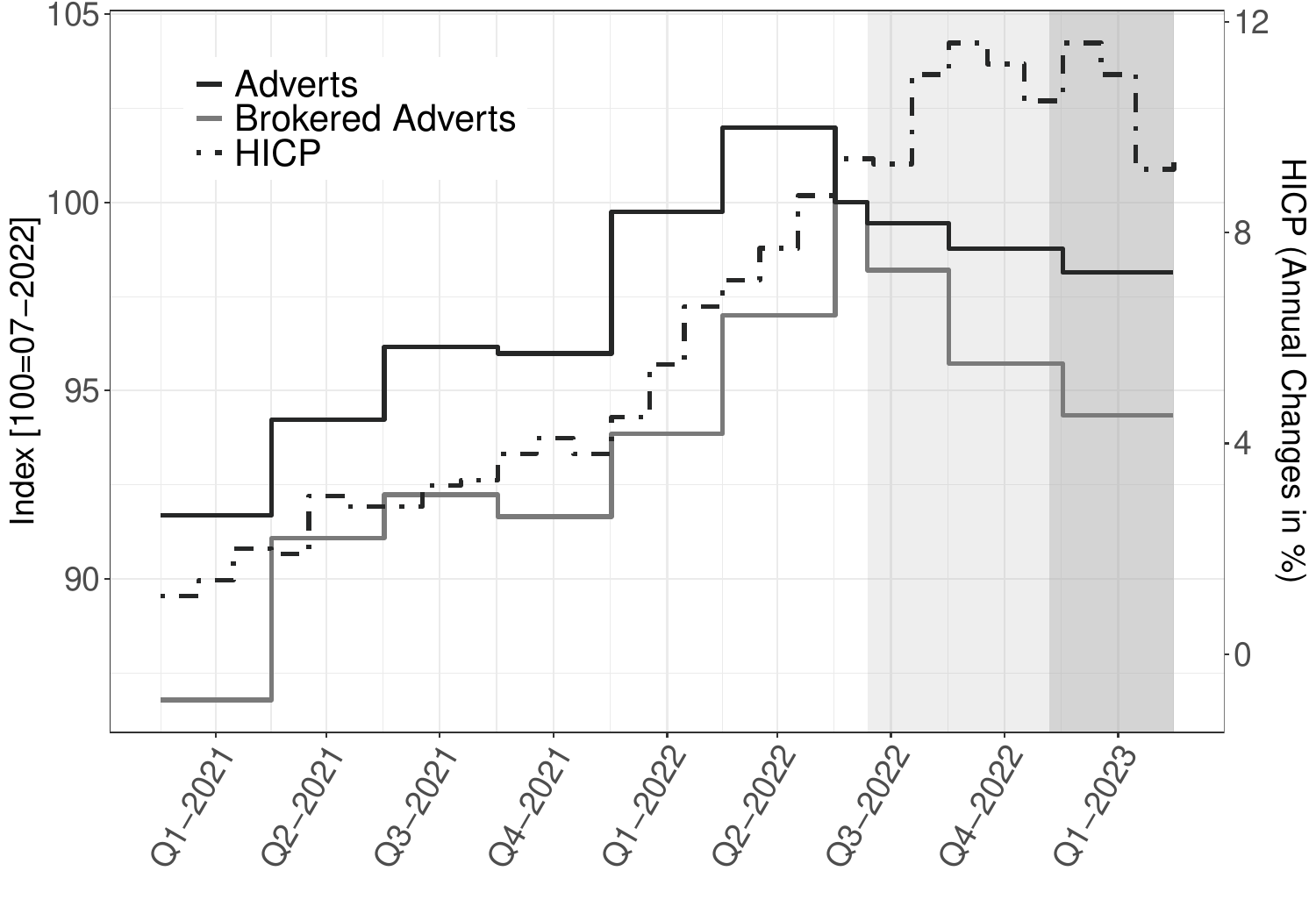}
\end{subfigure}
\hfill
\begin{subfigure}[t]{0.45\textwidth}
\centering
 \caption{Normalisation to Second Regime Switch}
    \label{fig:COL_prices_2}
    \includegraphics[width=\textwidth]{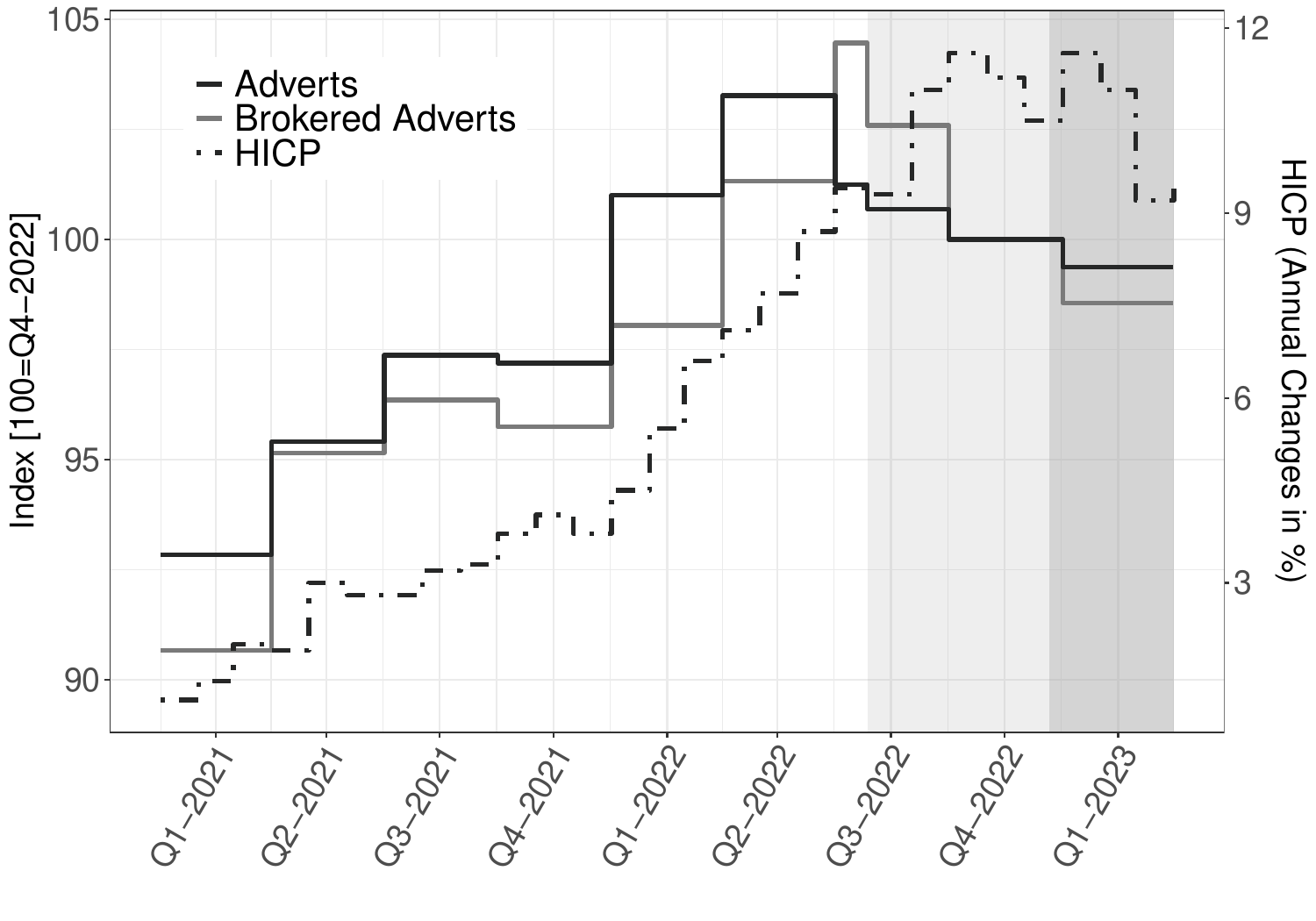}
\end{subfigure}
\end{center}
\begin{footnotesize}
    \emph{Notes:} The Figures contrast inflation in Austria (HICP year-to-year changes) and RPPIs. RPPI time-dummies are split to match changes in regimes. RPPIs in \autoref{fig:COL_prices_1} are normalised to 01-26 July 2022, and in \autoref{fig:COL_prices_2} to the entire Q4-2022 as the rate was increased on 21st of December 2022 only leaving too few observations for a stable estimate. The shaded area highlights interest rate regimes.\\
\end{footnotesize}
\end{figure}

\autoref{fig:COL_prices} shows price effects. When comparing the correlations of house prices with changes in the HICP between $(A^B)$ and $(A)$, we find again stronger relations for $(A^B)$ indicating that the demand-side and supply-side perceptions deviated, and were over large parts of this period more pronounced in models estimated on $(A^B)$. This suggests that the demand side lead this effect as we have already shown for counts: decreased demand for properties also means that final prices measured by $(A^B)$ would gradually fall -- or at least not increase any more. The supply-side did not only keep posting advertisements as suggested before, sellers were also not willing to offer properties at the by then already decreased market price yet a certain share of sellers was apparently willing to accept nominal price decreases.

When comparing changes in house price indices with changes in overall consumer prices, it is clear that housing leads and indicates an earlier turning point. This is not surprising as the HICP in entire Europe does not account for changes in owner-occupied housing costs and by that does not react at all to housing booms-bust cycles \citep[see][for a discussion]{hill2023owner}. We conclude that a decrease in demand did also lead to lower prices while sellers sentiments were more sticky confirming item 3 of \autoref{hyp:prices.inflation}.

\autoref{fig:COL_prices} shows this explicitly: 
RPPIs extracted from $(A)$ and $(A^B)$ are shown together with the HICP. Again, we split quarterly time-dummies to match the timing of hikes in the policy rate. We hypothesised that due to shrinking demand, $(A^B)$ would react first. However, it turns out that $(A)$ falls first. This is likely because at this date the KIM-VO had already been announced but not yet enforced, and this period may have still encouraged some credit-constrained households to bid up prices in Q3-2022 before restrictions became binding. Still, price decreases following the first policy rate hike were much larger in magnitude in $(A^B)$ than in $(A)$ as shown in \autoref{fig:COL_prices_1}. This suggests that the demand-side was in the mid-term the driving force behind this change. When moving from the first to the second interest rate regime, price decreases were again larger in $(A^B)$ than in $(A)$ as shown in \autoref{fig:COL_prices_2} indicating that the demand-side could bid down prices which fits the demonstrated decrease in demand measured by the number of final transactions in \autoref{sec:results.quantities.inflation}.

\begin{table}[h]
\begin{center}
\caption{Correlations between Mortgage Rates and RPPIs by Subregime}
\label{tab:corr_mortgage_rate_prices}
\begin{tabular}{rrr | rrr|}
&\multicolumn{2}{c}{ }& \multicolumn{3}{c}{\MakeUppercase{Mortgage Rates}}\\
&\multicolumn{2}{c}{ }& \multicolumn{1}{c}{Low Rate} & \multicolumn{1}{c}{First Increase}  & \multicolumn{1}{c}{High Rate}  \\ 
\cline{4-6}
\multirow{6}{*}{\begin{turn}{90} RPPI\end{turn}}&\multirow{2}{4em}{\hfil Subregime 1} & $(A)$ & $0.643^{**}$  & -- & -- \\ 
                & & $(A^B)$             & $0.682^{***}$ & -- & --\\
&\multirow{2}{4em}{\hfil Subregime 2} & $(A)$   & -- & $-0.200$ &--\\ 
                            && $(A^B)$   & -- & $-0.879^{*}$  &--\\
&\multirow{2}{4em}{\hfil Subregime 3} & $(A)$  &   -- & -- & $0.579$   \\
                            && $(A^B)$  & -- & -- & $-0.400$  \\ 
                            \cline{4-6}
\end{tabular}
\end{center}
\begin{footnotesize}
    \emph{Notes:} The table reports Pearson correlation coefficients between mortgage interest rates and RPPIs computed from $(A)$ and $(A^B)$, respectively. 
    Significance codes refer to a one-sided paired Pearson correlation test with $H_0: \; \rho(\text{RPPI,Mortgage Rate} ) = 0$ vs. $H_1: \; \rho (\text{RPPI,Mortgage Rate} ) > 0$ for Subregime 1 and  $H_0: \; \rho(\text{RPPI,Mortgage Rate} ) = 0$ vs. $H_1: \; \rho (\text{RPPI,Mortgage Rate} )<0$ for Subregimes 2 and 3 and follow standard notation: $^{***}p<0.001$; $^{**}p<0.01$; $^{*}p<0.05$.
\end{footnotesize}
\end{table}

Further, we assess again average granted mortgage rates as a proxy for the cost of purchasing.
We correlate mortgage rates with month-to-month changes in price indices extracted from our baseline model (\autoref{app:baseline.model}). \autoref{tab:corr_mortgage_rate_prices} reports results: The correlation differs in sign and significance across the three regimes. While there was a strong co-movement between mortgage rates and both, advertised and final prices, this correlation turns negative thereafter for $(A^B)$. 
Results for $(A)$ do not consistently follow this change in trends. 
Complementary full regression results are reported in \autoref{tab:CoL_price_effects} in the Appendix.


\section{Conclusions}\label{sec:conclusions}

This study investigates the impact of using various types of real estate price and quantity data to identify the effect of exogenous shocks on housing markets. We argue that at least three dimensions need to be differentiated for such assessments: prices versus quantities, immediate versus gradually evolving, and led by the supply (sellers) or demand side.
For enabling these tests, we develop a hierarchical hedonic price model as well as a complementary count model fit to describe an entire county's housing market.

Making use of a rich data pool describing the Austrian housing market and a sequence of three housing-external shocks, we demonstrate how an unsuitable data type or testing strategy may lead to misleading conclusions and provide guidelines how to a priori judge which data source to prefer. We claim that a discussion about the ideal data type for a planned analysis along at least the three dimensions discussed in this paper should be part of any pre-analysis plan involving identification using real estate data. This would increase reliability. In the long term, such a framework should also enable better comparability of research results across single studies and, by that, would allow the research community to draw meta-conclusions from the bulk of scientific studies available across time and space.
Further, results from our testing framework suggest to extended the list of measures used to monitor housing markets by supervision bodies for financial stability purposes.



\bibliographystylelatex{apalike}


\newpage
\appendix
\section*{Appendix}\label{appendix}

\setcounter{page}{1}
\setcounter{figure}{0}
\setcounter{table}{0}

\setcounter{footnote}{0} 

\renewcommand\thepage{\arabic{page}}
\renewcommand\thefootnote{A.\arabic{footnote}}
\renewcommand\thefigure{A.\arabic{figure}}
\renewcommand\thetable{A.\arabic{table}}
\renewcommand\thesection{A.\arabic{section}}

\section{Data and Baseline Models}
\subsection{Descriptive Statistics}\label{app:summary}

\begin{table}[h]
\begin{center}
\caption{Summary Statistics -- $(D)$}
\label{tab:summary_deeds}
\begin{tabular}{lrr}
  \toprule
  \toprule
 &  Count & \quad Share [\%] \\
 \cmidrule{2-3}
Object & \\ 
  \quad Apartments & 193,800 & 73.30 \\ 
  \quad Single-family houses & 70,457 & 26.70 \\ 
Federal States && \\ 
  \quad Burgenland & 6,909 & 2.60 \\ 
  \quad Carinthia & 16,568 & 6.30 \\ 
  \quad Lower Austria & 47,022 & 17.80 \\ 
  \quad Upper Austria & 38,290 & 14.50 \\ 
  \quad Salzburg & 19,919 & 7.50 \\ 
  \quad Styria & 36,138 & 13.70 \\ 
  \quad Tyrol & 26,235 & 9.90 \\ 
  \quad Vienna & 57,706 & 21.80 \\ 
    \quad Vorarlberg & 15,470 & 5.90 \\ 
Area classification & &\\ 
  \quad Urban & 127,032 & 48.10 \\
  \quad Largely urban & 33,551 & 12.70 \\ 
  \quad Regional & 52,469 & 19.90 \\ 
  \quad Rural & 51,205 & 19.40 \\ 
  \midrule
  Number of Observations & 264,257 & 100\\
   \bottomrule
   \bottomrule
\end{tabular}
\end{center}
\begin{footnotesize}
    \emph{Notes:} The table reports counts and shares of categorical characteristics reported in notary deeds.\\
    \emph{Source:} Land Registry
\end{footnotesize}
\end{table}
\begin{table}
\begin{center}
\caption{Summary Statistics: Categorical Variables -- $(A^B)$ and $(A)$ }\label{tab:summarystats_cat}
\begin{scriptsize}
\begin{tabular}{l cccc}
 \toprule 
 \toprule
 & \multicolumn{2}{c}{Brokered Advertisements $(A^B)$} & \multicolumn{2}{c}{Advertisements $(A)$} \\
 &  Count & Share [\%] &  Count & Share [\%] \\
\cmidrule{2-5}
Type  \\
    \quad Apartments & 28569 & 71.10 & 52397 & 78.80 \\
    \quad Single-family houses & 11613 & 28.90 & 14132 & 21.20 \\ 
Federal State &&&& \\
    \quad Burgenland & 833 & 2.10 & 1073 & 1.60 \\ 
    \quad Carinthia & 2066 & 5.10 & 2153 & 3.20 \\ 
    \quad Lower Austria & 8178 & 20.40 & 12076 & 18.20 \\ 
    \quad Salzburg & 3122 & 7.80 & 3100 & 4.70 \\ 
    \quad Styria & 4356 & 10.80 & 5847 & 8.80 \\ 
    \quad Tyrol & 2888 & 7.20 & 2881 & 4.30 \\ 
    \quad Upper Austria & 4606 & 11.50 & 4388 & 6.60 \\ 
    \quad Vienna & 13033 & 32.40 & 34380 & 51.70 \\ 
    \quad Vorarlberg & 1100 & 2.70 & 631 & 0.90 \\ 
Location Classification &&&& \\ 
    \quad urban & 23333 & 58.10 & 48214 & 72.50 \\ 
    \quad largely urban & 4847 & 12.10 & 5330 & 8.00 \\ 
    \quad regional & 6532 & 16.30 & 7770 & 11.70 \\ 
    \quad rural & 5470 & 13.60 & 5215 & 7.80 \\ 
No. of Rooms &&&& \\ 
    \quad 1 & 1416 & 3.50 & 2964 & 4.50 \\
    \quad 2 & 9223 & 23.00 & 18369 & 27.60 \\ 
    \quad 3 & 11535 & 28.70 & 19761 & 29.70 \\ 
    \quad $>$3 & 18008 & 44.80 & 25435 & 38.20 \\ 
Age &&&& \\ 
    \quad 0 (New) & 9206 & 22.90 & 18448 & 27.70 \\ 
    \quad 1-5 & 4697 & 11.70 & 10359 & 15.60 \\ 
    \quad 6-10 & 1534 & 3.80 & 2262 & 3.40 \\ 
    \quad 11-20 & 3330 & 8.30 & 4599 & 6.90 \\ 
    \quad 21-40 & 6222 & 15.50 & 8004 & 12.00 \\ 
    \quad post war period & 10555 & 26.30 & 13515 & 20.30 \\ 
    \quad 1914-45 & 1149 & 2.90 & 1639 & 2.50 \\ 
    \quad vintage & 3489 & 8.70 & 7703 & 11.60 \\ 
Renovated &&&& \\ 
    \quad No & 29640 & 73.80 & 51268 & 77.10 \\ 
    \quad Yes & 10542 & 26.20 & 15261 & 22.90 \\ 
Open Space &&&& \\ 
    \quad None & 8500 & 21.20 & 14931 & 22.40 \\ 
    \quad $<$15m$^2$ & 14651 & 36.50 & 23103 & 34.70 \\ 
    \quad $>$15m$^2$ & 17031 & 42.40 & 28495 & 42.80 \\ 
Basement &&&& \\
    \quad No & 20216 & 50.30 & 40527 & 60.90 \\ 
    \quad Yes & 19966 & 49.70 & 26002 & 39.10 \\ 
Parking &&&&\\ 
    \quad not included & 17515 & 43.60 & 32273 & 48.50 \\ 
    \quad  included & 22667 & 56.40 & 34256 & 51.50 \\ 
Air Conditioning &&&& \\ 
    \quad not available & 38331 & 95.40 & 61979 & 93.20 \\ 
    \quad available & 1851 & 4.60 & 4550 & 6.80 \\ 
Barrier-free Accessible &&&& \\ 
    \quad No & 32956 & 82.00 & 50590 & 76.00 \\ 
    \quad Yes & 7226 & 18.00 & 15939 & 24.00 \\ 
Wellness Facilities &&&& \\ 
    \quad No & 38858 & 96.70 & 64119 & 96.40 \\ 
    \quad Yes & 1324 & 3.30 & 2410 & 3.60 \\ 
Accessibility & & & &  \\ 
   \quad 1 (least remote) & 7732 & 19.20 & 14007 & 21.10 \\ 
   \quad 2 & 6696 & 16.70 & 14324 & 21.50 \\ 
   \quad 3 & 6262 & 15.60 & 13985 & 21.00 \\ 
   \quad 4 & 8679 & 21.60 & 13167 & 19.80 \\ 
   \quad 5 (most remote) & 10813 & 26.90 & 11046 & 16.60 \\ 
General Condition &&&& \\ 
    \quad First-time occupancy & 8762 & 21.80 & 26616 & 40.00 \\ 
    \quad As new & 13002 & 32.40 & 19701 & 29.60 \\ 
    \quad Poor & 1751 & 4.40 & 3335 & 5.00 \\ 
    \quad Unclassified & 16667 & 41.50 & 16877 & 25.40 \\ 
    \midrule
    Number of Observations & 40,182 & 100 & 66,529 & 100\\
\bottomrule 
\bottomrule
\end{tabular}
\end{scriptsize}
\end{center}
\begin{scriptsize}
   \emph{Notes:} Counts and percentage of hedonic characteristics for brokered adverts ($A^B$) and adverts ($A$). Time period: 01.01.2019 -- 31.03-2023.
    \end{scriptsize}
\end{table}

\autoref{tab:summary_deeds}, \autoref{tab:summarystats_cat} and \autoref{tab:summarystats_num} provide summary statistics for notary deeds, realised (`brokered') adverts and adverts.

\begin{figure}[h]
\begin{center}
  \caption{Regional Distribution of Deeds and Population} 
  \label{fig:map_deeds_pop}
  \subfloat[(a)][]{\label{fig:map_deeds}
\includegraphics[width=0.48\textwidth]{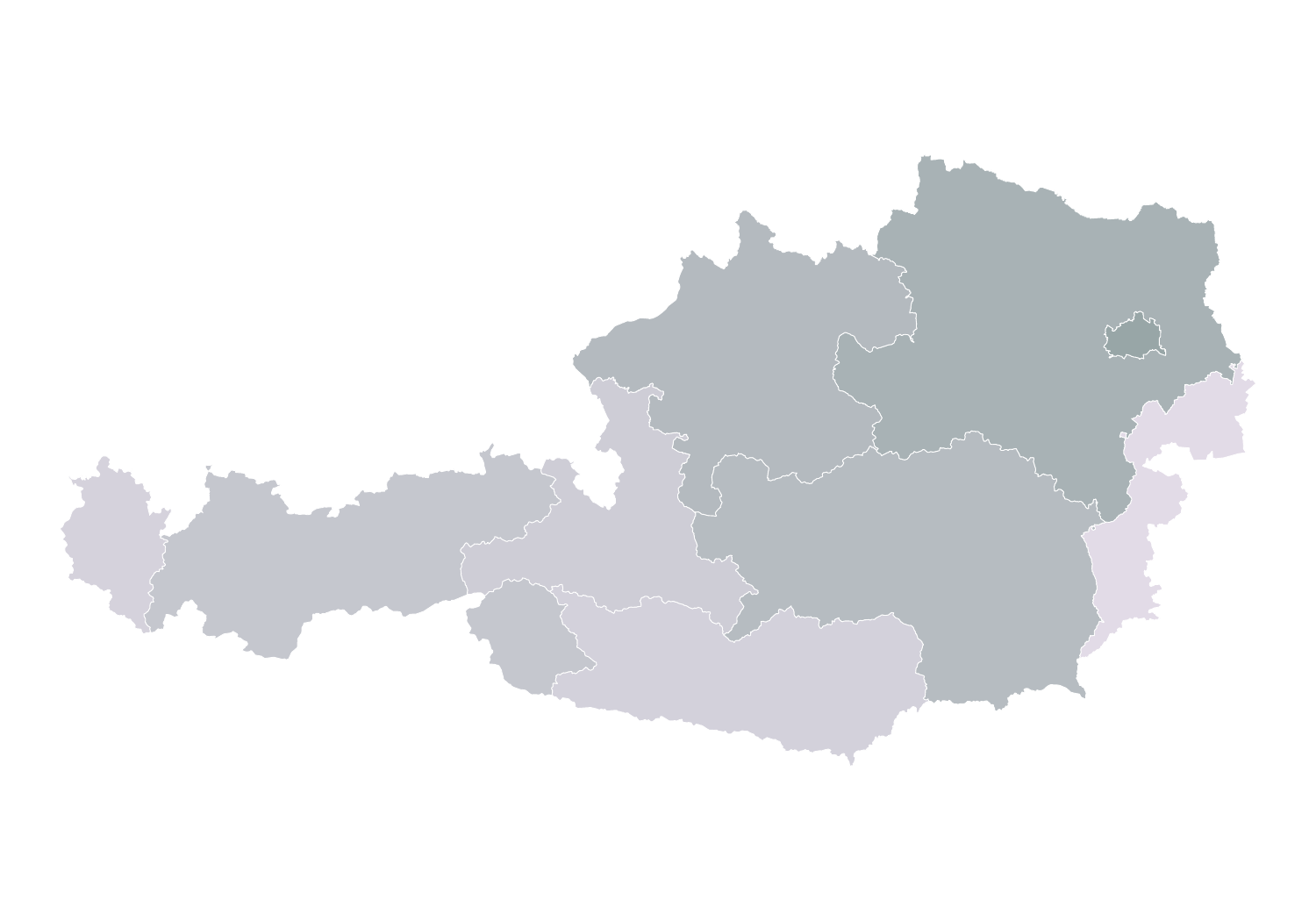}
}
\hfill
  \subfloat[(b)][]{\label{fig:map_pop}
\includegraphics[width=0.48\textwidth]{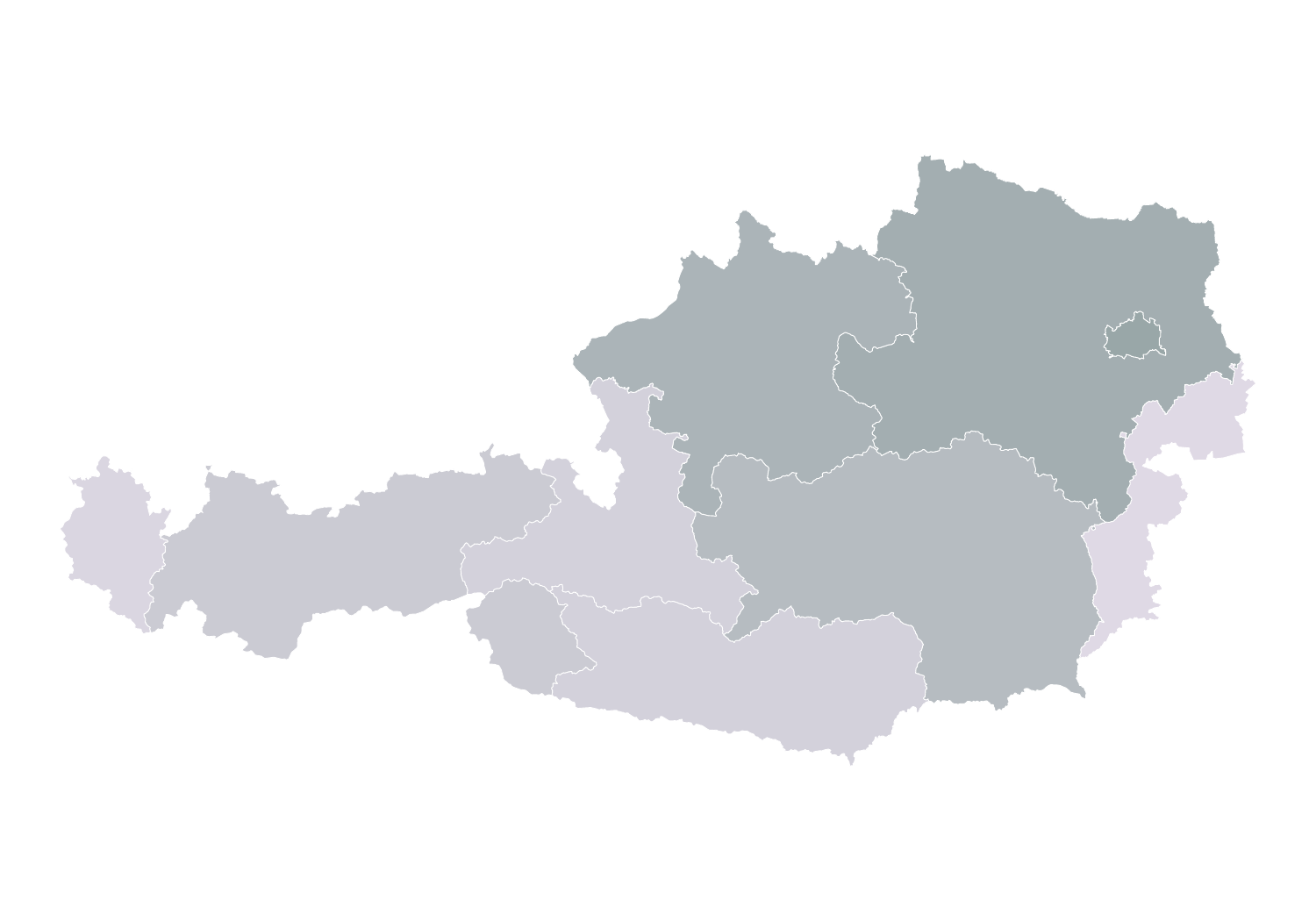}
}
\end{center}
\begin{center}
    \includegraphics[width=0.2\textwidth]{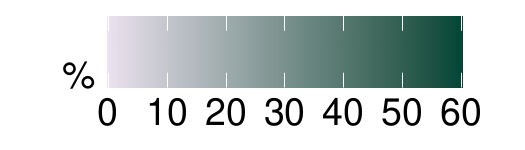}
\end{center}
\begin{footnotesize}
    \emph{Notes:} \autoref{fig:map_deeds} depicts the regional distribution of recorded deeds and \autoref{fig:map_pop} the share of the total Austrian population registered per federal state.\\
    \emph{Source:} Land registry \& Statistik Austria
\end{footnotesize}
\end{figure}

\begin{figure}[h]
\begin{center}
  \caption{Differences in Price Levels on Federal States and District Level ($A^B$)} 
  \label{fig:price_bl}
    \begin{subfigure}[t]{0.45\textwidth}
    \centering
    \caption{}
    \label{fig:price_bl:a}
\includegraphics[width=1\textwidth]{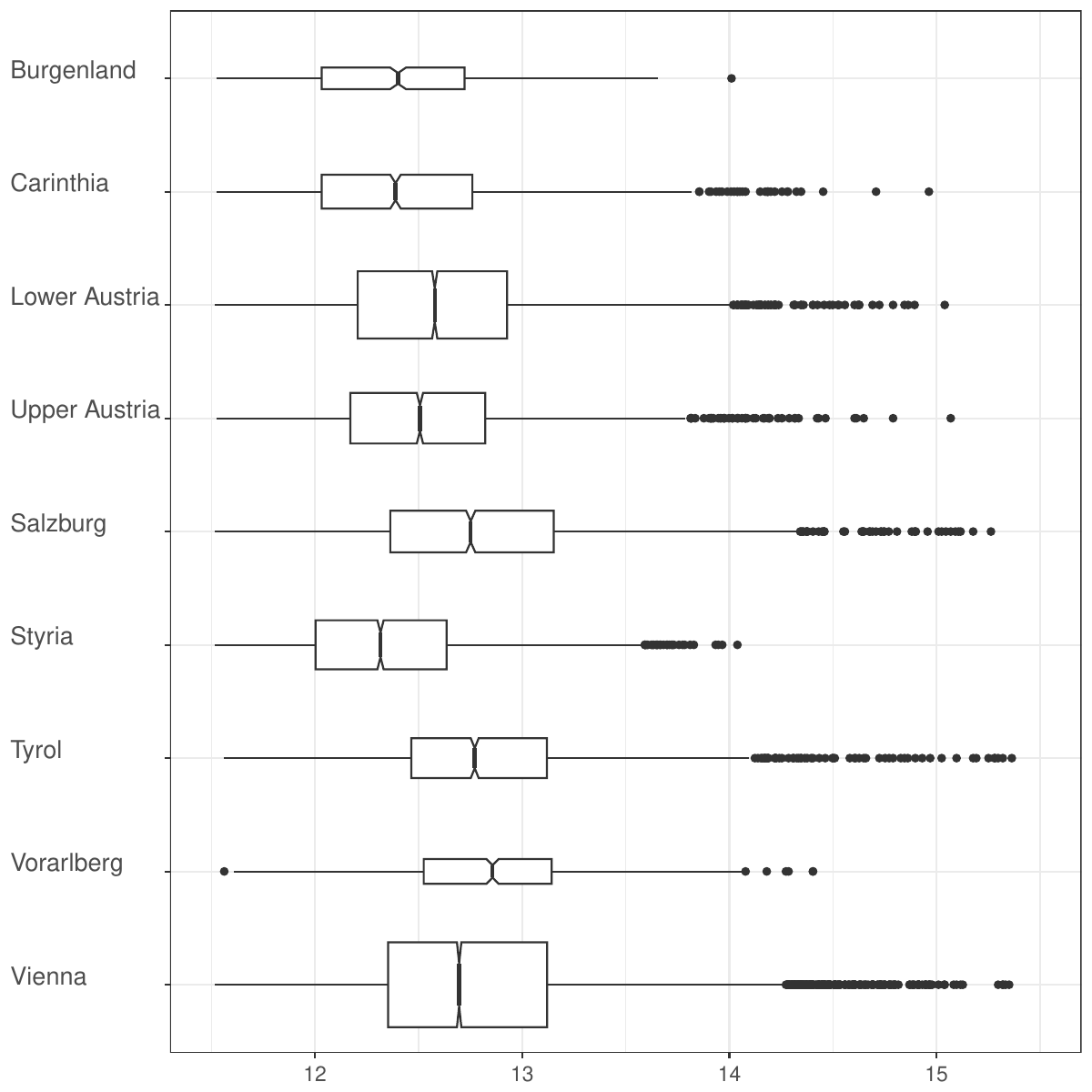}
    \end{subfigure}
        \begin{subfigure}[t]{0.45\textwidth}
    \centering
    \caption{}
    \label{fig:price_bl:b}
\includegraphics[width=1\textwidth]{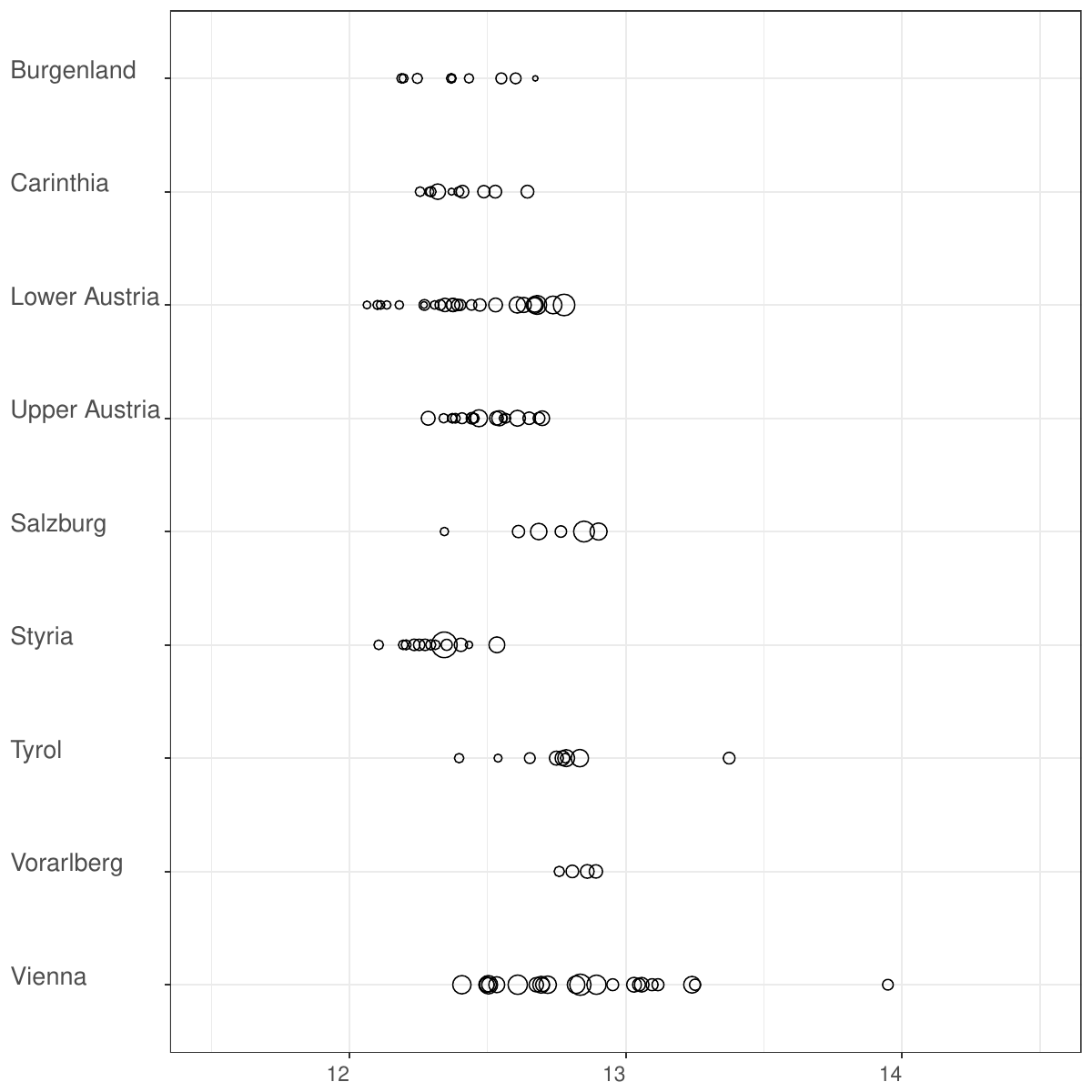}
\end{subfigure}
\end{center}
\begin{footnotesize}
    \emph{Notes:} \autoref{fig:price_bl:a} depicts boxplot with binned logarithmic prices by federal states (the width corresponds to square root of number of observations on federal states level) and \autoref{fig:price_bl:b} mean logarithmic prices on district level for federal states (the size corresponds to the number of observations on district level). Time period: 01.01.2019 -- 31.03.2023. \\
    \emph{Source:} DataScience Service GmbH
\end{footnotesize}
\end{figure}

\begin{table}[h]
\begin{center}
\caption{Summary Statistics: Continuous Variables -- $(A^B)$, $(A)$}
\label{tab:summarystats_num}
\begin{tabular}{c ccccc}
\toprule
\toprule
 & Mean & Median & Std. Dev. & Q1 & Q3  \\ 
 \cmidrule{2-6}
\multicolumn{5}{l}{\textbf{Deeds $(D)$}} \\
 Price [EUR] & 368,274.86 & 297,000.00 & 285,661.06 & 206,000.00 & 430,800.00  \vspace*{2mm}\\
\multicolumn{5}{l}{\textbf{Brokered Adverts $(A^B)$}}  \\
Price [EUR] & 368,370.41 & 297,000.00 & 286,337.19 & 206,204.75 & 431,268.25  \\
Living Area [sqmt]  & 100.02 & 86.92 & 54.11 & 64.10 & 121.64 \\ 
Land price  [EUR/sqmt]& 599.56 & 500.00 & 503.73 & 200.00 & 900.00  \vspace*{5mm}\\
\multicolumn{5}{l}{\textbf{Adverts ($A$)} }\vspace*{2mm}
\\
Price [EUR] & 452,458.65 & 336,900.00 & 411,873.22 & 232,550.00 & 510,840.00  \\
Living Area [sqmt] & 97.57 & 83.24 & 56.69 & 60.61 & 118.70  \\ 
Land price [EUR/sqmt] & 744.09 & 675.00 & 557.66 & 322.50 & 1000.00  \\
\bottomrule
\bottomrule
\end{tabular}
\end{center}
\begin{footnotesize}
    \emph{Notes:} The table reports summary statistics for continuous variables. Time period: 01.01.2019 -- 31.03.2023.
\end{footnotesize}
\end{table}

Non-surprisingly, we observe heterogeneous price levels across federal states (\autoref{fig:price_bl:a}) indicating high-price areas in touristic hotspots in Salzburg and Tyrol, as well as the capital Vienna.
Assessing prices on the lower district-level, \autoref{fig:price_bl:b} reveals that the data exhibits substantial heterogeneity in both the level and variability, whereby the most expensive districts in Vienna (1st district -- Innere Stadt/Inner City) and Tyrol (Kitzbühel) reach the highest levels in the country.

\subsection{Baseline Models}\label{app:baseline.model}
\autoref{tab:baseline_price} and \autoref{tab:baseline_count} report estimation results for our baseline price and count models. These models include the respective (hedonic) characteristics yet leave out regime and contextual variables. In the remainder of the article, only estimation results for these additional variables used for identification and interpretation are reported. 

We distinguish between apartments and single houses in the models as these two types exhibit different market characteristics. 
In the price model, location is modelled in two hierarchical layers: federal states and  district within states. 
We complement these information by further locational variables, i.e., the land price and the classification of districts to urban and rural areas based on the official statistical categorisation informed by the degree of urbanisation, accessibility and tourism activities (see \autoref{sec:stataut} for details). 
Further, we consider standard hedonic variables describing the structure of the apartments and houses. 
In the count model, location is considered in the form of federal state fixed effects and urban-rural classifiers. Also, a cyclical trend is modelled to capture seasonal patterns. Both models include time dummies.

\begin{longtable}{lcc}
\caption{Baseline Price Model} \label{tab:baseline_price}\\
\toprule
\toprule
& \multicolumn{2}{c}{Price}\\
 & $(A)$ & $(A^B)$\\
 & (1) & (2)\\
\cmidrule{2-3}
(Intercept)                       & $7.315^{***}$  & $8.013^{***}$  \\
                                  & $(0.043)$      & $(0.046)$      \\
Log(Area)                         & $0.872^{***}$  & $0.688^{***}$  \\
                                  & $(0.004)$      & $(0.005)$      \\
Apartment (Ref.: House)           & $-0.093^{***}$ & $-0.181^{***}$ \\
                                  & $(0.004)$      & $(0.005)$      \\
No. of Rooms (Ref.: Studio)\\
\; 2 Rooms                        & $-0.002$       & $0.049^{***}$  \\
                                  & $(0.005)$      & $(0.008)$      \\
\; 3 Rooms                        & $0.040^{***}$  & $0.135^{***}$  \\
                                  & $(0.006)$      & $(0.008)$      \\
\; 4 Rooms                        & $0.065^{***}$  & $0.180^{***}$  \\
                                  & $(0.007)$      & $(0.009)$      \\
Age of Dwelling (Ref.: New)\\
\; Age (1-5Y)                     & $-0.036^{***}$ & $-0.020^{***}$ \\
                                  & $(0.003)$      & $(0.005)$      \\
\; Age (6-10Y)                    & $-0.063^{***}$ & $-0.082^{***}$ \\
                                  & $(0.006)$      & $(0.008)$      \\
\; Age (11-20Y)                   & $-0.141^{***}$ & $-0.151^{***}$ \\
                                  & $(0.005)$      & $(0.006)$      \\
\; Age (21-40Y)                   & $-0.257^{***}$ & $-0.271^{***}$ \\
                                  & $(0.005)$      & $(0.005)$      \\
\; Age (Post War)                 & $-0.343^{***}$ & $-0.390^{***}$ \\
                                  & $(0.004)$      & $(0.005)$      \\
\; Age (Wars)                     & $-0.272^{***}$ & $-0.346^{***}$ \\
                                  & $(0.010)$      & $(0.013)$      \\
\; Age (Vintage)                  & $-0.233^{***}$ & $-0.282^{***}$ \\
                                  & $(0.006)$      & $(0.009)$      \\    
Renovated (Ref.: Not Renovated)   & $-0.074^{***}$ & $-0.090^{***}$ \\
                                  & $(0.007)$      & $(0.012)$      \\
Open Space Facilities (Ref.: No)\\
\; Open Space (0,15]              & $0.062^{***}$  & $0.062^{***}$  \\
                                  & $(0.003)$      & $(0.004)$      \\
\; Open Space (15, Inf]           & $0.129^{***}$  & $0.130^{***}$  \\
                                  & $(0.003)$      & $(0.004)$      \\
Basement (Ref.: No Basement)      & $0.016^{***}$  & $0.019^{***}$  \\
                                  & $(0.002)$      & $(0.003)$      \\
Parking (Ref.: No Parking)        & $-0.009^{***}$ & $0.019^{***}$  \\
                                  & $(0.002)$      & $(0.003)$      \\
Air Conditioning (Ref.: No AC)    & $0.083^{***}$  & $0.115^{***}$  \\
                                  & $(0.004)$      & $(0.006)$      \\
Step-free Access (Ref.: No Step-free Access)              & $0.009^{***}$  & $0.023^{***}$  \\
                                  & $(0.003)$      & $(0.004)$      \\
Wellness Facilities (Ref.: No Wellness Area)                    & $0.138^{***}$  & $0.146^{***}$  \\
                                  & $(0.006)$      & $(0.007)$      \\
General Condition (Ref.: Unclassified)\\
\; First Time Occupation          & $0.057^{***}$  & $0.054^{***}$  \\
                                  & $(0.003)$      & $(0.004)$      \\

 \; As new              & $0.019^{***}$  & $0.047^{***}$  \\
                                  & $(0.003)$      & $(0.003)$      \\
\; Poor                     & $-0.164^{***}$ & $-0.133^{***}$ \\
                                  & $(0.005)$      & $(0.007)$      \\
Location Classification (Ref.: Regional) \\
\; Urban                          & $-0.001$       & $0.018^{*}$    \\
                                  & $(0.006)$      & $(0.007)$      \\
\; Largely Urban                  & $-0.012$       & $-0.024^{***}$ \\
                                  & $(0.007)$      & $(0.007)$      \\
\; Rural                          & $0.035^{***}$  & $0.030^{***}$  \\
                                  & $(0.006)$      & $(0.006)$      \\

Log(Mean Plot Price)              & $0.255^{***}$  & $0.251^{***}$  \\
                                  & $(0.003)$      & $(0.004)$      \\
Accessibility Level (Ref.: 1-least remote)\\
\; Accessibility Level 2          & $-0.042^{***}$ & $-0.035^{***}$ \\
                                  & $(0.004)$      & $(0.006)$      \\
\; Accessibility Level 3          & $-0.074^{***}$ & $-0.046^{***}$ \\
                                  & $(0.005)$      & $(0.007)$      \\
\; Accessibility Level 4          & $-0.097^{***}$ & $-0.056^{***}$ \\
                                  & $(0.007)$      & $(0.009)$      \\
\; Accessibility Level 5 (Most Remote) & $-0.131^{***}$ & $-0.082^{***}$ \\
                                  & $(0.008)$      & $(0.010)$      \\
Age $\times$ Renovated (Ref.: New $\times$ Not Renovated)\\
\; Age (1-5Y) $\times$ Renovation & $0.021$        & $0.035$        \\
                                  & $(0.011)$      & $(0.018)$      \\
\; Age (6-10Y) $\times$ Renovation& $0.006$        & $0.085^{**}$   \\
                                  & $(0.019)$      & $(0.027)$      \\
\; Age (11-20Y) $\times$ Renovation& $0.053^{***}$  & $0.080^{***}$  \\
                                  & $(0.013)$      & $(0.018)$      \\
\; Age (21-40Y) $\times$ Renovation& $0.091^{***}$  & $0.102^{***}$  \\
                                  & $(0.010)$      & $(0.014)$      \\
\; Age (Post War) $\times$ Renovation& $0.107^{***}$  & $0.139^{***}$  \\
                                  & $(0.008)$      & $(0.013)$      \\
\; Age (Wars) $\times$ Renovation & $0.132^{***}$  & $0.140^{***}$  \\
                                  & $(0.014)$      & $(0.019)$      \\
\; Age (Vintage) $\times$ Renovation & $0.137^{***}$  & $0.158^{***}$  \\
                                  & $(0.009)$      & $(0.015)$      \\
\midrule
Time Dummies                & \cmark & \cmark\\
\midrule
BIC                               & $8,688$     & $6,125$     \\
Num. of Obs.                      & $66,529$        & $40,182$        \\
\bottomrule
\bottomrule
\end{longtable}
\begin{scriptsize}
   \emph{Notes:} Models are estimated for the time period 01.01.2019 -- 31.03.2023. $^{***}p<0.001$; $^{**}p<0.01$; $^{*}p<0.05$
    \end{scriptsize}

\begin{table}
\begin{center}
\caption{Baseline Quantity Model.}
\label{tab:baseline_count}
\small
 \resizebox{\textwidth}{!}{
\begin{tabular}{l cc cc}
\toprule
\toprule
 & $(D)$ & $(D)$ & $(A)$ & $(A)$\\
 & (1) & (2) & (3) & (4) \\
\cmidrule{2-5}
(Intercept)               & $-0.130^{***}$ & $-0.130^{***}$ & $-1.233^{***}$ & $-1.233^{***}$ \\
                          & $(0.037)$      & $(0.032)$      & $(0.054)$      & $(0.045)$      \\
Apartment (Ref.: House)   & $0.797^{***}$  & $0.797^{***}$  & $0.612^{***}$  & $0.612^{***}$  \\
                          & $(0.012)$      & $(0.012)$      & $(0.019)$      & $(0.019)$      \\
Federal State (Ref.: Lower Austria)\\ 
\; Burgenland             & $0.343^{***}$  & $0.343^{***}$  & $0.787^{***}$  & $0.787^{***}$  \\
                          & $(0.028)$      & $(0.028)$      & $(0.046)$      & $(0.046)$      \\
\; Carinthia              & $0.026$        & $0.026$        & $-0.193^{***}$ & $-0.193^{***}$ \\
                          & $(0.024)$      & $(0.024)$      & $(0.034)$      & $(0.034)$      \\
\; Upper Austria          & $-0.146^{***}$ & $-0.146^{***}$ & $-0.749^{***}$ & $-0.749^{***}$ \\
                          & $(0.023)$      & $(0.023)$      & $(0.029)$      & $(0.029)$      \\
\; Salzburg               & $0.229^{***}$  & $0.229^{***}$  & $0.124^{***}$  & $0.124^{***}$  \\
                          & $(0.025)$      & $(0.025)$      & $(0.035)$      & $(0.035)$      \\
\; Styria                 & $-0.121^{***}$ & $-0.121^{***}$ & $-0.353^{***}$ & $-0.353^{***}$ \\
                          & $(0.023)$      & $(0.023)$      & $(0.029)$      & $(0.029)$      \\
\; Tyrol                  & $0.041$        & $0.041$        & $-0.317^{***}$ & $-0.317^{***}$ \\
                          & $(0.023)$      & $(0.023)$      & $(0.034)$      & $(0.034)$      \\
\; Vorarlberg             & $0.149^{***}$  & $0.149^{***}$  & $-0.290^{***}$ & $-0.290^{***}$ \\
                          & $(0.025)$      & $(0.025)$      & $(0.057)$      & $(0.057)$      \\
\; Vienna                 & $0.797^{***}$  & $0.797^{***}$  & $1.517^{***}$  & $1.517^{***}$  \\
                          & $(0.036)$      & $(0.036)$      & $(0.037)$      & $(0.037)$      \\
Location Classification (Ref.: Urban) \\
\; Rather urban           & $-0.458^{***}$ & $-0.458^{***}$ & $-0.086^{**}$  & $-0.086^{**}$  \\
                          & $(0.019)$      & $(0.019)$      & $(0.030)$      & $(0.030)$      \\
\; Regional               & $-0.085^{***}$ & $-0.085^{***}$ & $0.139^{***}$  & $0.139^{***}$  \\
                          & $(0.017)$      & $(0.017)$      & $(0.027)$      & $(0.027)$      \\
\; Urban                  & $0.224^{***}$  & $0.224^{***}$  & $0.561^{***}$  & $0.561^{***}$  \\
                          & $(0.017)$      & $(0.017)$      & $(0.026)$      & $(0.026)$      \\
\midrule
$\beta_1$ ($f_{cycle}$)   & $-0.085^{***}$ &                & $-0.044$       &                \\
                          & $(0.021)$      &                & $(0.030)$      &                \\
$\beta_2$ ($f_{cycle}$)   & $0.055^{**}$   &                & $-0.102^{**}$  &                \\
                          & $(0.021)$      &                & $(0.031)$      &                \\
                                                    \midrule
Date of Minimum $f_{cycle}$  & November& & January & \\
Date of Maximum $f_{cycle}$  &  May & & July & \\
                          \midrule
Time Dummies                & \cmark & \cmark& \cmark& \cmark  \\
Seasonal Trend as Offset    & \xmark & \cmark & \xmark & \cmark \\
\midrule
BIC                       & $100,779$   & $100,759$   & $44,082$    & $44,064$    \\
Num. of obs.                 & $14,795$        & $14,795$        & $9,312$         & $9,312$         \\
\bottomrule\bottomrule
\end{tabular}
}
\end{center}
\begin{scriptsize}
   \emph{Notes:} Models are estimated for the time period 01.01.2019 -- 31.03.2023. $^{***}p<0.001$; $^{**}p<0.01$; $^{*}p<0.05$
    \end{scriptsize}
\end{table}


\subsection{Supplemental Data}
\label{sec:stataut}

To supplement our price and count baseline models, we use data on population and degree of urbanisation provided by Statistik Austria. 
The count model uses population per federal state%
\footnote{The data can be retrieved via \url{https://www.statistik.at/statistiken/bevoelkerung-und-soziales/bevoelkerung/bevoelkerungsstand/bevoelkerung-zu-jahres-/-quartalsanfang}, last accessed February 2024.} 
as exposure variable (see \autoref{sec:count_model}). Additionally, population counts are used to express the COVID-19 mortality rate in per capita levels. 

The urban-rural typology by Statistik Austria%
\footnote{See \url{https://www.statistik.at/atlas/?mapid=topo_stadt_land&layerid=layer1}, last accessed in March 2024, for more information.} 
informs our classification of regions into \textit{urban} (region type 101), \textit{largely urban} (region types 102 and 103), \textit{regional} (region types 210-330) and \textit{rural} (region types 410-430). Those classifications are used as explanatory variables in both models. \autoref{tab:summary_deeds} and \autoref{tab:summarystats_cat} provide an overview of the spatial distribution of the respective real estate data across regions.


\setcounter{figure}{0}
\setcounter{table}{0}

\setcounter{footnote}{0} 

\renewcommand\thepage{\arabic{page}}
\renewcommand\thefootnote{B.\arabic{footnote}}
\renewcommand\thefigure{B.\arabic{figure}}
\renewcommand\thetable{B.\arabic{table}}
\renewcommand\thesection{B.\arabic{section}}

 \section{Additional Tables and Figures}\label{app:additional_tables}

\begin{table}[h]
    \begin{center}
        \caption{Regime Variables and Event-Specific Periods of Assessment.}
    \label{tab:timeline}
\adjustbox{max width=\textwidth}{%
    \begin{tabular}{l c ccc c ccc}
    \toprule
    \toprule
          && \multicolumn{3}{c}{Regime} && \multicolumn{3}{c}{Model Coverage}\\
    Event && Start && End\\
       \midrule
       Pandemic               && 16.03.2020  &--& 30.06.2023 && 01.01.2019 & -- & 30.06.2022\tnote{a}  \\
       Bank Lending Standards && 13.12.2021  & -- 01.08.2022 -- & 30.06.2025 && 01.01.2021 & -- & 31.03.2023 \\
       
       Cost of Living Crisis  && 27.07.2022 & -- 21.12.2022 --           & 12.06.2024   && 01.01.2021 & -- & 31.03.2023 \\
       \bottomrule
       \bottomrule
    \end{tabular}
    }
    \end{center}
\vspace{-0.5cm}
 \begin{footnotesize} \singlespacing
 \emph{Notes:} The table reports the timing of events. The model coverage ensures a sufficiently long pre-regime period and ends with Q1/2023. Due to a post-COVID normalisation, we restrict the model coverage for the pandemic to 30 June 2022. As some variables were only available during the initial phase of the pandemic, models including workplace mobility are restricted to the period 01.01.2020 - 30.06.2022 and models including COVID-19 mortality rates to the period 09.03.2020 - 30.06.2020.
   \end{footnotesize}   
\end{table}

\begin{table}
\begin{center}
\caption{Lockdown Periods}
\label{tab:lockdown_periods}
\resizebox{\linewidth}{!}{
\begin{tabular}{l c rcl c c} 
\toprule
\toprule
1st Lockdown && 16 March 2020 &--& 13 April 2020 && national\\
& \\
2nd Lockdown && 17 November 2020 &--& 6 December 2020 && national\\
& \\
3rd Lockdown &&  26 December 2020 &--& 7 February 2021 && national \\
& \\
Regional Lockdown I   && 1 April 2021     &--& 18 April 2021 && Burgenland  \\
& \\
Regional Lockdown II  && 1 April 2021  &--& 2 May 2021 && Vienna \& Lower Austria \\
& \\
4th Lockdown && 22 November 2021 &--& 11 December 2021 && national\\
\bottomrule
\bottomrule
\end{tabular}
}
\end{center}
\begin{scriptsize}
\emph{Notes:} The table summarises the timing of legally binding lockdown periods by geographic scope.
\end{scriptsize}
\end{table}

\begin{table}
\begin{center}
\caption{$(A)$ and $(A^B)$: Pandemic Immediate Price Effects}
\label{tab:reduced_mob_prices}
\resizebox{1.01\linewidth}{!}{
\begin{tabular}{l ccccc c ccccc}
\toprule
\toprule
 & $(A)$ & $(A)$ & $(A)$ & $(A)$ & $(A)$ && $(A^B)$ & $(A^B)$ & $(A^B)$ & $(A^B)$ & $(A^B)$ \\
 & $(1)$ & $(2)$ & $(3)$ & $(4)$ & $(5)$ && $(6)$ & $(7)$ & $(8)$ & $(9)$ & $(10)$\\
\cmidrule{2-6} \cmidrule{8-12} 
COVID $\times$ Lockdown (Ref.: Pre-COVID)                                       & $0.065^{***}$  &                &                &                &                && $0.080^{***}$  &                &                &                &                \\
                                         & $(0.004)$      &                &                &                &                && $(0.005)$      &                &                &                &                \\
COVID $\times$ No Lockdown                                       & $0.109^{***}$  &                &                &                &                && $0.115^{***}$  &                &                &                &                \\
                                         & $(0.002)$      &                &                &                &                && $(0.003)$      &                &                &                &                \\
Reduced Workplace Mobility                   &                & $-0.075^{***}$ &                &                &                &&                & $-0.045^{**}$  &                &                &                \\
                                                      &                & $(0.012)$      &                &                &                &&                & $(0.016)$      &                &                &                \\
Reduced Workplace Mobility $\times$ Lockdown   &                &                & $-0.075^{***}$ &                &                &&                &                & $-0.046^{**}$  &                &                \\
                                                      &                &                & $(0.012)$      &                &                &&                &                & $(0.016)$      &                &                \\
Reduced Workplace Mobility $\times$ No Lockdown &                &                & $-0.071^{***}$ &                &                &&                &                & $-0.032$       &                &                \\
                                                      &                &                & $(0.015)$      &                &                &&                &                & $(0.019)$      &                &                \\
Mortality                                 &                &                &                & $0.018$        &                &&                &                &                & $0.041$        &                \\
                                                      &                &                &                & $(0.033)$      &                &&                &                &                & $(0.030)$      &                \\
Mortality $\times$ Lockdown               &                &                &                &                & $-0.038$       &&                &                &                &                & $-0.009$       \\
                                                      &                &                &                &                & $(0.043)$      &&                &                &                &                & $(0.043)$      \\
Mortality $\times$ No Lockdown              &                &                &                &                & $0.079$        &&                &                &                &                & $0.076^{*}$    \\
                                                      &                &                &                &                & $(0.044)$      &&                &                &                &                & $(0.037)$      \\
\midrule
Housing Characteristics                          & \checkmark     & \checkmark         & \checkmark        & \checkmark     &     \checkmark         && \checkmark    & \checkmark     & \checkmark         & \checkmark        & \checkmark      \\
Time Dummies                          & \xmark     & \checkmark         & \checkmark        & \checkmark     &     \checkmark         && \xmark    & \checkmark     & \checkmark         & \checkmark        & \checkmark          \\
Location Fixed Effects                          & \checkmark     & \checkmark         & \checkmark        & \checkmark     &     \checkmark         && \checkmark    & \checkmark     & \checkmark         & \checkmark        & \checkmark          \\
Location Random Effects                          & \checkmark     & \checkmark         & \checkmark        & \checkmark     &     \checkmark         && \checkmark    & \checkmark     & \checkmark         & \checkmark        & \checkmark          \\
\midrule
BIC                                                   & $8,114$     & $4,162$     & $4,180$     & $4,198$     & $4,208$     && $5,485$     & $4,000$     & $4,015$     & $4,004$     & $4,016$     \\
Num. obs.                                             & $58,181$        & $37,726$        & $37,726$        & $37,726$        & $37,726$        && $34,548$        & $24,138$        & $24,138$        & $24,138$        & $24,138$        \\
\bottomrule
\bottomrule
\multicolumn{11}{l}{\footnotesize{$^{***}p<0.001$; $^{**}p<0.01$; $^{*}p<0.05$}. Mobility and mortality data have been normalised between 0 and 1.}
\end{tabular}
}
\end{center}
\end{table}  

\begin{table}
\begin{center}
\caption{Bank Lending Standards: Price Effects}
\label{tab:lending_price_effects}
\renewcommand{\arraystretch}{1}
\adjustbox{width=0.9\textwidth}{%
\begin{tabular}{l cc  c  cc}
\toprule
\toprule
 & \multicolumn{2}{c}{$(A)$} && \multicolumn{2}{c}{$(A^B)$}\\
  & $(1)$ & $(2)$  && $(3)$ & $(4)$ \\
\cmidrule{2-3} \cmidrule{5-6}
KIM-VO (pre-announcement)               & $-0.062^{***}$  &                  && $-0.056^{***}$   &                                 \\
                                        & $(0.003)$       &                  && $(0.004)$        &                \\
\emph{Ref.:KIM-VO (announcement)}       &                 &                  &&                  &                \\
                                        &                 &                  &&                  &                 \\
KIM-VO (enactment)                      & $-0.020^{***}$  &                  && $0.003$          &                \\
                                        & $(0.004)$       &                  && $(0.005)$        &                \\
Q1 2021 (Pre-KIM-VO)                    &                 & $-0.093^{***}$   &&                  & $-0.148^{***}$   \\
                                        &                 & $(0.010)$        &&                  & $(0.011)$        \\
Q2 2021 (Pre-KIM-VO)                    &                 & $-0.066^{***}$   &&                  & $-0.099^{***}$   \\
                                        &                 & $(0.010)$        &&                  & $(0.011)$        \\
Q3 2021 (Pre-KIM-VO)                    &                 & $-0.046^{***}$   &&                  & $-0.087^{***}$   \\
                                        &                 & $(0.010)$        &&                  & $(0.011)$        \\
Q4 2021 (Pre-KIM-VO)                    &                 & $-0.047^{***}$   &&                  & $-0.093^{***}$   \\
                                        &                 & $(0.010)$        &&                  & $(0.012)$        \\
Q4 2021 (KIM-VO announcement)           &                 & $-0.048^{***}$   &&                  & $-0.092^{***}$   \\
                                        &                 & $(0.014)$        &&                  & $(0.015)$        \\
Q1 2022 (KIM-VO announcement)           &                 & $-0.009$         &&                  & $-0.069^{***}$ \\
                                        &                 & $(0.010)$        &&                  & $(0.011)$       \\
Q2 2022 (KIM-VO announcement)           &                 & $0.013$          &&                  & $-0.036^{**}$  \\
                                        &                 & $(0.010)$        &&                  & $(0.011)$       \\
\emph{Ref.: Q3 2022 (KIM-VO announcement)} &              &                  &&                  &                 \\
                                        &                 &                  &&                  &                 \\
Q3 2022 (KIM-VO enactment)              &                 & $-0.016$         &&                  & $-0.028^{*}$    \\
                                        &                 & $(0.011)$        &&                  & $(0.012)$     \\
Q4 2022 (KIM-VO enactment)              &                 & $-0.019$         &&                  & $-0.050^{***}$   \\
                                        &                 & $(0.010)$        &&                  & $(0.012)$    \\
Q1 2023 (KIM-VO enactment)              &                 & $-0.025^{**}$    &&                  & $-0.064^{***}$  \\
                                        &                 &  $(0.010)$       &&                  & $(0.012)$    \\
\midrule
Housing Characteristics                 & \checkmark & \checkmark  && \checkmark & \checkmark     \\
Time Dummies                            & \xmark & \checkmark && \xmark & \checkmark       \\
Location Fixed Effects                  & \checkmark & \checkmark && \checkmark & \checkmark   \\
Location Random Effects                 & \checkmark & \checkmark && \checkmark & \checkmark \\
\midrule
BIC                                     & $4,503$     & $4,524$    &   & $4,830$     & $4,807$   \\
Num. obs.                               & $32,308$        & $32,308$       &  & $22,080$        & $22,080$       \\
\bottomrule
\bottomrule
\end{tabular}
}
\end{center}
\begin{footnotesize}
    \emph{Notes:} The quarterly time index is split at the announcement of the new bank lending standards on 13 December 2021 and at the enactment on 1 August 2022. The base level of the index variable is set to Q3 2022 (KIM-VO announcement), i.e., the period from 01 July 2022 to 31 July 2022 before the KIM-VO was enacted.\end{footnotesize}
\end{table}

\begin{table}
\begin{center}
\caption{Cost of Living Crisis: Price Effects $(A)$ and $(A^B)$}
\label{tab:CoL_price_effects}
\renewcommand{\arraystretch}{1}
\adjustbox{width=0.9\textwidth}{%
\begin{tabular}{l c c c c c }
\toprule
\toprule
&  $(A)$ & $(A)$ &&   $(A^B)$ & $(A^B)$\\\
 & $(1)$ & $(2)$    && $(3)$ & $(4)$ \\
\cmidrule{2-3} \cmidrule{5-6}
Q1 2021 $\times$ HICP                             &                $-0.013$       &             &&                $0.010$        &                \\
                                               &                $(0.007)$      &                &&                 $(0.009)$      &                \\
Q2 2021 $\times$ HICP                             &               $0.003$        &             &&                $0.023^{***}$  &                \\
                                                               & $(0.004)$      &                &&                 $(0.005)$      &                \\
Q3 2021 $\times$ HICP                             &                $0.009^{*}$    &             &&                 $0.023^{***}$  &                \\
                                                        & $(0.004)$      &                &&                 $(0.005)$      &                \\
Q4 2021 $\times$ HICP                             &                $0.006^{*}$    &             &&                $0.017^{***}$  &                \\
                                                              & $(0.003)$      &                &&                $(0.004)$      &                \\
Q1 2022 $\times$ HICP                                           & $0.012^{***}$  &             &&                $0.016^{***}$  &                \\
                                               &                 $(0.002)$      &                &&                 $(0.003)$      &                \\
Q2 2022 $\times$ HICP                             &                 $0.011^{***}$  &             &&                $0.016^{***}$  &                \\
                                               &                $(0.001)$      &                &&                 $(0.002)$      &                \\
Q3 2022 $\times$ HICP                                           & $0.006^{***}$  &             &&                 $0.014^{***}$  &                \\
                                                             & $(0.001)$      &                &&                 $(0.001)$      &                \\
Q4 2022 $\times$ HICP                                            & $0.005^{***}$  &             &&                 $0.010^{***}$  &                \\
                                                              & $(0.001)$      &                &&                 $(0.001)$      &                \\
Q1 2023 $\times$ HICP                             &                $0.004^{***}$  &             &&                $0.009^{***}$  &                \\
                                                              & $(0.001)$      &                &&                $(0.001)$      &                \\
Mortgage Rate $\times$ Low policy rate environment                           &  & $0.106^{***}$   &&                               &                \\
                                               &                                & $(0.010)$      &&                               &                \\
Mortgage Rate $\times$ First policy rate increase                      &       & $0.062^{***}$   &&                                &                \\
                                               &                                & $(0.005)$      &&                                &                \\
Mortgage Rate $\times$ High policy rate environment           &                 & $0.043^{***}$        &&                                &                \\
                                               &                &                 $(0.004)$      &&                                &                \\
Mortgage Rate (3m lag) $\times$ Low policy rate environment              &                &                &&      &                $-0.105^{***}$ \\
                                               &                                &                &&                               & $(0.025)$      \\
Mortgage Rate (3m lag) $\times$ First policy rate increase &                &                                &&    &                 $-0.047^{**}$ \\
                                               &                &                                &&                                & $(0.017)$      \\
Mortgage Rate (3m lag) $\times$ High policy rate environment                 &                &   &&                             & $-0.038^{***}$ \\
                                               &                                &                &&                               & $(0.011)$      \\
\midrule
Housing Characteristics                                  & \checkmark     &   \checkmark     &       & \checkmark  & \checkmark    \\ 
Time Dummies                                  & \xmark    & \xmark   &&  \xmark  & \xmark   \\
Location Fixed Effects & \checkmark  & \checkmark  &         & \checkmark  & \checkmark   \\
Location Random Effects                        &  \checkmark    & \checkmark  & &  \checkmark   & \checkmark   \\
\midrule
BIC                                            &  $4,541$     & $4,727$     &&  $4,827$     & $4,998$    \\
Num. obs.                                      &  $32,308$        & $32,308$        &&  $22,080$        & $22,080$       \\
\bottomrule \bottomrule
\multicolumn{3}{l}{\scriptsize{$^{***}p<0.001$; $^{**}p<0.01$; $^{*}p<0.05$}}
\end{tabular}
}
\end{center}
\vspace{-0.4cm}
{\scriptsize
\emph{Notes:} The table reports coefficients associated with time-dummies interacted with the Austrian HICP (1) and (3). Further, models (2) and (4) include average newly granted mortgage rates per sub-regime. Due to the delay between fixing an interest rate and recording a transaction, a lag of three months is respected for $(D)$. Results are robust to changes of 1-2 months in the lag length. 
}
\end{table}

\begin{figure}[h]
\begin{center}
\captionsetup{font=large}
  \caption{Bank Lending Standards and Cost of Living Crisis: Gradual Quantity Effects $(D)$ and $(A)$} \label{fig:gradual_counts_II}
  \begin{subfigure}[t]{0.45\textwidth}
  \caption{$(D)$: KIM-VO regimes}
  \label{fig:quantity_gradual_deeds_top10_KIM}
\includegraphics[width=\textwidth]{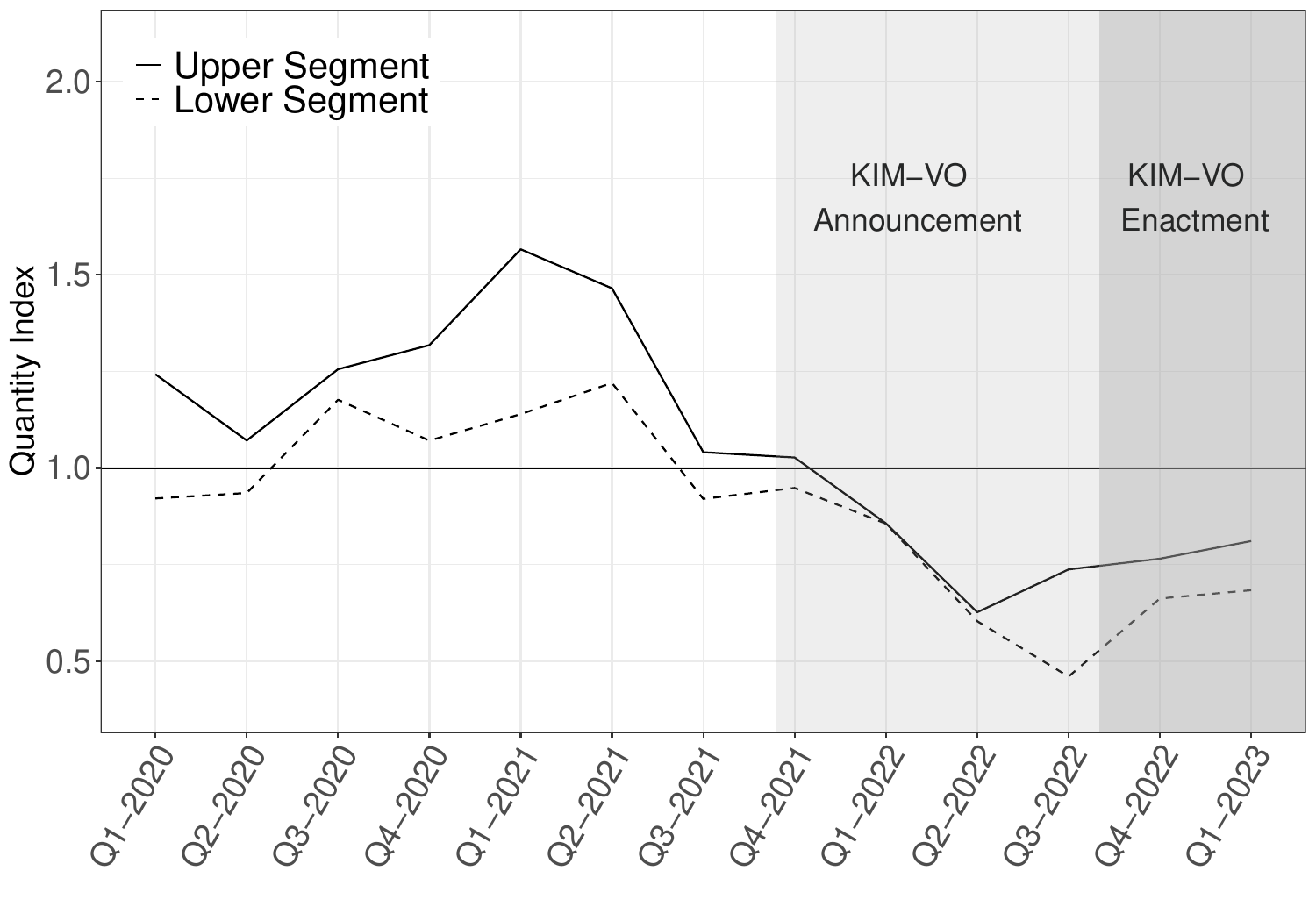}
  \end{subfigure}
  \hfill
\begin{subfigure}[t]{0.45\textwidth}
\caption{$(A)$: KIM-VO regimes}
\label{fig:quantity_gradual_adverts_top10_KIM}
\includegraphics[width=\textwidth]{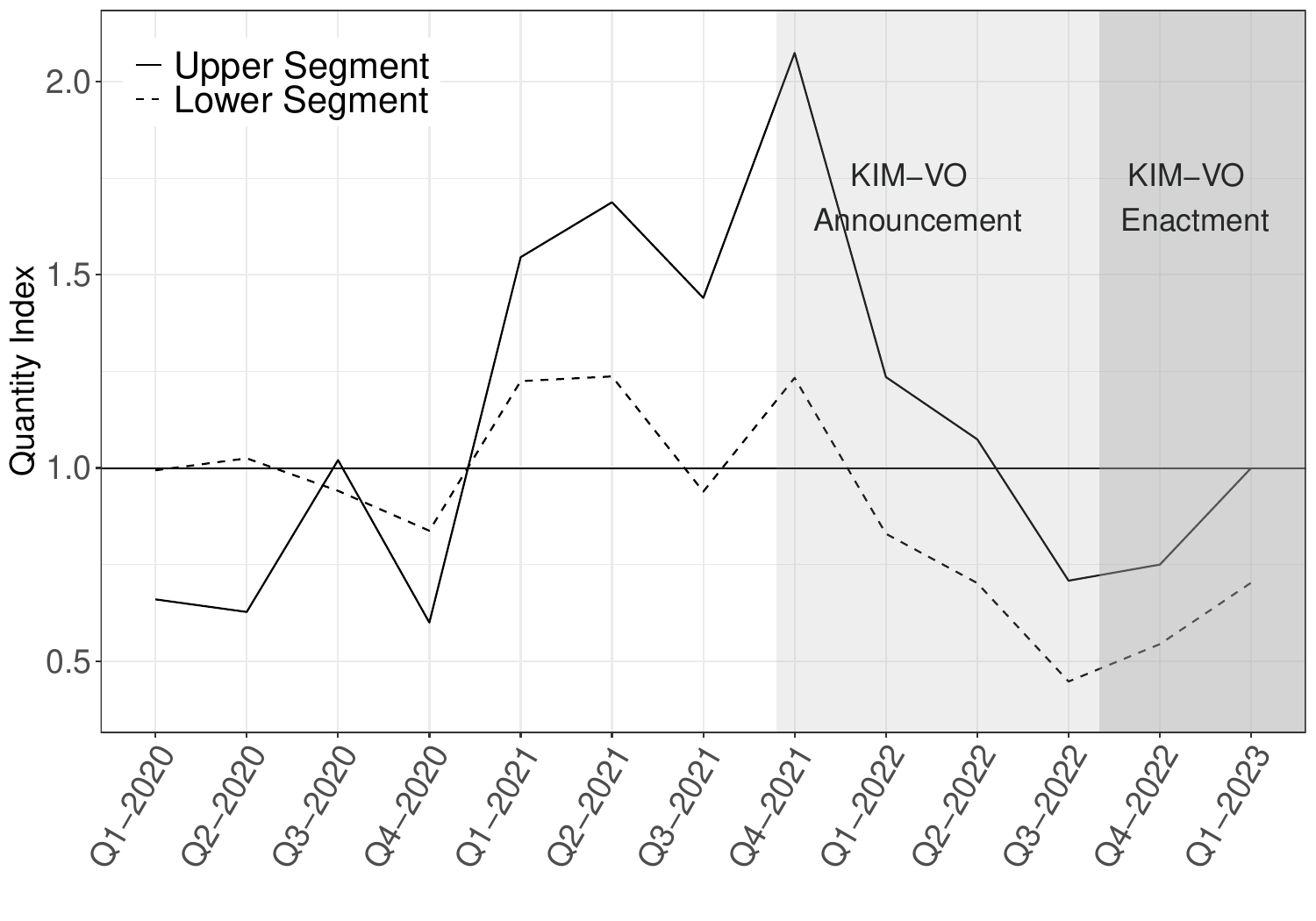}
\end{subfigure}

  \begin{subfigure}[t]{0.45\textwidth}
  \caption{$(D)$: Interest rate regimes}
  \label{fig:quantity_gradual_deeds_top10_COL}
\includegraphics[width=\textwidth]{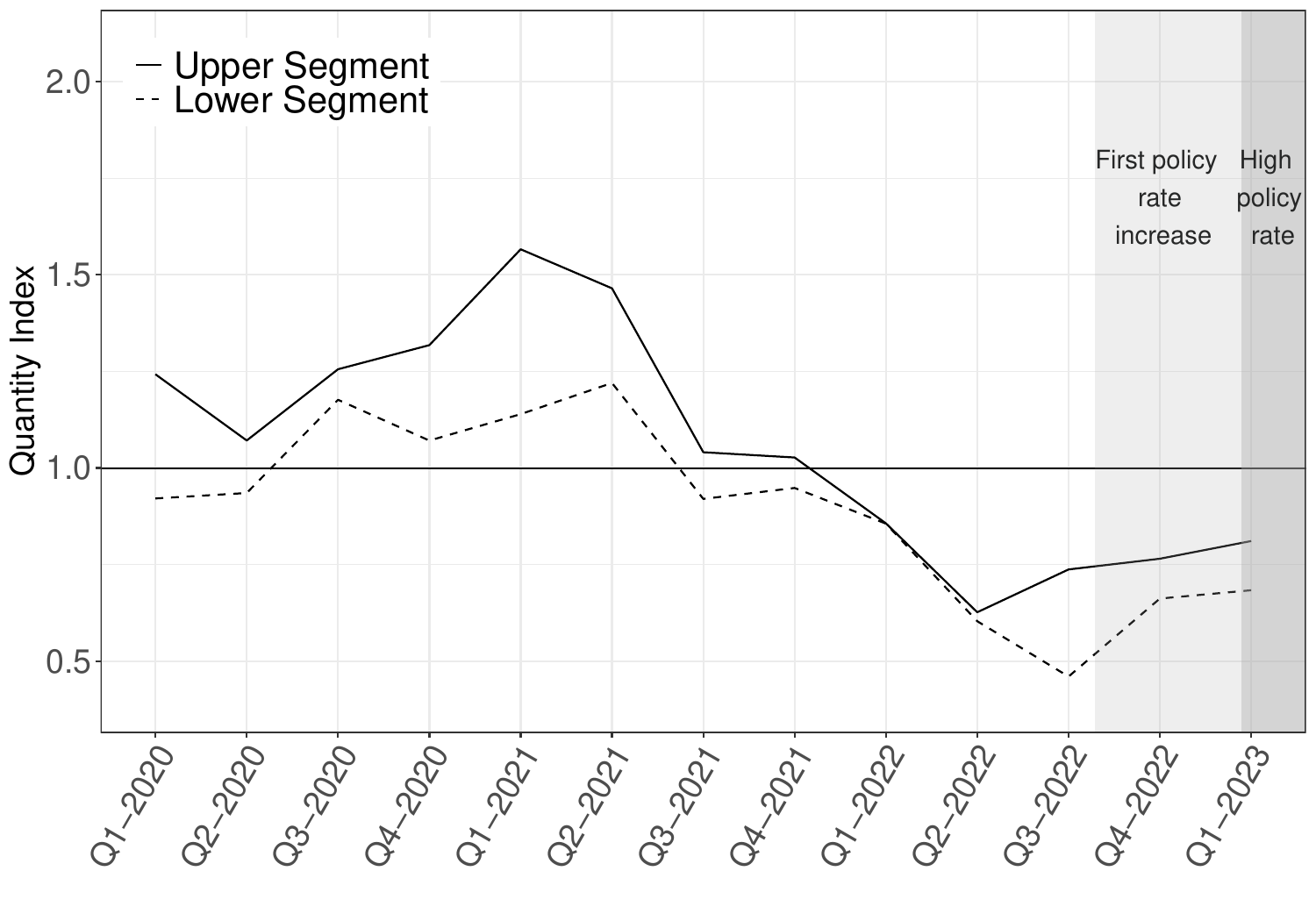}
  \end{subfigure}
  \hfill
\begin{subfigure}[t]{0.45\textwidth}
\caption{$(A)$: Interest rate regimes}
\label{fig:quantity_gradual_adverts_top10_COL}
\includegraphics[width=\textwidth]{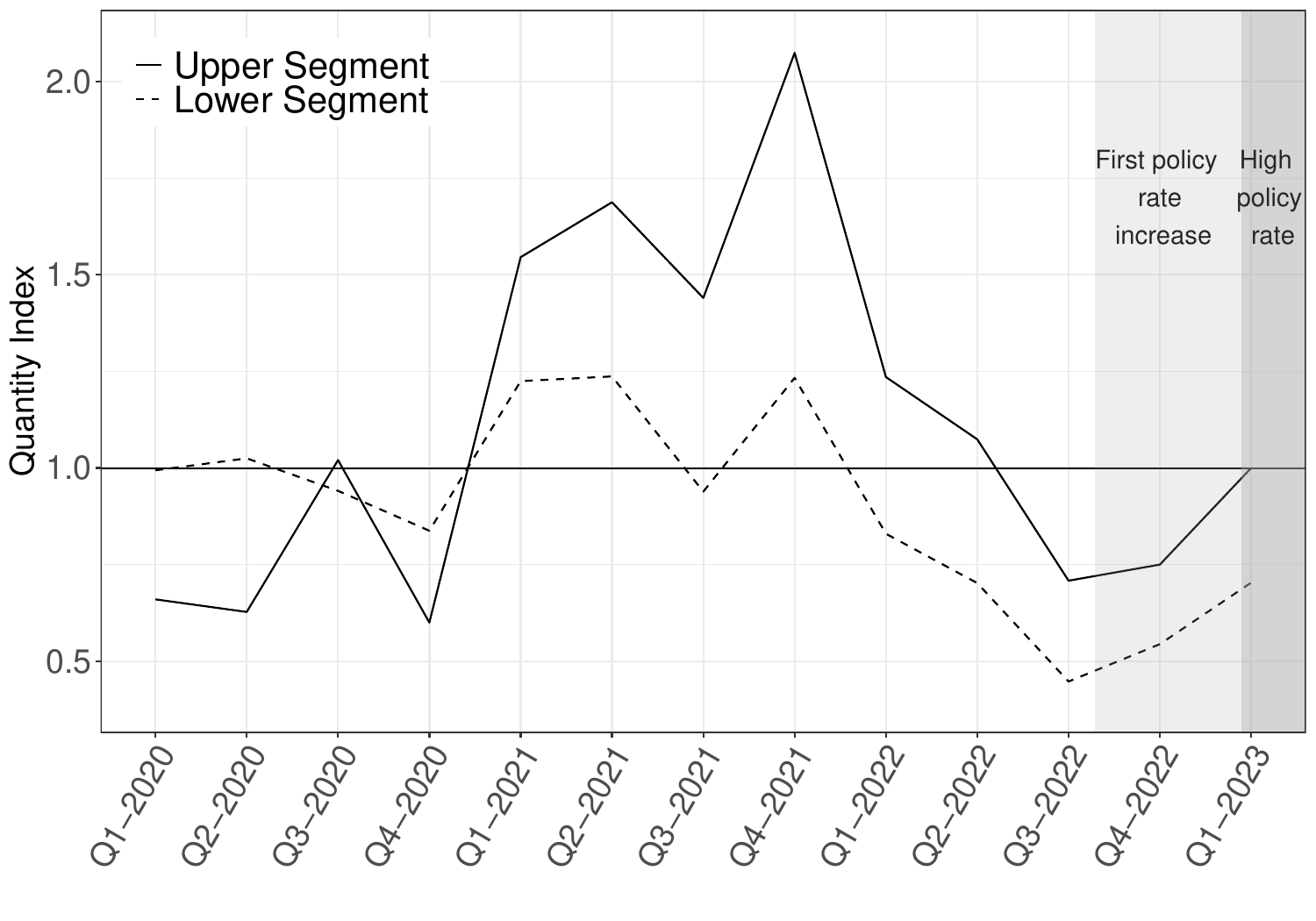}
\end{subfigure}

  \begin{subfigure}[t]{0.45\textwidth}
  \caption{$(D)$: Interest rate regimes}
  \label{fig:quantity_gradual_deeds_all_COL}
\includegraphics[width=\textwidth]{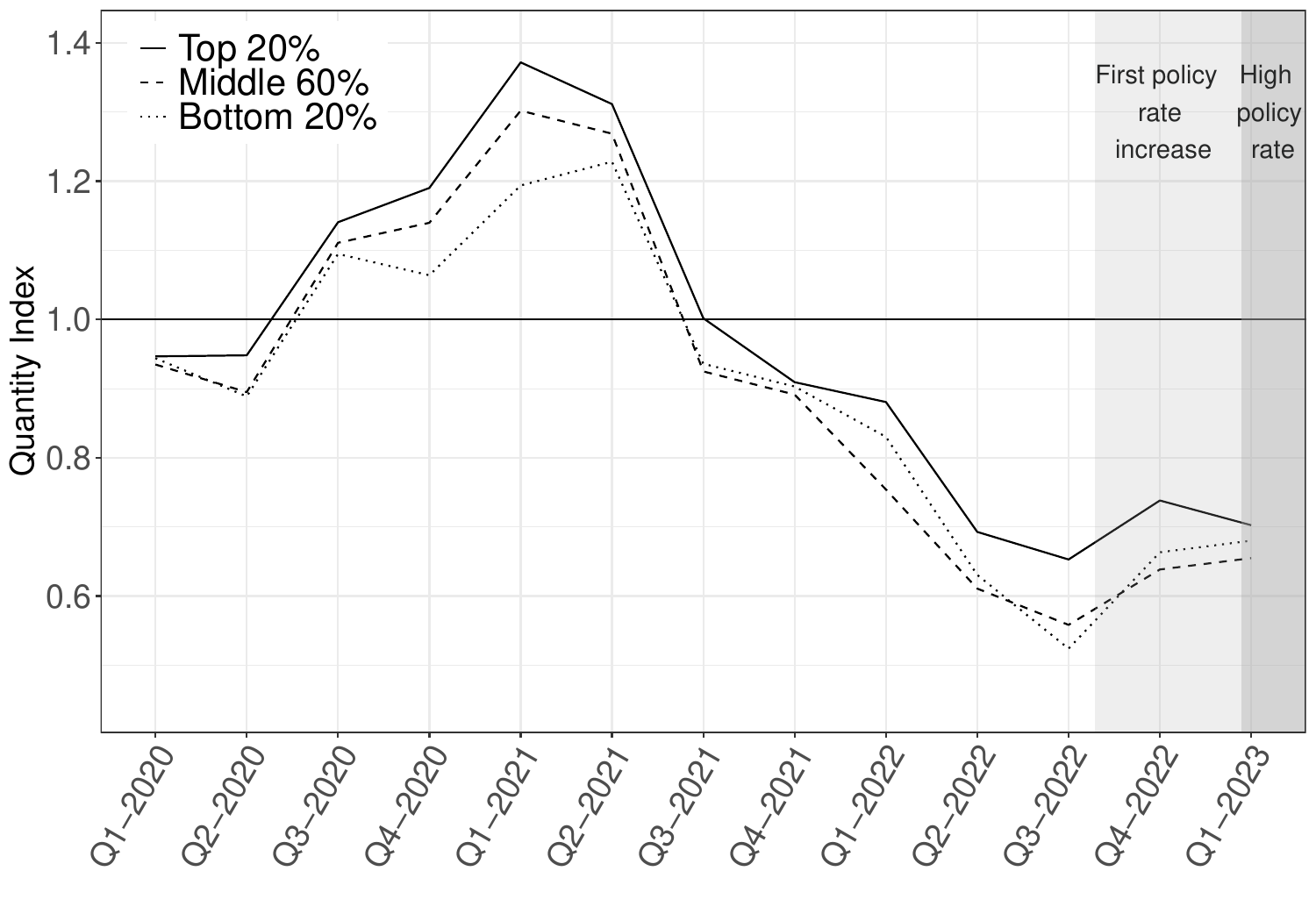}
  \end{subfigure}
  \hfill
\begin{subfigure}[t]{0.45\textwidth}
\caption{$(A)$: Interest rate regimes}
\label{fig:quantity_gradual_adverts_all_COL}
\includegraphics[width=\textwidth]{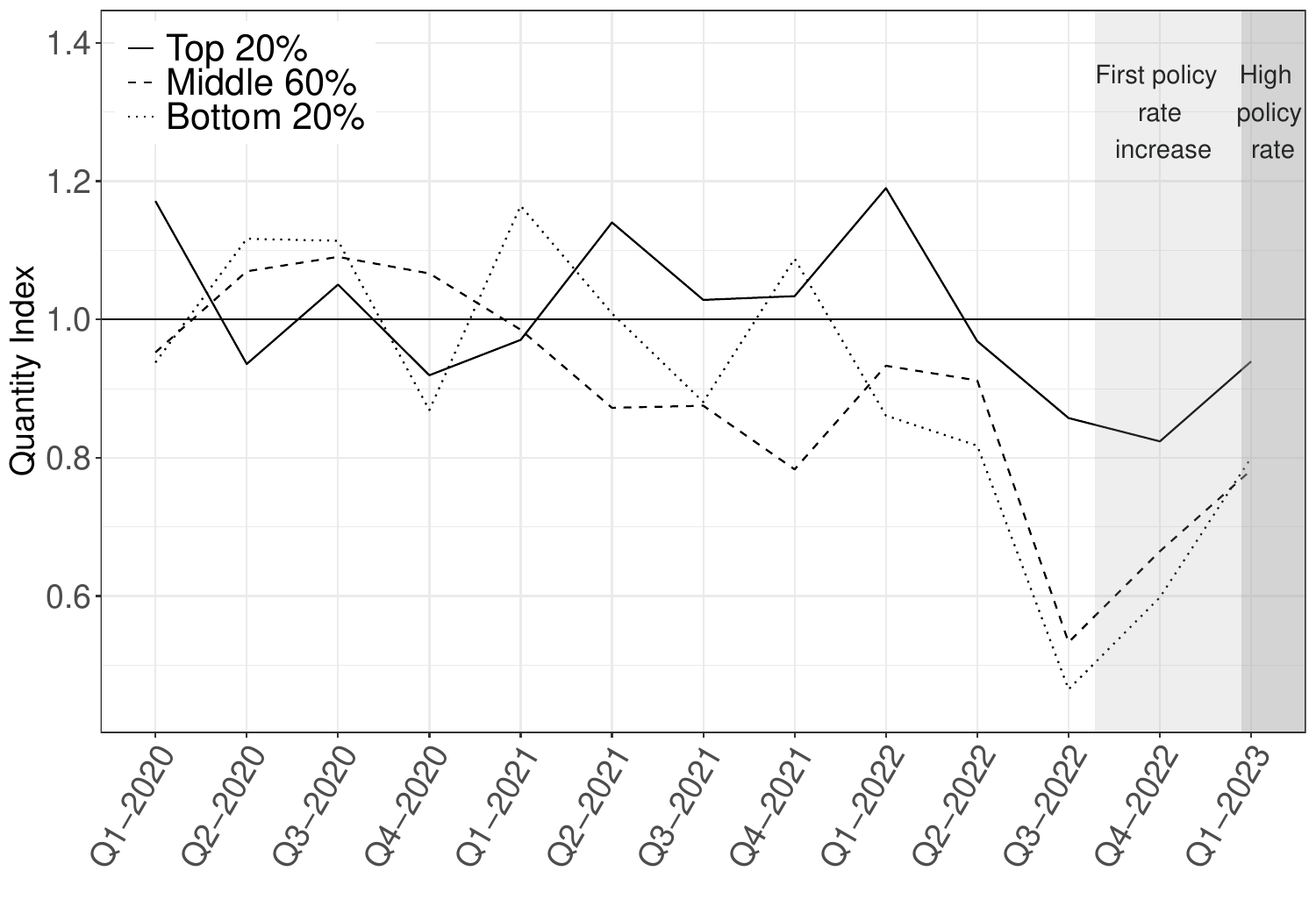}
\end{subfigure}
   \end{center}
\begin{scriptsize}
\emph{Notes:} The Figure shows gradually evolving quantity effects by market segments as quarterly count indices expressing changes relative to the same quarter in the preceding year. Segments are created using the 10 most and least expensive districts for $(D)$ and $(A^B)$, respectively  (\autoref{fig:quantity_gradual_deeds_top10_KIM}, \autoref{fig:quantity_gradual_adverts_top10_KIM}, \autoref{fig:quantity_gradual_deeds_top10_COL}, \autoref{fig:quantity_gradual_adverts_top10_COL}) and the full distribution (\autoref{fig:quantity_gradual_deeds_all_COL}, \autoref{fig:quantity_gradual_adverts_all_COL}) complementing \autoref{fig:KIM_gradual_counts}.
\end{scriptsize}
\end{figure}



\end{document}